\documentclass[11pt, a4paper, oneside]{Thesis} 

\graphicspath{{Pictures/}} 

\usepackage[square,sort,compress,numbers]{natbib}
\usepackage{cite}
\usepackage{amsmath,latexsym} 
\usepackage{float}
\usepackage[bottom]{footmisc}
\renewcommand{\thefootnote}{\arabic{footnote}}
\newtheorem*{theorem*}{\bf{Theorem}}
\usepackage{mathtools}
\usepackage{tasks}
\usepackage{acronym}
\usepackage{nomencl}
\usepackage{caption}
\usepackage{subfig,graphicx}
\usepackage{blindtext} 
\usepackage{epigraph}
\usepackage{dcolumn}

\title{\ttitle} 
\DeclareUnicodeCharacter{2212}{-}
\DeclareUnicodeCharacter{2212}{+}
\newcommand\blfootnote[1]{%
     \begingroup
     \renewcommand\thefootnote{}\footnote{#1}%
     \addtocounter{footnote}{-1}%
      \endgroup
    }
\begin{document}

\setstretch{1.7} 

\fancyhead{} 
\rhead{\thepage} 
\lhead{} 

%

\thesistitle{Accelerated Expansion of the Universe in Nonmetricity-based Modified Gravity}
\documenttype{\textbf{THESIS}}
\supervisor{\textbf{Prof. Pradyumn Kumar Sahoo}}
\supervisorposition{\textbf{Professor}}
\supervisorinstitute{\textbf{BITS-Pilani, Hyderabad Campus}}
\examiner{}
\degree{Ph.D. Research Scholar}
\coursecode{\textbf{DOCTOR OF PHILOSOPHY}}
\coursename{Thesis}
\authors{\textbf{GAURAV NARAYANRAO GADBAIL}}
\IDNumber{\textbf{2020PHXF0401H}}
\addresses{}
\subject{}
\university{\texorpdfstring{\href{http://www.bits-pilani.ac.in/} 
                {Birla Institute of Technology and Science, Pilani}} 
                {Birla Institute of Technology and Science, Pilani}}
\UNIVERSITY{\texorpdfstring{\href{http://www.bits-pilani.ac.in/} 
                {\textbf{BIRLA INSTITUTE OF TECHNOLOGY AND SCIENCE, PILANI}}} 
                {BIRLA INSTITUTE OF TECHNOLOGY AND SCIENCE, PILANI}}



\department{\texorpdfstring{\href{http://www.bits-pilani.ac.in/pilani/Mathematics/Mathematics} 
                {Mathematics}} 
                {Mathematics}}
\DEPARTMENT{\texorpdfstring{\href{http://www.bits-pilani.ac.in/pilani/Mathematics/Mathematics} 
                {Mathematics}} 
                {Mathematics}}
\group{\texorpdfstring{\href{Research Group Web Site URL Here (include http://)}
                {Research Group Name}} 
                {Research Group Name}}
\GROUP{\texorpdfstring{\href{Research Group Web Site URL Here (include http://)}
                {RESEARCH GROUP NAME (IN BLOCK CAPITALS)}}
                {RESEARCH GROUP NAME (IN BLOCK CAPITALS)}}
\faculty{\texorpdfstring{\href{Faculty Web Site URL Here (include http://)}
                {Faculty Name}}
                {Faculty Name}}
\FACULTY{\texorpdfstring{\href{Faculty Web Site URL Here (include http://)}
                {FACULTY NAME (IN BLOCK CAPITALS)}}
                {FACULTY NAME (IN BLOCK CAPITALS)}}

\maketitle

\clearpage
\setstretch{1.3} 

\pagestyle{empty} 
\pagenumbering{gobble}


\addtocontents{toc}{\vspace{2em}} 

\frontmatter 
\Certificate
\Declaration

\begin{acknowledgements}

Without the continuous guidance and support of numerous individuals, the journey towards completing my PhD would not have been possible. I would like to express my heartfelt gratitude to each of them.

I would like to express my sincere thanks and gratitude to my supervisor, \textbf{Prof. Pradyumn Kumar Sahoo}, Professor, Department of Mathematics, BITS-Pilani, Hyderabad Campus, Hyderabad, for his unwavering support, guidance, and vast experience throughout my Ph.D. work. I am grateful to him for his endless patience and faith in me, which encouraged me to complete this thesis by developing a deeper understanding of the subject and an aptitude for scientific research. 

I sincerely thank my Doctoral Advisory Committee (DAC) members, \textbf{Prof. Bivudutta Mishra} and \textbf{Prof. K. Venkata Ratnam} for their guidance and valuable suggestions to improve my research work. 

I am privileged to extend my gratitude to the \textbf{Head} of the Mathematics Department, the \textbf{DRC} convener, \textbf{faculty} members, and my \textbf{colleagues} for their help, support, and encouragement in this amazing journey. I convey special thanks to my co-authors for their valuable discussions, suggestions, and collaborations. 
 
I would like to express my sincere gratitude to \textbf{BITS-Pilani, Hyderabad Campus} for providing the essential facilities for my research. I am also deeply thankful to the \textbf{University Grants Commission (UGC)}, New Delhi, India, for awarding me the Junior and Senior Research Fellowship (UGC-Ref. No.: 201610122060), which has been instrumental in supporting my research endeavors.

Finally, I thank my \textbf{parents} and \textbf{grandparents} for their unconditional love, support, and encouragement and for always being with me. 

Last but not the least, my heartiest thanks to my friends \textbf{Sanjay}, \textbf{Simran}, \textbf{Amit}, \textbf{Zinnat}, \textbf{Raja}, and \textbf{Prayas} for their wholehearted support over the years.  

\vspace{1.4 cm}
Gaurav N. Gadbail,\\
ID: 2020PHXF0401H

\end{acknowledgements}

\begin{abstract} 

In this thesis, we investigate the cosmological implications of modified gravity theory, specifically nonmetricity-based gravity such as $f(Q)$ gravity, as an alternative to the standard $\Lambda$ cold dark matter ($\Lambda$CDM) model to explain the accelerated expansion of the universe. Chapter \ref{Chapter1} introduces the foundational concepts required to explore nonmetricity-based modified gravity theories in cosmology. It outlines the mathematical preliminaries and provides a detailed review of General Relativity (GR) and the $\Lambda$CDM model. Also, it discusses the shortcomings of the $\Lambda$CDM model. This chapter establishes the groundwork for investigating 
$f(Q)$ gravity as a promising alternative to address the challenges faced by GR and the 
$\Lambda$CDM model.\\
 In chapter \ref{Chapter2}, we construct explicit cosmological reconstructions for $f(Q)$ gravity within the background of the Friedmann-Lemaître-Robertson-Walker (FLRW) evolution history. We derive generalized functions of the nonmetricity scalar $Q$ that reproduce the exact $\Lambda$CDM expansion history. Further, we reconstruct the FLRW cosmology in terms of the e-folding parameter, demonstrating that specific $f(Q)$ models can mimic well-known cosmological evolutions including the $\Lambda$CDM model. However, chapter \ref{Chapter3} addresses the limitations arising from arbitrary functional forms of the $f(Q)$ Lagrangian. To resolve this, we employ a model-independent Gaussian process reconstruction using direct Hubble measurements. This approach allows us to reconstruct the Hubble parameter \( H(z) \) and its derivatives, leading to a data-driven reconstruction of the $f(Q)$ region. Motivated by the quadratic behavior, we propose a new parametrization \( f(Q) = -2\Lambda + \epsilon Q^2 \), where \(\epsilon\) quantifies deviations from $\Lambda$CDM. Additionally, power-law and exponential $f(Q)$ models are tested against the reconstructed region, refining parameter constraints.\\
In chapter \ref{Chapter4}, we incorporate a quintessence scalar field into the $f(Q)$ gravity framework to address both inflation and late-time acceleration. Using the GP with the OHD data, we reconstruct the scalar field potential \( V(\phi) \) in a model-independent manner. This approach allows us to obtain a suitable quintessence scalar field model that aligns with the OHD under the framework of power-law $f(Q)$ gravity. Our analysis shows that while early DE has minimal influence on the present accelerated expansion, the reconstructed quintessence models offer new insights into the dynamics of cosmic acceleration under power-law $f(Q)$ gravity. Chapter \ref{Chapter5} examines the interacting dark energy and dark matter within the power-law $f(Q)$ model using dynamical system analysis. Two interaction forms are examined, and the evolution of cosmological parameters, including \(\Omega_m\), \(\Omega_r\), \(\Omega_{DE}\), \(q\), and \(\omega\) are studied for different values of the model parameter and the interaction parameter. Stability analysis identifies fixed points corresponding to a de Sitter and quintessence-like solutions, influenced by model parameters and interaction terms. Finally, Chapter \ref{Chapter6} presents concluding remarks and outlines future directions.
\end{abstract}

\Dedicatory{\bf \begin{LARGE}
Dedicated to
\end{LARGE} 
\\
\vspace{0.2cm}
\it My Family Members\\}



\lhead{\emph{Contents}} 
\tableofcontents 
\addtocontents{toc}{\vspace{1em}}
\lhead{\emph{List of Tables}}
\listoftables 
\addtocontents{toc}{\vspace{1em}}
\lhead{\emph{List of Figures}}
\listoffigures 
\addtocontents{toc}{\vspace{1em}}




\lhead{\emph{List of Symbols and Abbreviations}}
\listofsymbols{ll}{
\begin{tabular}{lcl}
$\mathbb{R}^{n}$ &:& n-dimensional real space\\
$\otimes$ &:& Tensor product\\
$g_{\mu\nu}$ &:& Lorentzian metric\\
$g$ &:& Determinant of $g_{\mu\nu}$\\
$\Gamma^{\lambda}_{\,\,\,\mu\nu}$ &:& General affine connection\\
$\{^{\,\lambda}_{\,\,\,\mu\nu}\}$&:& Levi-Civita connection\\
$\nabla_{i}$&:& Covariant derivative \\
$R^{\sigma}_{\,\,\lambda \mu\nu}$ &:& Riemann tensor \\
$R_{\mu\nu}$&:& Ricci tensor \\
$R$ &:& Ricci scalar \\
$z$&:& Redshift\\
$q$&:& Deceleration parameter\\
$\mathcal{T}$ &:& Torsion\\
$Q$ &:& Nonmetricity\\
$\Lambda$ &:& Cosmological constant\\
$T_{\mu\nu}$ &:& Stress-energy tensor\\
$G$ &:& Newton's gravitational constant\\
$\Omega$ &:& Density parameter\\
GR &:& General Relativity\\
$\Lambda$CDM &:& $\Lambda$ Cold Dark Matter\\
EoS &:& Equation of State\\
OHD &:&  Observational Hubble Data\\
CC &:& Cosmic Chronometer \\
BAO &:& Baryon Acoustic Oscillations\\
DE &:& Dark Energy\\
DM &:& Dark Matter\\
TEGR &:& Teleparallel Equivalent to GR\\
STEGR &:& Symmetric teleparallel Equivalent to GR\\
GP &:& Gaussian Process
\end{tabular}
}

\addtocontents{toc}{\vspace{2em}}

%
%


\clearpage 





\mainmatter 

\pagestyle{fancy} 


 \chapter{Introduction} 
\label{Chapter1}
\epigraph{\justifying \textit{``No one will be able to read the great book of the Universe if he does not understand its language which is that of mathematics."}}{\textit{Galileo Galilei}}

\lhead{Chapter 1. \emph{Introduction}} 
Cosmology is interdisciplinary science that explores the origin, evolution, current state, and ultimate fate of the Universe, requiring insights from nearly every branch of physics. It spans a vast range of scales, from the microscopic to the cosmic, using specialized units like megaparsecs (Mpc) for distances, gigayears (Gyr) for time, and electron-volts (eV) for energy, reflecting its close ties to particle physics. One of the most significant achievements in cosmology is the development of the hot Big Bang model, which describes the Universe's beginning as a hot, dense state around 13.8 billion years ago. This theory has been remarkably successful in explaining key observations, including the cosmic microwave background radiation (a relic from the early Universe), the abundance of light elements through primordial nucleosynthesis, and the large-scale structure of galaxies. The Big Bang model suggests that the Universe has been expanding and cooling since its inception, leading to the formation of stars, galaxies, and other cosmic structures. Although Big Bang theory has endured numerous observational tests, it leaves some fundamental questions unanswered, such as the nature of the initial singularity and what triggered the expansion. Advances in observational cosmology have enabled precise measurements that further refine our understanding of this model. Observations of the cosmic microwave background and large-scale structure provide insight into the composition, expansion history, and early conditions of the Universe. This ongoing interplay between theory and observation is critical for unraveling the mysteries of the Universe.

The foundation of modern cosmology was laid in 1915 when Albert Einstein proposed the General Relativity (GR), revolutionizing our understanding of gravity. GR introduced a novel perspective, where gravity is no longer viewed as a force but as a manifestation of the curvature of spacetime caused by mass and energy. This curvature is described mathematically by the metric tensor \( g_{\mu\nu} \), and the dynamics of spacetime is governed by Einstein's field equations, a set of partial differential equations formulated using the language of differential geometry. The first experimental confirmation of GR came in 1919, when Sir Arthur Eddington \cite{Eddington/1920} observed the bending of starlight by the Sun during a total solar eclipse. This phenomenon is consistent with GR predictions, which demonstrate how massive objects curve spacetime and influence the trajectory of light. Since then, GR has been rigorously tested and has proven to be the most accurate description of gravitational phenomena. It explains planetary orbits, including the anomaly in Mercury's orbit, and has been further validated by the study of distant binary millisecond pulsars \cite{Damour/1999}. The solutions to Einstein's equations have provided insights into the structure of stars, black holes, and the collapse of spherical objects, significantly advancing our understanding of gravity and the Universe \cite{Schwarzschild/1916a,Schwarzschild/1916b,Reissner/1916,Nordstrom/1918,Oppenheimer/1939}.  

However, while GR excels at describing gravity, its integration into cosmology has revealed profound mysteries, particularly regarding the accelerated expansion of the Universe. The historical narrative of this cosmic acceleration began in 1917 when Albert Einstein introduced the cosmological constant (\( \Lambda \)) in his field equations of GR \cite{einstein/1917}. At this time, the prevailing view of the Universe was that of a static, homogeneous, and isotropic Universe with spherical geometry. To align his theory with this notion, Einstein incorporated \( \Lambda \) as a repulsive force to counteract gravitational collapse, ensuring a static Universe. However, this addition struggled to explain small-scale perturbations and remained a purely theoretical concept, lacking observational support.

However, Alexander Friedmann's work \cite{friedman/1922,friedman/1924} in the 1920s demonstrated that Einstein’s field equations naturally allowed dynamic, non-static solutions. His work revealed that the Universe could expand or contract, contradicting Einstein’s assumptions of a static universe. 
A pivotal moment occurred in 1929 when Edwin Hubble provided observational confirmation of an expanding Universe \cite{Hubble/1929}. By measuring the redshifts of distant galaxies, Hubble established the relationship between a galaxy’s recession velocity (\( v \)) and its distance from Earth (\( d \)), encapsulated in the equation \( v = H_{0} \, d \), where \( H_{0} \) is the Hubble constant \cite{Hubble/1931}. This discovery fundamentally altered our understanding of the Universe, shifting the focus from a static Universe to one in constant expansion. The idea of an expanding Universe was further explored in the 1930s by Georges Lemaître \cite{Lemaître/1927}, Howard P. Robertson \cite{Robertson1,Robertson2,Robertson3}, and Arthur Geoffrey Walker \cite{Walker/1937}. Their collective efforts resulted in the exact solutions that are now recognized as the Friedmann-Lemaître-Robertson-Walker (FLRW) metric. This metric forms the foundation of modern cosmology, describing a Universe that evolves with time. In light of these findings, Einstein abandoned the cosmological constant in 1931, calling it his ``greatest blunder". Attention then turned to understanding the dynamics of an expanding Universe driven primarily by matter and radiation.

Decades later, in 1998, the cosmological landscape changed dramatically with the discovery of the accelerating expansion of the Universe. Using Type Ia supernovae as standard candles, independent studies by Adam Riess \cite{Riess/1998}, Saul Perlmutter \cite{Perlmutter/1999}, and their respective teams revealed that the expansion of the Universe was not slowing due to gravitational pull, as previously assumed, but accelerating. This groundbreaking discovery revived the cosmological constant as a viable explanation, now reinterpreted as the energy density of empty space, which we call dark energy (DE). The implications of this finding were profound, as it indicated that DE constitutes approximately 68\% of the energy content of the Universe, fundamentally reshaping our understanding of the evolution of the Universe. The significance of this discovery was globally recognized in 2011 when Saul Perlmutter, Brian Schmidt, and Adam Riess were awarded the Nobel Prize in Physics. Their work not only confirmed the existence of DE but also opened new avenues for exploring the Universe's ultimate fate and the underlying physics driving its acceleration. 


DE remains one of the greatest mysteries in modern cosmology. Numerous theoretical models have been proposed to explain DE, ranging from a cosmological constant to dynamic scalar fields. However, none of these models have achieved universal acceptance, as many face significant theoretical and observational challenges. The most widely accepted framework for describing the dynamics of the Universe is the \(\Lambda\) CDM model, which combines Einstein’s GR with two primary components: the cosmological constant (\(\Lambda\)) and cold dark matter (CDM). The cosmological constant represents a constant energy density permeating space, acting as DE and driving the accelerated expansion of the Universe. CDM accounts for the unseen matter that explains the formation of large-scale cosmic structures. Together, \(\Lambda\)CDM provides an elegant yet simplified model of the evolution and structure of the Universe. It has successfully explained several observations, including the cosmic microwave background (CMB), large-scale structure distribution, and accelerated expansion. Despite its success, the \(\Lambda\)CDM model faces critical challenges. One of the main issues is the ``cosmological constant problem", which asks why the observed value of \(\Lambda\) is so small compared to theoretical predictions \cite{Velten/2014}. Additionally, the Hubble tension \cite{Valentino/2021}—a persistent discrepancy between the locally measured Hubble constant and its value inferred from the early Universe—raises concerns about the model's completeness. 

Relevant to these challenges, the question of cosmic acceleration has motivated a number of modified gravity models that attempt to produce such an acceleration without relying on a cosmological constant. These models, often referred to as self-accelerating, aim to address the issue by modifying the underlying gravitational framework. In most cases, these models do not solve the cosmological constant problems directly but operate under the hope that mechanisms such as gravitational screening or cancellations may suppress the contribution of vacuum energy to gravitational and cosmological dynamics. Some modified gravity models propose partial gravitational screening mechanisms; however, none have yet succeeded fully. Nevertheless, these alternative approaches open exciting avenues for addressing the DE problem and deepening our understanding of cosmic acceleration.

 Motivated by these challenges, researchers have explored modifications to the symmetric teleparallel theory, a framework equivalent to GR that replaces curvature with nonmetricity. This thesis investigates one such extension: $f(Q)$ gravity, a modification of symmetric teleparallel gravity, to explore its cosmological implications.  Specifically, we investigate the cosmological implications of nonmetricity-based \( f(Q) \) gravity as an alternative to the standard \(\Lambda\)CDM model for explaining the accelerated expansion of the Universe. Using a combination of analytical reconstructions, observational data analysis, and dynamical system analysis, we explore various formulations of \( f(Q) \) gravity. The study aims to address critical issues, including Hubble tension, the nature of DE, and the mechanisms driving cosmic acceleration, contributing to our understanding of the Universe's evolution and physics beyond GR.

The Introduction begins with Section \ref{section 1.1}, which covers the mathematical preliminaries necessary to understand cosmology. This includes the metric tensor, Christoffel symbols, covariant derivatives, geodesics, and curvature tensors such as the Riemann curvature tensor, Ricci tensor, and Einstein tensor, essential tools for describing the geometry of spacetime. Section \ref{section 1.2} shifts focus to GR, introduced by Einstein in 1915, highlighting its field equations and the energy-momentum tensor for the perfect fluid, which form the foundation of modern gravitational theory. Section \ref{section 1.3} explores the progression from GR to standard cosmology through the FLRW metric, describing a homogeneous and isotropic Universe. This section also introduces the cosmological standard model (\(\Lambda\)CDM), which integrates the cosmological constant and CDM. Despite its success, \(\Lambda\)CDM faces challenges, such as the unresolved Hubble tension and the mysterious nature of DE, which require alternative approaches. Section \ref{section 1.4} introduces modified gravity theories, emphasizing their potential to address these challenges. Section \ref{section 1.5} focuses on modified gravity based on non-metricity, particularly the symmetric teleparallel equivalent of GR (STEGR), which reformulates gravity using non-metricity instead of curvature. STEGR extensions are presented as promising alternatives to the \(\Lambda\)CDM model, offering new insights into the accelerated expansion of the Universe. In addition, this section includes a comprehensive review of the literature, situating the research within the broader context of existing studies. Section \ref{section 1.7} discusses the Gaussian process methodology used for the reconstruction process from observational Hubble data. Finally, Section \ref{section 1.8} concludes the chapter by summarizing the key discussions.

\section{Mathematical Preliminaries}\label{section 1.1}

GR is a geometric theory that can be formulated using geometric and invariant objects defined on a smooth differentiable manifold $\mathcal{M}$, which will later be identified as the spacetime manifold. The manifold structure allows us to construct local coordinate systems and apply differential calculus at points on the manifold. The types of manifold of interest are called Lorentzian manifolds, which locally look like a flat Minkowski space. In GR, the dimension of the manifold is fixed to be four, but for generality, we work on an arbitrary number of dimensions. The geometric objects of importance are the tensors, which are multilinear maps that live in the tensor product space built from the tangent space and the cotangent space at each point $p$ on the manifold. For a rank $(m,n)$ tensor, this space is
\begin{equation}
    T^{(m,n)}_p(\mathcal{M})= \underbrace{T_p(\mathcal{M})\otimes T_p(\mathcal{M})\otimes...\otimes T_p(\mathcal{M})}_{m\,\,times}\otimes\underbrace{T_p^*(\mathcal{M})\otimes...\otimes T_p^*(\mathcal{M})}_{n\,\,times},
\end{equation}
with $T_p(\mathcal{M})$ and $T_p^*(\mathcal{M})$ the tangent and cotangent\footnote{The dual space to the tangent space $T_p(\mathcal{M})$ is called a cotangent space $T_p^*(\mathcal{M})$.} space at $p$ respectively. The set of tensor product spaces at all points in the manifold $p\in \mathcal{M}$ is the tensor bundle $T^{(m,n)}_p(\mathcal{M})$, from which we get tensor fields as sections of the tensor bundle. Hence, we see that the tensor fields are indeed geometric objects and do not depend on coordinates.\\
However, as we like to work in coordinates when performing calculations, it is also useful to have a coordinate expression for these objects. A rank $(m,n)$ tensor can be written in local coordinates $x^{\mu}$ as
\begin{equation}
    T^{(m,n)}_p(\mathcal{M})=T^{\mu_1\mu_2...\mu_m}_{\,\,\,\,\,\,\,\,\,\,\,\,\,\,\,\,\,\,\,\,\,\,\,\,\,\,\,\,\nu_1\nu_2...\nu_n}\frac{\partial}{\partial x^{\mu_1}}\otimes \frac{\partial}{\partial x^{\mu_2}}\otimes...\otimes\frac{\partial}{\partial x^{\mu_m}}\otimes dx^{\nu_1}\otimes dx^{\nu_2}\otimes ...\otimes dx^{\nu_n},
\end{equation}
where $\frac{\partial}{\partial x^{\mu}}$ and $dx^{\nu}$ are the vector and one-form coordinate bases, respectively.\\
An important property of geometric objects is how their components change when moving from one set of coordinates ${x^{\mu}}$ to another ${y^{\mu}}$. The components of tensors satisfy the following transformation rule under a change of coordinates $x^{\mu}\to\hat{x}^{\mu}(x)$
\begin{equation}
T^{\mu_1\mu_2...\mu_m}_{\,\,\,\,\,\,\,\,\,\,\,\,\,\,\,\,\,\,\,\,\,\,\,\,\,\,\,\,\nu_1\nu_2...\nu_n}(x)\longmapsto \hat{T}^{\mu_1\mu_2...\mu_m}_{\,\,\,\,\,\,\,\,\,\,\,\,\,\,\,\,\,\,\,\,\,\,\,\,\,\,\,\,\nu_1\nu_2...\nu_n}(\hat{x})=\frac{\partial \hat{x}^{\mu_1}}{\partial x^{\alpha_1}}...\frac{\partial \hat{x}^{\mu_m}}{\partial x^{\alpha_m}} \frac{\partial x^{\beta_1}}{\partial \hat{x}^{\nu_1}}...\frac{\partial x^{\beta_n}}{\partial \hat{x}^{\nu_n}}T^{\alpha_1\alpha_2...\alpha_m}_{\,\,\,\,\,\,\,\,\,\,\,\,\,\,\,\,\,\,\,\,\,\,\,\,\,\,\,\,\beta_1\beta_2...\beta_n}(x).
\end{equation}
Together with the corresponding transformation rules for the bases, the object $T^{(m,n)}$ is invariant\footnote{An invariant is a property of a mathematical object that remains unchanged under certain operations or transformations applied to the object.}.

\subsection{Metric tensor}
An important tensor in curved space-time is the metric tensor, by which one can define the square of the distance between two neighboring points. Thus, the metric tensor $g_{\mu\nu}$ of rank $(0,2)$, with its inverse denoted by $g^{\mu\nu}$ of rank $(2,0)$, should be a symmetric and non-degenerate tensor, implying that it has ten components. The metric tensor defines an inner product in the tangent space $T_p(\mathcal{M})$ at each point on the manifold $p\in \mathcal{M}$. Given two vectors $U,V\in T_p(\mathcal{M})$ the metric is the map $g(U,V)=U.V=V.U=g(V,U)\in \mathbb{R}$. This can be thought of as the generalization of the dot product in Euclidean space and gives a notion of distances on the manifold. The metric is also known as the spacetime interval, which we write in local coordinates as
\begin{equation}
    ds^2=g_{\mu\nu}\,dx^{\mu}\,dx^{\nu}.
\end{equation}
In addition, we can treat the metric tensor $g_{\mu\nu}$ as a matrix, and the determinant of the matrix is $g=det(g_{\mu\nu})$.\\
The norm or magnitude of a vector $V^{\mu}$ is defined as an inner product of itself and is defined as $l^2=g_{\mu\nu}\,V^{\mu}\,V^{\nu}$. A spacetime vector $V^{\mu}$ is classified as 
\begin{equation}
    g_{\mu\nu}\,V^{\mu}\,V^{\nu} \begin{cases}
        <0,\,\,\,\,  V^{\mu}\,\, \text{is timelike}, \\
        =0,\,\,\,\,  V^{\mu}\,\, \text{is null or lightlike},\\
        >0,\,\,\,\,  V^{\mu}\,\, \text{is spacelike.}
    \end{cases}
\end{equation}
Another tensor is the Kronecker delta, denoted by \(\delta^{\mu}_{\sigma}\), which is a tensor of order \((1,1)\). It is defined in terms of the metric tensor by the relation \(g^{\mu\nu}g_{\nu\sigma}=g_{\sigma\nu}g^{\nu\mu} = \delta^{\mu}_{\sigma}\).

\subsection{Christoffel symbol, covariant derivative, and geodesics}
Curvature in geometry and physics is fundamentally tied to a concept known as a connection. A connection provides a method for comparing vectors that reside in the tangent spaces of nearby points on a surface or a manifold. This comparison is crucial because it allows us to transport vectors smoothly across different points, making it possible to define and understand curvature.\\ 
The connection that naturally arises from the metric of a space is of particular importance. The metric itself describes the way distances and angles are measured on the manifold and, from this, a unique connection can be derived. This connection is expressed mathematically through what is known as the \textbf{Christoffel symbol}, and it is written as
\begin{equation}
\label{LCC}
    \{^{\,\lambda}_{\,\,\,\mu\nu}\}=\frac{1}{2}g^{\lambda\sigma}\left(\partial_{\mu}g_{\sigma\nu}+\partial_{\nu}g_{\mu\sigma}-\partial_{\sigma}g_{\mu\nu}\right).
\end{equation}
The Christoffel symbol serves as a set of coefficients that describe how the basis vectors of the tangent space change as one moves from point to point on the manifold. It is central to the mathematical description of curvature because it directly enters the equations that define how curvature behaves in a given space. This connection derived from the metric is fundamental to conventional GR. It is referred to by several names: the Christoffel connection, the Levi-Civita connection, or the Riemannian connection. The field that studies manifolds equipped with a metric and their associated connections is known as the Riemannian geometry. When the metric has a Lorentzian signature, which is typical in the context of spacetime in GR, the study is referred to as pseudo-Riemannian geometry.\\
The fundamental use of a connection is to take a \textbf{covariant derivative} $\nabla_{\lambda}$\footnote{A generalization of partial derivative.}, which for an arbitrary tensor is given as
\begin{multline}
\nabla_{\lambda} T^{\mu_{1} \mu_{2}....\mu_{m}}_{~~~~~~~~~~~\nu_{1} \nu_{2}....\nu_{n}} = \partial_{\lambda}T^{\mu_{1} \mu_{2}....\mu_{m}}_{~~~~~~~~~~~\nu_{1} \nu_{2}....\nu_{n}} +  \{^{\,\mu_1}_{\,\,\,\lambda\sigma}\}  T^{\sigma \mu_{2}....\mu_{m}}_{~~~~~~~~~~~\nu_{1} \nu_{2}....\nu_{n}}  + \{^{\,\mu_2}_{\,\,\,\lambda\sigma}\}  T^{\mu_{1} \sigma....\mu_{m}}_{~~~~~~~~~~~\nu_{1} \nu_{2}....\nu_{n}}+\\.....+ \{^{\,\mu_m}_{\,\,\,\lambda\sigma}\}  T^{\mu_{1} \mu_2....\mu_{m-1}\sigma}_{~~~~~~~~~~~~~~~\nu_{1} \nu_{2}....\nu_{n}}
 - \{^{\,\sigma}_{\,\,\,\lambda\nu_1}\}   T^{\mu_{1} \mu_{2}....\mu_{m}}_{~~~~~~~~~~~\sigma \nu_{2}....\nu_{n}} - \{^{\,\sigma}_{\,\,\,\lambda\nu_2}\}  T^{\mu_{1} \mu_{2}....\mu_{m}}_{~~~~~~~~~~~\nu_{1}\sigma....\nu_{n}}-\\ .....
 -  \{^{\,\sigma}_{\,\,\,\lambda\nu_n}\}  T^{\mu_{1} \mu_{2}....\mu_{m}}_{~~~~~~~~~~~\nu_{1}\nu_2....\nu_{n-1}\sigma}.
\end{multline}
Consequently, this demonstrates that the Christoffel or Levi-Civita, connection is both metric-compatible and torsion-free. Metric compatibility means that the covariant derivative of the metric tensor vanishes, expressed as \(\nabla_{\lambda} g_{\mu\nu} = 0\). The connection being torsion-free is indicated by its symmetry in the lower indices. Additionally, for a scalar \(\phi\), we have \(\nabla_{\lambda} \phi = \partial_{\lambda} \phi\), reflecting the fact that the covariant derivative of a scalar reduces to its partial derivative.\\
Given the concept of a covariant derivative, we can introduce the idea of parallel transport for a tensor. In flat space-time, parallel transport involves moving a vector along a curve while ``keeping it constant". Specifically, a vector $T^{\mu}$ is considered constant along a curve $x^{\mu}(\lambda)$ if its components do not vary with the parameter $\lambda$,
\begin{equation}
\label{PT1}
    \frac{dT^{\mu}}{d\lambda}=\frac{dx^{\nu}}{d\lambda}\partial_{\nu}T^{\mu}=0.\,\,\,\,\,\,(flat\,\,spacetime)
\end{equation}
We generalized this to curved spacetime by replacing the partial derivative in \eqref{PT1} by a covariant derivative. This gives the so-called \textbf{directional covariant derivative}\footnote{This operator combines the tangent vector $\frac{dx^{\alpha}}{d\lambda}$ and the covariant derivative $\nabla_{\alpha}$ into a single operator that gives the covariant derivative along the path parameterized by $\lambda$.} as
\begin{equation}
    \frac{D}{d\lambda}=\frac{dx^{\alpha}}{d\lambda}\nabla_{\alpha}.
\end{equation}
A vector is parallel transported along the curve $x^{\mu}(\lambda)$, with $\lambda$ being the affine parameter, as the condition that the directional covariant derivative of vector $T^{\mu}$ along the curve vanishes,
\begin{equation}
  \frac{D}{d\lambda}T^{\mu}=\frac{dx^{\alpha}}{d\lambda}\nabla_{\alpha} T^{\mu}=0.\,\,\,\,\,\,(curved\,\,spacetime)
\end{equation}
With parallel transport defined, the next logical step is to introduce the concept of a geodesic\footnote{a generalization of the notion of a straight line.}. The Levi-Civita connection also appears in the definition of geodesic. A parameterized curve \( x^{\mu}(\lambda) \) is called a \textbf{geodesic} if its tangent vector \( \frac{dx^{\alpha}}{d\lambda} \) is parallel transported along the curve. This condition is expressed as
\begin{equation}
    \frac{D}{d\lambda}\frac{dx^{\alpha}}{d\lambda}=0,
\end{equation}
which implies the \textbf{geodesic equation}
\begin{equation}
    \frac{d^2x^{\mu}}{d\lambda^2}+\{^{\,\mu}_{\,\,\,\alpha\beta}\}\frac{dx^{\alpha}}{d\lambda}\frac{dx^{\beta}}{d\lambda}=0.
\end{equation}

\subsection{Riemann curvature tensor, Ricci tensor, and Einstein tensor}
We now introduce the Riemann curvature tensor \( R^{\alpha}_{\,\,\,\,\beta\mu\nu} \), which plays a crucial role in describing the curvature of spacetime at each point within a given manifold. This tensor summarizes how vectors are parallelly transported around an infinitesimal loop, revealing the intrinsic curvature of the underlying space. Mathematically, the Riemann curvature tensor is defined through the commutator of covariant derivatives acting on a vector field \( T^{\alpha} \). Specifically, this relationship is expressed as
\begin{equation}
[\nabla_{\mu},\nabla_{\nu}]\,T^{\alpha} =(\nabla_{\mu}\nabla_{\nu} - \nabla_{\nu}\nabla_{\mu})\,T^{\alpha} = R^{\alpha}_{\,\,\,\,\beta\mu\nu} \, T^{\beta},
\end{equation}
where the \textbf{Riemann curvature tensor} or simply \textbf{Riemann tensor} is explicitly expressed in terms of the Levi-Civita connection as follows
\begin{equation}
    R^{\alpha}_{\,\,\,\,\beta\mu\nu}=\partial_{\mu}\{^{\,\alpha}_{\,\,\,\nu\beta}\}-\partial_{\nu}\{^{\,\alpha}_{\,\,\,\mu\beta}\}+\{^{\,\alpha}_{\,\,\,\mu q}\}\{^{\,q}_{\,\,\,\nu\beta}\}-\{^{\,\alpha}_{\,\,\,\nu q}\}\{^{\,q}_{\,\,\,\mu\beta}\}.
\end{equation}
The Riemann tensor has some algebraic properties:
\begin{itemize}
\item $R^{\alpha}_{\,\,\,\,\beta\mu\nu} = - R^{\alpha}_{\,\,\,\,\beta\nu\mu}$, \hspace{0.2in} ($Antisymmetric$)
\item $R^{\alpha}_{\,\,\,\,\beta\mu\nu} + R^{\alpha}_{\,\,\,\,\mu\nu\beta} + R^{\alpha}_{\,\,\,\,\nu\beta\mu} =0 $, \hspace{0.2in} ($Cyclic$)
\item $\nabla_{\sigma}R^{\alpha}_{\,\,\,\,\beta\mu\nu}+ \nabla_{\mu}R^{\alpha}_{\,\,\,\,\beta\nu\sigma}+\nabla_{\nu} R^{\alpha}_{\,\,\,\,\beta\sigma\mu}=0$. \hspace{0.2in} ($Bianchi$  $Identity$)
\end{itemize}
Further, the \textbf{Ricci tensor} is derived by contracting the Riemann tensor in a specific manner
\begin{equation}
    R_{\mu\nu}=R^{\alpha}_{\,\,\,\,\mu\alpha\nu}.
\end{equation}
An important property of the Ricci tensor is that it is symmetric, which means that it satisfies the relation $R_{\mu\nu}=R_{\nu\mu}$.
The trace of the Ricci tensor is the \textbf{Ricci scalar} and it is expressed as
\begin{equation}
    R=R^{\mu}_{\nu}=g^{\mu\nu}R_{\mu\nu}.
\end{equation}
Moreover, the \textbf{Einstein tensor} is defined using the Ricci tensor and Ricci scalar as
\begin{equation}
    G_{\mu\nu}=R_{\mu\nu}-\frac{1}{2}g_{\mu\nu}\,R.
\end{equation}
Additionally, the Bianchi identity that is twice contracted implies that the covariant divergence of the Einstein tensor vanishes, that is, $\nabla^{\mu}G_{\mu\nu}=0$.

\section{General Relativity}\label{section 1.2}
The most advanced current theory of gravitation and spacetime is Einstein’s theory of GR, a classical field theory in which gravity is described through geometry and curvature. Despite being over a century old, GR has withstood every experimental test, from solar system test to astrophysical and cosmological experiments. The mathematics of GR is expressed through differential geometry, interpreting gravity as the curvature of spacetime itself. It is remarkable that such a broad range of phenomena and precise predictions emerge from a theory that is both conceptually simple and mathematically elegant. The dynamics of GR are governed by the Einstein field equations, which relate the geometry of spacetime to the distribution of matter and energy. This theory not only revolutionized our understanding of gravity but also provided the framework for modern cosmology and the study of black holes, gravitational waves, and the expansion of the Universe.
\subsection{Einstein's field equations}
GR explores the interplay between spacetime curvature and matter, showing how spacetime curvature influences the motion of matter and, conversely, how matter affects the curvature of spacetime \cite{Misner/1973}. This interaction is formalized in the Einstein-Hilbert action, which includes both curvature and matter components. The action is expressed as
\begin{equation}
\label{Action}
S= \frac{1}{2\kappa^2}\int R\,\sqrt{-g}\, d^{4}x + S_{matter},
\end{equation}
where $S_{matter}=\int\sqrt{-g}\mathcal{L}_m\,d^4x$ is the action for the matter fields, and $\mathcal{L}_m$ is the matter Lagrangian density. Here, \(\kappa^2 = 8\pi G/c^4\), where \(G\) is the Newtonian gravitational constant and $c$ is the speed of light. This expression accounts for a difference in the relative normalization between the gravitational action and the matter action. The term \(d^4x\) represents the four-dimensional volume element in spacetime.\\
By varying the Einstein-Hilbert action \eqref{Action} with respect to the metric tensor $g_{\mu\nu}$, one derives the Einstein field equations, which are the fundamental equations of GR as
\begin{equation}
\label{EF}
G_{\mu \nu}= R_{\mu \nu} -\frac{1}{2} g_{\mu \nu}\,R = \frac{8 \pi G}{c^{4}} T_{\mu \nu},
\end{equation}
where $T_{\mu\nu}$ is the energy-momentum tensor, which describes the matter distribution for a perfect fluid and is defined as
\begin{equation}
\label{EMT1}
T_{\mu \nu}= -\frac{2}{\sqrt{-g}} \frac{\delta S_{matter}}{\delta g^{\mu \nu}}.
\end{equation}
In the Einstein field equation \eqref{EF}, the left-hand side represents the ``geometry and curvature" of spacetime, while the right-hand side represents ``energy and momentum". Using these equations, along with the fact that the covariant derivative of the Einstein tensor vanishes, leads to the conservation equation for the energy-momentum tensor in a curved spacetime
\begin{equation}
    \nabla^{\mu}T_{\mu\nu}=0.
\end{equation}
This conservation equation ensures that the total energy and momentum in any region of spacetime remain constant, even as spacetime itself evolves due to the presence of matter and energy. This is a generalization of the conservation laws familiar from classical mechanics adapted to the curved spacetime of the GR.\\
Shortly after developing GR, Einstein attempted to construct a static cosmological model. To achieve this, he needed to modify the field equation with an ordinary matter source by introducing an additional term on the left-hand side of the equation. This term took the form \(\Lambda g_{\mu\nu}\), where \(\Lambda\) is the cosmological constant. The resulting modified form of the Einstein's equation is
\begin{equation}
\label{MEE}
     R_{\mu \nu} -\frac{1}{2} g_{\mu \nu}\,R+\Lambda g_{\mu\nu} = \frac{8 \pi G}{c^{4}} T_{\mu \nu}.
\end{equation}
Including this term does not affect the conservation of the energy-momentum tensor, since the covariant derivative of the metric tensor is zero, \(\nabla_{\lambda} g_{\mu\nu} = 0\).\\
The action leading to Eq.\eqref{MEE} is 
\begin{equation}
\label{Action2}
S= \frac{1}{2\kappa^2}\int\sqrt{-g}\,(R-2\Lambda)\,d^{4}x + S_{matter}.
\end{equation}
Here, we observe that the cosmological constant corresponds to a term purely associated with the volume in the action.

\subsection{Energy momentum tensor for perfect fluid}
In cosmological models, we assume the Universe to be homogeneous, isotropic, and filled with a perfect fluid. This framework allows us to derive the energy-momentum tensor, which is crucial for describing the dynamics of the Universe. Describing spacetime is essential, particularly since it is not "empty." In the context of a perfect fluid—a fluid fully characterized by its rest-frame mass density \(\rho\) and isotropic pressure \(p\), the energy-momentum tensor is given by
\begin{equation}
\label{EMT}
T_{\mu\nu} = (\rho + p)\,u_{\mu}u_{\nu} + p\,g_{\mu\nu}.
\end{equation}
Here, \(u_{\mu}\) represents the four-velocity of the particles. The energy density \(\rho\), is expressed as \(\rho = mn\), where \(n\) is the number of particles per unit volume and \(m\) is the mass of a particle. In the rest frame, the four-velocity is proportional to \((1, 0, 0, 0)\).\\
This equation represents the relationship between the fluid properties and the curvature of spacetime. The term $(\rho + p)\,u_{\mu}u_{\nu}$ describes the contribution of the fluid energy density and pressure in the direction of fluid flow, while the term $p\,g_{\mu\nu}$ accounts for the isotropic contribution of the pressure to spacetime.
At the rest frame, the energy-momentum tensor of a perfect fluid can take the following matrix form
\begin{equation}
\begin{pmatrix}
    (\rho+p)+p\,g_{00} & 0 & 0 & 0\\
    0 & p\,g_{11} & 0 & 0\\
     0 & 0 & p\,g_{22} & 0\\
    0 & 0 & 0 & p\,g_{33} 
\end{pmatrix}.
\end{equation}
 For a symmetric energy-momentum tensor, the first diagonal component corresponds to the conservation of energy, while the remaining diagonal components correspond to the conservation of momentum. In Minkowski space, which describes flat spacetime, both energy and momentum are conserved. This conservation law is mathematically expressed as \(\partial_{\mu}T^{\mu}_{\,\,\,\nu}=0\).\\
However, as discussed in the earlier sections, the partial derivative is insufficient for describing changes in curved spacetime, which is the framework of GR. Despite this, we still require the conservation of energy and momentum to hold, even in a curved space-time. In GR, this requirement is fulfilled by promoting the conservation law to the covariant form
\begin{equation}
\label{CE}
\nabla_{\mu}T^{\mu}_{\,\,\,\nu}=0.
\end{equation}
Here, \(\nabla_{\mu}\) represents the covariant derivative, which accounts for the curvature of spacetime, ensuring that conservation laws are maintained in all geometries.\\
 With all the essential concepts in place to understand and solve Einstein's equation, we can now shift our focus to cosmology. By adopting certain common assumptions about the fundamental characteristics of the Universe, Einstein's equations can be effectively solved, allowing us to explore the dynamics and structure of the cosmos.
 
\section{From GR to Standard Cosmology}\label{section 1.3}
When Einstein first applied GR to cosmology in 1917 \cite{einstein/1917}, he derived a static and closed solution to his equations. Soon after, Willem de Sitter proposed a solution without matter, now called the de Sitter solution \cite{Sitter/1917}. Later, Alexander Friedmann \cite{friedman/1922,friedman/1924} introduced solutions describing an expanding Universe, which were observationally supported by Edwin Hubble’s 1929 discovery of the redshift-distance relation \cite{Hubble/1929}. In the 1930s, the expanding Universe concept was refined through the work of Georges Lemaître \cite{Lemaître/1927}, Howard P. Robertson \cite{Robertson1,Robertson2,Robertson3}, and Arthur G. Walker \cite{Walker/1937}, culminating in the development of the Friedmann-Lemaître-Robertson-Walker (FLRW) metric.  

\subsection{The FLRW metric}
The FLRW metric is based on the cosmological principle, which asserts that regardless of the direction or location in the Universe, the observations will be broadly consistent. This implies that the matter distribution will be approximately uniform and that the size of the Universe will not appear to vary based on the observer's position within it \cite{Bergstrom/2004}. Specifically, the cosmological principle encompasses two key properties: homogeneity and isotropy.\\ 
\textbf{Homogeneity} means that the Universe exhibits the same large-scale structure everywhere, allowing the space to be described by a single metric. \textbf{Isotropy} implies that this uniformity is independent of the direction in which observations are made. The cosmic microwave background radiation (CMBR) strongly supports this description, as it is remarkably evenly distributed throughout the sky \cite{Carroll/2004}.\\ 
The FLRW metric describes a Universe that is both homogeneous and isotropic, with matter and energy uniformly distributed as a perfect fluid. It can be expressed in spherical polar coordinates system as\footnote{There are two types of signature conventions generally used by researchers to explore the fate of the Universe: \((+,-,-,-)\) and \((-,+,+,+)\). Throughout this thesis, we adopt the second signature convention, \((-,+,+,+)\), for the metric.}
\begin{equation}
\label{FRWmetric}
ds^{2}= -c^{2}\,dt^{2} + a^{2}(t) \left[\frac{dr^{2}}{1-k\,r^{2}} + r^{2} d\theta^{2} + r^{2} sin^{2}\theta\,d\phi^{2}\right].
\end{equation}
The function \( a(t) \) is the time-dependent cosmic scale factor, which describes the relative change in the size of the Universe with time. The constant \( k \) represents the curvature of space, it can take one of three values: \( k = -1 \) corresponds to a negatively curved (open) Universe, \( k = 0 \) corresponds to a flat Universe, and \( k = 1 \) corresponds to a positively curved (closed) Universe. The physical interpretation of these scenarios becomes clearer by using a different form of the metric, which is achieved by introducing a new radial coordinate \(\chi\) defined as
\begin{equation}
    d\chi=\frac{dr}{\sqrt{1-k\,r^{2}}}.
\end{equation}
Integration of this gives
\begin{equation}
    r=f_k(\chi),
\end{equation}
where \( f_k(\chi) \) is defined as
\begin{equation}
   f_k(\chi)\equiv\begin{cases}
        sinh(\chi),\,\,\,\,k=-1 \\
        \chi,\,\,\,\,\,\,\,\,\,\,\,\,\,\,\,\,\,\,\,\,\,k=0\\
        sin(\chi),\,\,\,\,\,\,\,\,k=1.
    \end{cases}
\end{equation}
With this substitution, the line element becomes
\begin{equation}
    ds^{2}= -c^{2}\,dt^{2} + a^{2}(t) \left[d\chi^2 + f_k^2(\chi)\left(d\theta^{2} + sin^{2}\theta\,d\phi^{2}\right)\right].
    \end{equation}
It can also be useful to switch from the cosmic time to conformal time, defined by \( d\eta \equiv \frac{dt}{a(t)} \). With this change, the line element becomes
\begin{equation}
    ds^{2}= a^2(\eta)\left[-c^{2}\,d\eta^{2} + d\chi^2 + f_k^2(\chi)\left(d\theta^{2} + sin^{2}\theta\,d\phi^{2}\right)\right].
\end{equation}
The metric has been separated into a static component and a time-dependent conformal factor \( a(\eta) \). This structure is especially useful for analyzing the propagation of light, where \( ds^2 = 0 \). Additionally, conformal time can be a valuable variable transformation for deriving certain exact solutions to the Einstein equations.

\subsection{Cosmological standard model}
The FLRW metric applies for any time-dependent behavior of the scale factor \( a(t) \). Our next task is to insert this metric into Einstein’s field equations to obtain the Friedmann equations, which connect the scale factor to the energy-momentum distribution in the Universe. We will represent the matter and energy content as a perfect fluid.\\
When the FLRW metric is substituted into Eq.\eqref{EMT}, the energy-momentum tensor for a perfect fluid becomes
\begin{equation}
\label{EMTFRW}
    T^{\mu}_{\,\,\,\,\nu}=diag(-\rho,p,p,p),
\end{equation}
with the trace given by \( T = T^{\mu}_{\,\,\,\,\mu} = -\rho + 3p \).\\
We can now derive the conservation equation for the energy-momentum tensor in Eq.\eqref{CE}, which applies individually to each matter fluid. This leads to the following equation
\begin{equation}
\label{fluideq}
    \dot{\rho}+3\frac{\dot{a}}{a}(\rho+p)=0.
\end{equation}
For most cosmological fluids, the equation of state (EoS) can be expressed as $p=w\rho$, where $w$ is a constant. Substituting this into Eq.\eqref{fluideq} gives
\begin{equation}
    \frac{\dot{\rho}}{\rho}=-3(1+w)\frac{\dot{a}}{a}\,\,\,\,\implies\,\,\,\,\rho\propto a^{-3(1+w)}.
\end{equation}
This relation describes how the energy density decreases with the scale factor, depending on the equation of state. Next, we will discuss the three key types of fluids—radiation, matter, and vacuum—that significantly influence the evolution of the Universe.\\
\textbf{Radiation} has a pressure that is one-third of its energy density, \( p = \frac{1}{3} \rho \). A Universe dominated by radiation is known as a radiation-dominated Universe. For radiation, the energy density evolves as
\[
    \rho_r \propto a^{-4}.
\]
\textbf{Matter} consists of collision-free non-relativistic particles with negligible pressure, that is, \( p_m = 0 \). This type of matter is also referred to as ``dust", and a Universe where matter dominates the energy density is called a matter-dominated Universe. For matter, the energy density evolves as
\[
    \rho_m \propto a^{-3},
\]
indicating that the energy density decreases as the Universe expands.\\
The additional factor of $a^{-1}$ compared to matter arises in a radiation-dominated Universe because, in addition to the dilution due to the expanding volume (like in the matter case), the wavelength of the radiation also stretches as the Universe expands.\\
\textbf{Vacuum} energy, commonly associated with DE, behaves very differently from matter and radiation. Its EoS is given by $p_{\Lambda}=-\rho_{\Lambda}$, which means that the pressure is negative and is equal in magnitude to the energy density. This negative pressure leads to the peculiar property that the vacuum energy remains constant over time, regardless of the expansion of the Universe. In mathematical terms
\[
    \rho_{\Lambda} \propto a^0.
\]
This means that vacuum energy does not dilute as the Universe expands; instead, it remains constant. This constant energy density leads to the accelerated expansion of the Universe, as observed in the current era. A Universe dominated by vacuum energy is referred to as a vacuum-dominated or DE-dominated Universe.\\
We now focus on Einstein's equation, as presented in Eq.\eqref{MEE}. With all the necessary components in place, we are ready to solve Einstein's equation in the context of the FLRW metric.\\
From the FLRW metric, we derive two fundamental equations, known as the Friedmann equations, which govern the evolution of the scale factor $a(t)$ over time. These equations are expressed as
\begin{equation}
\label{FE1}
    \left(\frac{\dot{a}}{a}\right)^2=\frac{\kappa^2}{3}\rho-\frac{k}{a^2}+\frac{\Lambda}{3},
\end{equation}
\begin{equation}
\frac{\Ddot{a}}{a}=-\frac{\kappa^2}{6}(\rho+3p)+\frac{\Lambda}{3}, 
    \label{FE2}
\end{equation}
where \( \rho \) is the total energy density and \( p \) is the total pressure, each accounting for all components of the Universe. We recall that \( \kappa^2 = \frac{8\pi G}{c^4} \), and the dot notation represents derivatives with respect to cosmic time.\\
In the study of cosmology, various terms and parameters play a crucial role in describing the physical properties and evolution of the Universe. Here, we will introduce the key parameters. The rate of expansion of the Universe is described by the \textbf{Hubble parameter}, defined as
\begin{equation}
    H=\frac{\dot{a}}{a}.
\end{equation}
The current value of the Hubble parameter is called the \textbf{Hubble constant}, denoted as $H_0$. Another key parameter is the \textbf{deceleration parameter}, which measures how the expansion rate changes over time and is defined as
\begin{equation}
    q=-\frac{a\Ddot{a}}{\dot{a}^2}.
\end{equation}
This parameter helps to quantify whether the expansion of the Universe is accelerating or decelerating. Thus, \( q < 0 \) indicates an accelerating universe, whereas \( q > 0 \) signifies a decelerating universe. Furthermore, the \textbf{density parameter} is another useful quantity, defined as
\begin{equation}
    \Omega=\frac{\kappa^2}{3\,H^2}\rho=\frac{\rho}{\rho_c},
\end{equation}
where $\rho_c=\frac{3\,H^2}{\kappa^2}$ is the critical density, which helps normalize the energy density into a dimensionless fraction. Additionally, the contribution from the cosmological constant and spatial curvature can also be expressed using dimensionless density parameters as
\begin{equation}
    \Omega_{\Lambda}=\frac{\Lambda}{3H^2},\,\,\,\,\Omega_k=-\frac{k}{a^2\,H^2}.
\end{equation}
These parameters describe the relative contributions of DE ($\Omega_{\Lambda}$) and the curvature of the Universe ($\Omega_{k}$) to the overall dynamics of cosmic expansion.\\
Then the first Friedmann equation \eqref{FE1} takes the form 
\begin{equation}
\label{Hubb}
    \frac{H}{H_0}=\left[\Omega_{m0}\,a^{-3}+\Omega_{r0}\,a^{-4}+\Omega_{k0}\,a^{-2}+\Omega_{\Lambda 0}\right]^{1/2},
\end{equation}
where $\Omega_{i0}\,\,(i=m,r,k,\Lambda)$ represents the present-day values of the density parameters relative to the critical density. At the present time, when $a(t_0)=1$ (where $t_0$ is the present age of the Universe), this equation leads to the following constraint:
\begin{equation}
    \Omega_{m0}+\Omega_{r0}+\Omega_{k0}+\Omega_{\Lambda 0}=1.
\end{equation}
The parameter of the curvature density can be expressed as $\Omega_{k0}=1-\Omega_{m0}-\Omega_{r0}-\Omega_{\Lambda 0}$. For a flat Universe (where $k=0$), this simplifies to $\Omega_{m0}+\Omega_{r0}+\Omega_{\Lambda 0}=1$. This reflects the total energy density of the Universe being precisely equal to the critical density required to keep the Universe flat. This constraint is fundamental in cosmology as it relates the observed contents of the Universe to its overall geometry.\\
The current benchmark cosmological model is the $\Lambda$CDM model, also known as the concordance model. This model is the most accepted model of the Universe in modern cosmology due to its ability to correctly predict properties of the Universe. The $\Lambda$ CDM model is based on several fundamental assumptions, such as the evolution of the density distribution of the Universe over time. According to the $\Lambda$CDM model, the Universe has gone through distinct phases: an initial radiation-dominated era, followed by a matter-dominated phase, and finally a period dominated by DE. Presently, the model posits that the Universe's energy density is primarily composed of DE and CDM. DM was initially hypothesized to explain certain galactic phenomena, where observable matter alone could not account for the observed gravitational effects. As mentioned before, the DE $\Lambda$ was originally just a constant that Einstein introduced to get a static Universe. However, subsequent discoveries, particularly the observation of the Universe's expansion by Edwin Hubble, rendered the concept of a static Universe obsolete. Later, the discovery of an accelerated expansion in the late 1990s led cosmologists to revisit the cosmological constant. Now, $\Lambda$ is interpreted as the energy density of the empty space or DE, which drives the accelerated expansion of the Universe. According to current estimation, DE is thought to make up about 68\% of the total energy content of the Universe, while DM and ordinary (baryonic) matter contribute 26\% and 5\%, respectively, and other components such as neutrinos and photons are nearly negligible.\\
The $\Lambda$CDM model is strongly supported by observational evidence, which make this model the best description of the observed Universe. We will discuss the evidence for the standard cosmological model and present the most recent best-fit parameter estimations in support of it.\\
The exact shape of the Universe is still a matter of debate in physical cosmology, but experimental data from various independent sources studying the CMBR from WMAP \cite{WMAP/2011}, BOOMERanG \cite{Bernardis/2000}, and the Planck telescope \cite{Planck/2018} confirm that the Universe is flat. So, the
contribution of the curvature can effectively be ignored: $|\Omega_k|<<1$. 
Data from the CMB showing radiation density is past its epoch, i.e. $|\Omega_r|<<1$. This
means that we have $\Omega_m+\Omega_{\Lambda}=1$. By studying the CMBR, we can obtain best fit values where $\Omega_m\approx 0.3$ and
$\Omega_{\Lambda}\approx 0.7$. One of the more recent studies obtains for a flat $\Lambda$CDM cosmology, the CMB observations \cite{Planck/2018} constrain the
present-day values of the dimensionless density parameters, Hubble rate, and DE EoS parameter to be:
\begin{equation*}
\Omega_{m0}=0.315\pm0.007,\,\,\,\,\,\,\Omega_{\Lambda0}=0.685\pm 0.007,
\end{equation*}
\begin{equation*}
    w_0=-1.03\pm0.03,\,\,\,\,\,H_0=67.4\pm0.5\,\,\text{km s$^{-1}$ Mpc$^{-1}$ }.
\end{equation*}
With $\Omega_{k0}\approx 0 \approx \Omega_{r0}$ and $\Omega_{m0}+\Omega_{\Lambda0}=1$, we remark that for the $\Lambda$CDM model, the deceleration parameter at the present time is defined as
\begin{equation*}
    q_0\simeq\frac{1}{2}\Omega_{m0}-\Omega_{\Lambda0}\approx -0.5275.
\end{equation*}
The fact that these results come from widely different observational experiments and yet have a great concordance with one another gave $\Lambda$CDM the name concordance model.

\subsection{Drawbacks of the $\Lambda$CDM model}
Although the cosmological constant model ($\Lambda$CDM) provides a good fit to most observations, it faces several significant challenges.\\
\textbf{Theoretical Challenges:}\\
\textbf{The cosmological constant problem} is perhaps the most severe \cite{Weinberg/1989}. This problem is a major unresolved issue in theoretical physics, stemming from the immense discrepancy between the predicted and observed values of the $\Lambda$, which represents the energy density of empty space and drives the Universe's accelerated expansion.\\
According to quantum field theory (QFT), particles in the standard model contribute significantly to the value of $\Lambda$. The expected theoretical value is approximately: $\Lambda^{QFT} \approx 10^{-60} M_{Pl}^4$, where \( M_{Pl} \) is the Planck mass. However, when we calculate the vacuum energy density from quantum field theory, it is found to be  
\[
\rho_{\Lambda}^{QFT} \approx 10^{73} \, \text{GeV}^4.
\]
On the other hand, observational data suggest that the value of $\Lambda$ is around: $\Lambda^{Obs} \approx 10^{-120} M_{Pl}^4$,  
with its current energy density estimated as  
\[
\rho_{\Lambda}^{Obs} \approx 10^{-47} \, \text{GeV}^4.
\]  
This implies a discrepancy between the theoretical and observed values of the vacuum energy density  
\[
\frac{\rho_{\Lambda}^{QFT}}{\rho_{\Lambda}^{Obs}} \approx 10^{120}.
\]  
This enormous mismatch, spanning 120 orders of magnitude, is known as the cosmological constant problem.\\
Another challenge faced by the cosmological constant is the \textbf{Coincidence Problem} \cite{Velten/2014}, which arises from the observation that the energy densities of DE (associated with the cosmological constant, \( \Lambda \)) and matter are of the same order of magnitude in the current epoch, despite their vastly different evolutionary behaviors over cosmic time. While DE remains nearly constant as the Universe expands, the matter density decreases with expansion as \( \rho_m \propto a^{-3} \), where \( a \) is the scale factor. This leads to the question: \textit{Why do we observe this near-equality of these two energy densities at the present epoch?} This issue is termed the coincidence problem.\\  
The coincidence problem and the cosmological constant problem are significant theoretical challenges rooted in observations. Addressing these issues requires investigating frameworks that extend beyond GR and the \( \Lambda \)CDM model.\\
\textbf{Observational Challenges:}\\
Beyond the well-known challenges of cosmology, two increasingly significant issues have garnered attention in recent years: the \textbf{Hubble tension} and the \textbf{$\sigma_8$ tension}.\\ 
The Hubble constant ($H_0$) is the present-day value for the rate of expansion of the Universe. The Hubble tension refers to the persistent discrepancy between the $H_0$ values derived from early-Universe observations, such as the CMB measurements by Planck, and those obtained from late-time observations, particularly local distance ladder methods like those employed by the SH0ES collaboration \cite{Weinberg/2013,Valentino/2021,Riess/2019}. In CMB Planck measurement, the early Universe data is analyzed within the framework of the $\Lambda$CDM model. In contrast, the SH0ES measurements are conducted in a model-independent manner, relying on data from the late Universe. These early-time measurements provide a value\footnote{This value can vary slightly with the inclusion of additional datasets, such as BAO, BBN, or DESI \cite{Ivanov/2020,Valentino/2021a}.} of \( H_0 = 67.27 \pm 0.60 \, \text{km} \, \text{s}^{-1} \, \text{Mpc}^{-1} \) \cite{Planck/2018}, while late-time measurements yield \( H_0 = 73.04 \pm 1.04 \, \text{km} \, \text{s}^{-1} \, \text{Mpc}^{-1} \) \cite{Riess/2022}. Low-redshift measurements using alternative methods consistently produce similarly high values of \( H_0 \), in contrast to early-time estimates. In recent years, this tension has only intensified, becoming more statistically significant, and most current results disagree at a level of at least \( 5\sigma \) \cite{Valentino/2021,Riess/2019}. Furthermore, this discrepancy cannot be easily attributed to systematic errors, as resolving it would require the presence of multiple independent systematic effects \cite{Valentino/2021}.\\
Another significant challenge for $\Lambda$CDM cosmology is the tension $\sigma_8$ \cite{Valentino/2021,Valentino/2021a}, which arises from discrepancies in the inferred values of the density parameter of matter, $\Omega_m$, and the amplitude or growth rate of perturbations of matter, characterized by $\sigma_8$ and $f_{\sigma_8}$, in different observational datasets. These inconsistencies have been highlighted through analyses of Planck data in conjunction with results from weak lensing studies and redshift surveys. Within the $\Lambda$CDM framework, the Planck collaboration estimates \( S_8 = \sigma_8 \sqrt{\Omega_m/0.3} = 0.834 \pm 0.016 \), while local measurements suggest a lower value of \( S_8 = 0.745 \pm 0.039 \). This approximately \( 2\sigma \) discrepancy \cite{Joudaki/2017} has sparked interest in exploring alternatives to the $\Lambda$CDM model.

These limitations have prompted the exploration of alternative explanations. In the subsequent section, we review various theories that extend beyond the \(\Lambda\)CDM model.



\section{Modified Gravity Theory}\label{section 1.4}

In the preceding section, we discussed various shortcomings of GR that underscore the need for modification of GR and its equivalent theories. By altering the geometrical part of the Einstein-Hilbert action (see Eq.\eqref{Action}), 
we achieve modified theories of gravity. These theories are motivated by the desire to explain phenomena such as the accelerated expansion of the Universe, DM, DE, and discrepancies in galaxy rotation curves without invoking unknown forms of matter and energy. Notable theories include $f(R)$ gravity, which generalizes the Einstein-Hilbert action by replacing the Ricci scalar $R$ with a function $f(R)$, which thus becomes $S=\int(f(R)/2\kappa^2+L_m)\sqrt{-g}d^4x$ \cite{Buchdahl/1970,Barrow/1983}, aiming to explain cosmic acceleration without DE. A second approach to extending the Hilbert–Einstein action involves assuming a non-minimal coupling between geometry and matter. This research direction leads to distinct classes of gravitational theories. One such class is $f(R,T)$ gravity, described by the action $S=\int(f(R,T)/2\kappa^2+L_m)\sqrt{-g}d^4x$, where \( T \) is the trace of the energy-momentum tensor \cite{Harko/2011}. Another related theory is $f(R,L_m)$ gravity, with the action given by $S=\int f(R,L_m)\sqrt{-g}d^4x$ \cite{Bertolami/2007,Harko/2008}. Furthermore, a theoretical approach known as hybrid metric-Palatini gravity, which combines the metric and Palatini formalisms of modified gravity theories, was proposed in \cite{Harko/2012,Capozziello/2012} to construct a new type of gravitational Lagrangian.\\ 
Similarly, \(f(\mathcal{T})\) modified gravity theory generalizes the Teleparallel theory of gravity, which is equivalent to GR, by introducing an arbitrary function of the torsion scalar \(\mathcal{T}\), as originally introduced in references \cite{Moller/1961,Pellegrini/1963,Hayashi/1979}. In the teleparallel approach to gravity, the fundamental idea is to replace the metric \(g_{\mu\nu}\) of spacetime, which traditionally describes gravitational properties, with a set of tetrad vectors \(e_{\mu}^{i}\). The torsion generated by these tetrad fields can then be used to completely describe gravitational effects, replacing curvature with torsion. A significant advantage of \(f(\mathcal{T})\) gravity theory is that its field equations are of second order, unlike those of \(f(R)\) gravity, which are fourth order in the metric approach. For a detailed discussion of teleparallel theories, see \cite{Aldrovandi/2013}. Moreover, these theories can be extended by incorporating the trace of the energy-momentum tensor \(T\), leading to the \(f(\mathcal{T}, T)\) theory. These extensions are notable for generalizing the metric-affine theories, providing a rich structure capable of encompassing various gravitational phenomena within a single framework.\\
From the above presentation, it becomes evident that GR can be expressed in at least two equivalent geometric frameworks: the curvature representation, where both torsion and nonmetricity vanish, and the teleparallel representation, where both curvature and nonmetricity vanish. Additionally, a third equivalent representation is also feasible, which is fundamentally characterized by nonmetricity. This is called symmetric teleparallel representation, where both curvature and torsion vanish.\\
This thesis aims to investigate a third approach to modified gravity known as nonmetricity-based modified gravity. In this approach, the primary geometric variable that describes the properties of the gravitational interaction is given by the non-metricity \(Q\). This theory is discussed in detail in the following section.

\section{Nonmetricity-based Modified Gravity}\label{section 1.5}
The nonmetricity scalar, commonly denoted by $Q$, is a fundamental quantity in the context of metric-affine geometry, particularly in theories of gravity that extend GR. In the context of metric-affine geometry, it is not assumed that the connection $\Gamma_{\mu\nu}^{\lambda}$ is compatible with the metric $g_{\mu\nu}$. This leads to the concept of nonmetricity, defined as 
\begin{equation}
\label{Q}
    Q_{\lambda\mu\nu}=\nabla_{\lambda}g_{\mu\nu},
\end{equation}
 where $\nabla$ is the covariant derivative associated with the affine connection $\Gamma_{\mu\nu}^{\lambda}$.\\
 It should also be noted that nonmetricity is a quantity that depends on both the metric tensor and the connection. Specifically, by expanding Eq.\eqref{Q}, we obtain
\begin{equation}
     Q_{\lambda\mu\nu}=\nabla_{\lambda}g_{\mu\nu}=\partial_{\lambda}g_{\mu\nu}-\Gamma^{\beta}_{\lambda\nu}\,g_{\mu\beta}-\Gamma^{\beta}_{\lambda\mu}\,g_{\beta\nu}.
\end{equation}
 Let us explore the geometrical properties of nonmetricity in the next subsection.
 
\subsection{Geometrical interpretation of nonmetricity}
\begin{figure}[]
\centering
\includegraphics[width=5cm]{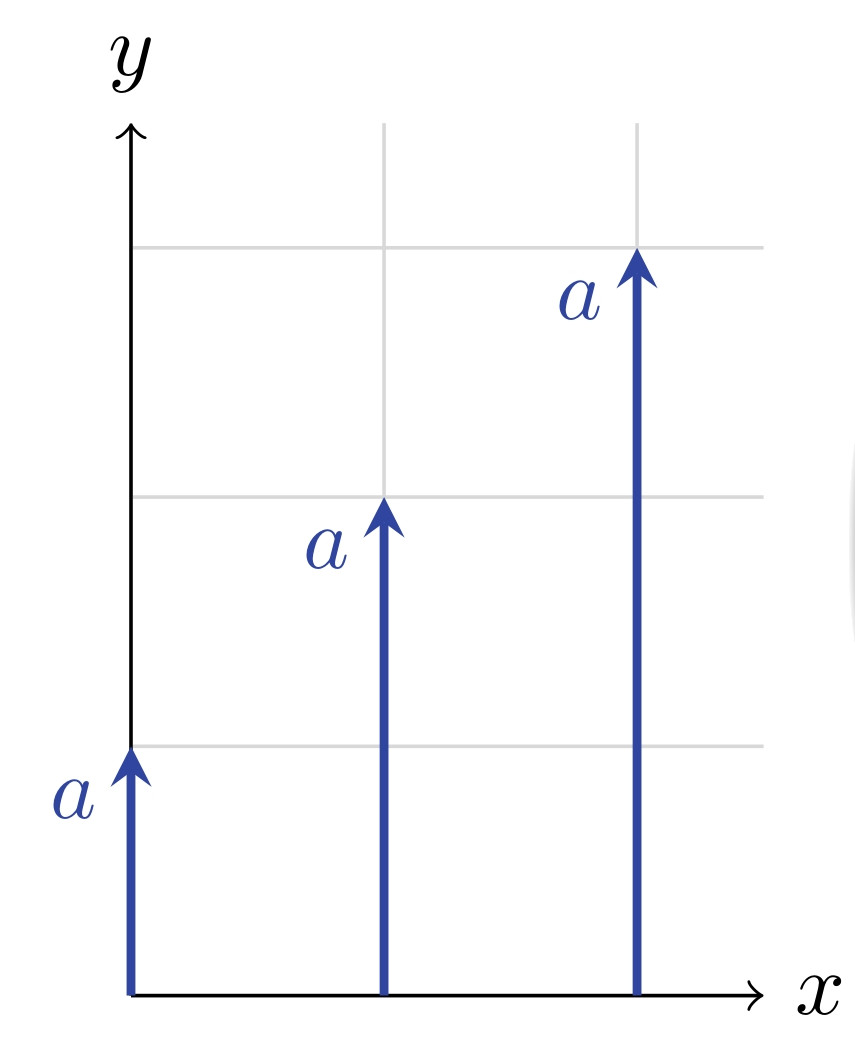}
\caption{Representation of nonmetricity in 2 dimensions. This figure is taken from the reference \cite{Saridakis/2021}.}
\label{Qrepresentation}
\end{figure}

To observe the effect of nonmetricity in the space, let us consider a vector \( a^{\mu} \) and define the magnitude of the vector \( a \) as \( a \cdot a = \| a \|^2 = g_{\mu\nu} a^{\mu} a^{\nu} \). Now, let us parallel transport the vector along a given curve \( C: x^{\mu}= x^{\mu}(\lambda) \). In a Riemannian space (where both torsion and nonmetricity vanish), we know that during such transportation, the inner product remains unchanged. Thus,
\begin{equation}
    \frac{D}{d\lambda}(\left\| a \right \|^2)=0.
\end{equation}
When nonmetricity is present a computation now reveals
\begin{equation}
    \frac{D}{d\lambda}(\left\| a \right \|^2)=\frac{D}{d\lambda}(g_{\mu\nu}\,a^{\mu}a^{\nu})=\frac{dx^{\alpha}}{d\lambda}(\nabla_{\alpha}g_{\mu\nu})\,a^{\mu}a^{\nu}+2\,g_{\mu\nu}\,\frac{dx^{\alpha}}{d\lambda}(\nabla_{\alpha}a^{\mu})\,a^{\nu}.
\end{equation}
Now, since $a^{\mu}$ is parallel transported along the curve, which means it satisfies the parallel transport equation
\begin{equation}
    \frac{dx^{\alpha}}{d\lambda}(\nabla_{\alpha}a^{\mu})=0.
\end{equation}
Thus, we arrive at
\[
    \frac{D}{d\lambda}(\left\| a \right \|^2) = Q_{\alpha\mu\nu}\frac{dx^{\alpha}}{d\lambda}\,a^{\mu}a^{\nu},
\]

where we have introduced the nonmetricity tensor \(Q_{\alpha\mu\nu} = \nabla_{\alpha}g_{\mu\nu}\). This expression indicates that the magnitude of a vector changes when it is parallel transported along a given curve. Consequently, nonmetricity relates to the property of vectors not preserving their magnitudes and inner products. The change in the magnitude of a vector during parallel transport along a curve is illustrated in Figure \ref{Qrepresentation}.

\subsection{Symmetric teleparallel theory equivalent to GR}
\begin{figure}[]
\centering
\includegraphics[width=16cm]{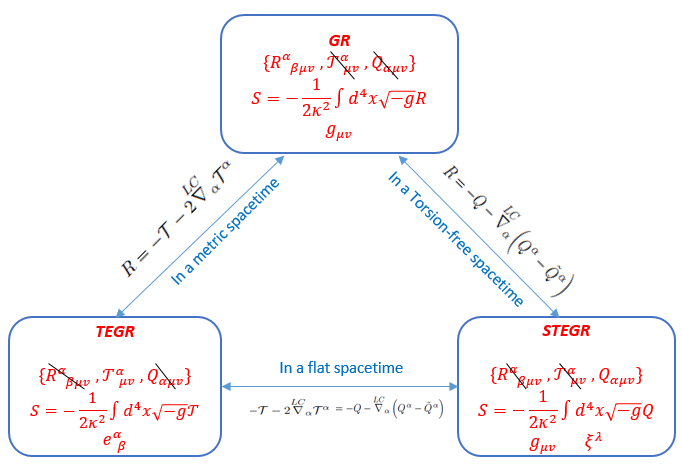}
\caption{Geometric trinity of gravity.}
\label{GT}
\end{figure}

 In the STEGR formulation, the properties of the gravitational interaction are described geometrically by the nonmetricity $Q$ of the metric, which defines the variation of the length of a vector during parallel transport around a closed loop. This formulation is referred to as symmetric teleparallel gravity (STG). It was first introduced by Nester and Yo \cite{Nester/1999}, where they emphasized that this approach provides a novel perspective on GR.  One of the important properties of STG is its ability to separate gravitational and inertial effects, which is not possible in GR \cite{Koivisto/2018}.\\
In differential geometry, the general affine connection can always be decomposed into three independent components \cite{Hehl/1995,Ortin/2015}
\begin{equation}
\Gamma^{\,\lambda}_{\,\,\,\mu\nu}=\{^{\,\lambda}_{\,\,\,\mu\nu}\}+K^{\,\lambda}_{\,\,\,\mu\nu}+L^{\,\lambda}_{\,\,\,\mu\nu},
\end{equation}
where the first term is the Levi-Civita connection of the metric $g_{\mu\nu}$, which is already defined in Eq.\eqref{LCC}.
The second term $K^{\,\lambda}_{\,\,\,\mu\nu}$ is the contortion
\begin{equation}
    K^{\,\lambda}_{\,\,\,\mu\nu}=\frac{1}{2}T^{\,\lambda}_{\,\,\,\mu\nu}+T^{\,\,\,\,\,\,\,\lambda}_{(\mu\,\,\,\,\,\,\nu)},
\end{equation}
with the torsion tensor defined as $T^{\,\lambda}_{\,\,\,\mu\nu}\equiv \Gamma^{\lambda}_{\,\,\,\,[\mu\nu]}$. The third term is the disformation
\begin{equation}
    L^{\,\lambda}_{\,\,\,\mu\nu}= -\frac{1}{2}g^{\lambda\sigma}\left(Q_{\mu\sigma\nu}+Q_{\nu\sigma\mu}-Q_{\sigma\mu\nu}\right),
\end{equation}
which is expressed in terms of the nonmetricity tensor in Eq.\eqref{Q}.\\ 
An interesting outcome of the interaction among different geometrical objects (such as curvature, torsion, and nonmetricity) is that it is possible to develop gravitational theories that are equivalent to GR without relying on the curvature. Instead, these theories utilize nonmetricity or torsion. In these theories, the curvature tensor $\mathcal{R}^{\alpha}_{\,\,\,\,\beta\mu\nu}$ of the affine connection can be defined, which is expressed in terms of the curvature tensor $R^{\alpha}_{\,\,\,\,\beta\mu\nu}$ of the Levi-Civita connection, together with contributions from torsion and nonmetricity. It can be written as follows
\begin{equation}
\label{R}
 \mathcal{R}^{\alpha}_{\,\,\,\,\beta\mu\nu}=R^{\alpha}_{\,\,\,\,\beta\mu\nu}+\mathcal{T}^{\lambda}_{\,\,\,\,\mu\nu}M^{\alpha}_{\,\,\,\,\lambda\beta}+2\overset{LC}{\nabla}_{[\mu}M^{\alpha}_{\,\,\,\,\nu]\beta}+2M^{\alpha}_{\,\,\,\,[\mu|\lambda|}M^{\lambda}_{\,\,\,\,\nu]\beta}.
\end{equation}
The symbol $\overset{LC}{\nabla}$ represents the covariant derivative with respect to the Levi-Civita connection. Furthermore, the expression $M^{\lambda}_{\,\,\,\,\mu\nu}=K^{\lambda}_{\,\,\,\,\mu\nu}+L^{\lambda}_{\,\,\,\,\mu\nu}$ encompasses all contributions that arise from both torsion and nonmetricity.\\
We contract the curvature tensor \eqref{R} to form the curvature scalar, and we get
\begin{equation}
\label{RS}
    \mathcal{R}=R+\mathcal{T}+Q+\mathcal{T}^{\lambda\mu\nu}Q_{\mu\nu\lambda}-\mathcal{T}^{\mu}\,Q_{\mu}+\mathcal{T}^{\mu}\,\tilde{Q}_{\mu}+\overset{LC}{\nabla}_{\alpha}\left(Q^{\,\alpha}-\tilde{Q}^{\,\alpha}+2\mathcal{T}^{\alpha}\right).
\end{equation}
It is clear that if we restrict the geometry to have zero torsion and nonmetricity, the curvature scalar \eqref{RS} is simply
\begin{equation}
    \mathcal{R}=R.
\end{equation}
This is reduced to the case of GR. If we constrain the geometry to a Weitzen\"{o}ck geometry, where both curvature and nonmetricity are set to zero, leaving torsion as the sole contributor to gravity, Eq.\eqref{RS} simplifies to
\begin{equation}
\label{T}
    R=-\mathcal{T}-2\overset{LC}{\nabla}_{\alpha}\mathcal{T}^{\alpha},
\end{equation}
where $\mathcal{T}$ is the torsion scalar and it can be written as
\begin{equation}
    \mathcal{T}\equiv\frac{1}{4}\mathcal{T}_{\lambda\mu\nu}\mathcal{T}^{\lambda\mu\nu}+\frac{1}{2}\mathcal{T}_{\lambda\mu\nu}\mathcal{T}^{\nu\mu\lambda}-\mathcal{T}_{\lambda}\mathcal{T}^{\lambda}.
\end{equation}
In Eq.\eqref{T}, we observe that the curvature scalar in GR is replaced by the torsion scalar in a Weitzen\"{o}ck geometry, along with an additional divergence term subject to a covariant derivative. This extra divergence term is integrated and evaluated at infinity, or the boundary, and does not affect the field equations. This implies that both geometrical frameworks are equivalent to each other. This gravitational theory, which depends on torsion rather than curvature, is equivalent to GR and is known as the Teleparallel Equivalent of GR (TEGR).\\
If we constrain the geometry to a symmetric teleparallel geometry, where both curvature and torsion are set to zero, and only nonmetricity remains as the determining factor of gravity. Then, by using Eq.\eqref{RS}, we can represent the Ricci scalar $R$ of the Levi-Civita connection in terms of the nonmetricity scalar and a divergence term, and it can be written as follows
\begin{equation}
\label{Q1}
    R=-Q-\overset{LC}{\nabla}_{\alpha}\left(Q^{\,\alpha}-\tilde{Q}^{\,\alpha}\right),
\end{equation}
where $Q$ is the nonmetricity scalar, which is given by
\begin{equation}
    Q=\frac{1}{4}Q_{\lambda\mu\nu}Q^{\lambda\mu\nu}-\frac{1}{2}Q_{\lambda\mu\nu}Q^{\nu\mu\lambda}-\frac{1}{4}Q_{\lambda}Q^{\lambda}+\frac{1}{2}Q_{\lambda}\tilde{Q}^{\lambda}.
\end{equation}
We define the two traces of nonmetricity tensor as
\begin{equation}
\label{2}
Q_{\lambda}=Q_{\lambda\,\,\,\,\mu}^{\,\,\,\,\mu}\, ,\,\,\,\,\,\,\,\,\tilde{Q}_{\lambda}=Q^{\mu}_{\,\,\,\,\lambda\mu}\,.
\end{equation}
Applying the same reasoning used when considering torsion instead of curvature, we observe that the curvature scalar of the Levi-Civita connection in Eq.\eqref{Q1} differs from the nonmetricity scalar only by a boundary term. This indicates that the theory is fundamentally equivalent to GR. This mutual equivalence, summarized in Figure \ref{GT}, is referred to as the geometric trinity of gravity. This non-metric theory of gravity, known as the Symmetric Teleparallel Equivalent of GR (STEGR), is described by the action
\begin{equation}
\label{STGaction}
    S_{STG}=-\frac{1}{2\kappa^2}\int d^4x \sqrt{-g}\,Q(g,\xi)+S_{matter}.
\end{equation}
Additionally, the conjugate of nonmetricity tensor $P_{\,\,\mu\nu}^{\lambda}$ is defined as 
\begin{equation}
\label{3}
4P_{\,\,\mu\nu}^{\lambda}=-Q^{\lambda}_{\,\,\,\,\mu\nu}+2Q^{\,\,\,\,\,\,\lambda}_{(\mu\,\,\,\,\nu)}-Q^{\lambda}g_{\mu\nu}-\tilde{Q}^{\lambda}g_{\mu\nu}-\delta^{\lambda}_{(\mu}\, Q\,_{\nu)},
\end{equation} 
obtaining a nonmetricity scalar as 
\begin{equation}
\label{4}
Q=-Q_{\lambda\mu\nu}P^{\lambda\mu\nu}.
\end{equation}
The nonmetricity scalar can also be expressed in terms of disformation tensor as
\begin{equation}
    Q=g^{\mu\nu}\left(L^{\,\alpha}_{\,\,\,\alpha\beta} L^{\,\beta}_{\,\,\,\mu\nu}- L^{\,\alpha}_{\,\,\,\beta\mu} L^{\,\beta}_{\,\,\,\nu\alpha}\right).
\end{equation}
Before we continue, it is important to mention that our geometrical framework has a flat and torsion-free connection. This means that it corresponds to a pure coordinate transformation from the trivial connection as described in \cite{Jimenez/2018}. More specifically, the connection can be parameterized with $\xi^{\sigma}$, which is a set of four arbitrary functions of the coordinates $x^{\sigma}$, as seen below
\begin{equation}
\label{Gamma}
    \Gamma^{\lambda}_{\,\,\,\mu\nu}=\frac{\partial x^{\lambda}}{\partial\xi^{\sigma}}\frac{\partial}{\partial x^{\mu}}\left( \frac{\partial \xi^{\sigma}}{\partial x^{\nu}}\right).
\end{equation}
Here, \(\frac{\partial x^{\lambda}}{\partial \xi^{\sigma}}\) represents the inverse of the Jacobian matrix \(\frac{\partial \xi^{\sigma}}{\partial x^{\lambda}}\). This implies that in any coordinate system \((x^0, x^1, x^2, x^3)\), we can select four independent functions \((\xi^0, \xi^1, \xi^2, \xi^3)\) such that the Jacobian matrix \(\frac{\partial \xi^{\sigma}}{\partial x^{\lambda}}\) is invertible, enabling the construction of a flat and torsionless connection via Eq.\eqref{Gamma}. Moreover, Eq.\eqref{Gamma} shows that the flat and torsionless connections possess a notable property: they can be globally set to zero through an appropriate choice of coordinates. Specifically, any flat and torsionless connection takes the form \eqref{Gamma} with some functions \(\xi^{\sigma}\). By choosing coordinates such that \(x^{\sigma} = \xi^{\sigma}\), the connection vanishes globally because \(\frac{\partial}{\partial x^{\mu}}\left( \frac{\partial \xi^{\sigma}}{\partial x^{\nu}} \right) = 0\). This coordinate choice is known as the coincident gauge. In this gauge, where the connection is globally zero, \(L^{\lambda}_{\ \mu\nu} = -\{^{\lambda}_{\ \mu\nu}\}\), allowing the nonmetricity scalar \(Q\) to be expressed in terms of the Levi-Civita connection as
\begin{equation}
    Q=g^{\mu\nu}\left(\{^{\,\alpha}_{\,\,\,\alpha\beta}\}\{^{\,\beta}_{\,\,\,\mu\nu}\}-\{^{\,\alpha}_{\,\,\,\beta\mu}\}\{^{\,\beta}_{\,\,\,\nu\alpha}\}\right).
\end{equation}
In this coincident gauge, it is clear that the action in \eqref{STGaction} can be expressed as
\begin{equation}
 S_{CGR}=S_{STG}[\Gamma=0]=\frac{1}{2\kappa^2}\int d^4x \sqrt{-g}\,g^{\mu\nu}\left(\{^{\,\alpha}_{\,\,\,\beta\mu}\}\{^{\,\beta}_{\,\,\,\nu\alpha}\}-\{^{\,\alpha}_{\,\,\,\alpha\beta}\}\{^{\,\beta}_{\,\,\,\mu\nu}\}\right)+S_{matter}.
\end{equation}
We refer to this as the action of Coincident General Relativity (CGR).
The right-hand side of this action corresponds to the quadratic component of the Hilbert action, which differs only from the curvature scalar of GR by a total derivative. It benefits from involving only the first derivatives of the metric, resulting in a well-posed variational principle without the need for Gibbons-Hawking-York (GHY) boundary terms. Consequently, we arrive at the remarkable conclusion that the theory described by \(Q\) is equivalent to an improved version of GR, where the boundary term is absent. The connection can be fully trivialized, offering a much simpler geometrical interpretation of gravity, where the origins of the tangent space and spacetime coincide. This theory is referred to as coincident GR.

\subsection{The extensions of STEGR}\label{section 1.6}
The standard STEGR provides a compelling framework to describe gravity through nonmetricity. However, various extensions of STEGR have been proposed to address unresolved questions in cosmology. These extensions introduce modifications by incorporating scalar fields or functional generalizations of the nonmetricity scalar. In the following subsections, we discuss two prominent extensions: scalar-nonmetricity theory and \(f(Q)\) gravity theory, both of which aim to provide a deeper insight into the observed cosmic phenomena.  
\subsubsection{Scalar-nonmetricity theory}
The scalar theory in nonmetricity gravity extends the framework of symmetric teleparallel gravity (STG), which is built on the concept of nonmetricity, by introducing scalar fields. Scalar fields in this framework serve as additional degrees of freedom that enrich the theory's structure and potential applications. Scalar fields in nonmetricity gravity interact with the space-time geometry characterized by the nonmetricity tensor. These interactions can be categorized into minimal and non-minimal coupling based on the relationship between the scalar field and the non-metricity scalar $Q$.\\
\textbf{Minimal coupling:} The scalar field does not pair directly with the nonmetricity scalar $Q$. It evolves independently, with its dynamics influenced only through the spacetime metric $g_{\mu\nu}$.\\
In this case, the action is given by 
\begin{equation}
\label{ph}
    S=-\frac{1}{2}\int \left[Q- g^{\mu\nu} \partial_{\mu}\phi\, \partial_{\nu}\phi - 2V(\phi)\right]\sqrt{-g}\,d^4x+S_{matter},
\end{equation}
where the second term represents the kinetic energy of the scalar field \(\phi\), and \(V(\phi)\) is its potential energy. Here, the scalar field evolves independently of the nonmetricity scalar $Q$, resulting in simpler dynamics similar to standard scalar field cosmology.\\   
\textbf{Non-minimal coupling:} The scalar field explicitly couples to the nonmetricity scalar \( Q \), modifying the gravitational dynamics and introducing richer interactions. In this framework, the non-minimal coupling introduces an additional interaction term, \( F(\phi)Q \), where the scalar field modulates the effective gravitational strength.\\
The action is given by \cite{Runkla/2018,Jara/2018}
\begin{equation}
\label{ph}
   S=-\frac{1}{2}\int \left[F(\phi)Q-G(\phi) g^{\mu\nu} \partial_{\mu}\phi\, \partial_{\nu}\phi - 2V(\phi)\right]\sqrt{-g}\,d^4x+S_{matter},
\end{equation}
where $F(\phi)$ is the coupling function between the scalar field and the gravitational scalar $Q$. This direct interaction allows for more complex gravitational effects and broader cosmological applications such as the early inflationary epoch \cite{Belinchon/2017} and the current accelerated expansion of the Universe \cite{Ade/2016}.

\subsubsection{$f(Q)$ gravity theory}
Now, we come to the extension of STEGR by replacing the nonmetricity $Q$ in action \eqref{STGaction} with an arbitrary function $f(Q)$. This extended version of STEGR is referred to as $f(Q)$ gravity theory proposed by \cite{Jimenez/2018}. The action functional of $f(Q)$ gravity is described as
\begin{equation}
    \label{f(Q)action}
    S_{f(Q)}=-\frac{1}{2\kappa^2}\int d^4x \sqrt{-g}\,f(Q)+S_{matter},
\end{equation}
where $f(Q)$ is the arbitrary function  of $Q$.\\
In \(f(Q)\) gravity, the equations of motion are obtained by varying the action with respect to the metric. This variation results in a set of modified field equations that include the influence of nonmetricity, offering a richer structure for describing gravitational phenomena. The motion equations are expressed as
\begin{eqnarray}
\label{fQ_FE}
\frac{2}{\sqrt{-g}}\nabla_{\sigma}\left(f_Q\sqrt{-g}\,P^{\sigma}_{\,\,\mu\nu}\right)-\frac{1}{2}f(Q)\,g_{\mu\nu}
+f_Q\left(P_{\mu\sigma\lambda}Q_{\nu}^{\,\,\,\sigma\lambda}-2Q_{\sigma\lambda\mu}P^{\sigma\lambda}_{\,\,\,\,\,\,\nu}\right)=\kappa^2 T_{\mu\nu},
\end{eqnarray}
where $f_Q=\frac{d f}{d Q}$. The energy-momentum tensor for matter is defined in Eq.\eqref{EMT1}.
Also, by varying action \eqref{f(Q)action} with respect to the connection results in
\begin{equation}
\label{conn}
\nabla_{\sigma}\lambda_{k}^{\,\,\,\mu\nu\sigma}+\lambda_{k}^{\,\,\,\mu\nu}=\sqrt{-g}f_Q\,P_{\,k}^{\,\,\,\mu\nu}+H_{\,k}^{\,\,\,\mu\nu},
\end{equation}
where $H_{\,k}^{\,\,\,\mu\nu}=-\frac{1}{2}\frac{\delta\mathcal{L}_m}{\delta\Gamma^{k}_{\,\,\,\mu\nu}}$ is the hypermomentum tensor density.

It is possible to simplify Eq.\eqref{conn} by taking into account the antisymmetry property of $\mu$ and $\nu$ in the Lagrangian multiplier coefficients
\begin{equation}
    \nabla_{\mu}\nabla_{\nu} \left(f_{Q}\sqrt{-g}\,P_{\,k}^{\,\,\,\mu\nu}+H_{\,k}^{\,\,\,\mu\nu}\right)=0.
\end{equation}
More specifically, the connection can be parameterized with a collection of functions $\xi^{\alpha}$ as Eq.\eqref{Gamma}. The connection equation of motion can be easily calculated by noticing that the variation of the connection with respect to $\xi^{\sigma}$ is equivalent to performing a diffeomorphism so that $\delta_{\xi}\Gamma^{\,\sigma}_{\,\,\,\mu\nu}=-\mathcal{L}_{\xi}\Gamma^{\,\sigma}_{\,\,\,\mu\nu}=-\nabla_{\mu}\nabla_{\nu}\xi^{\sigma}$, where we have used the connection to be flat and torsion-free \cite{Jimenez/2020}. Furthermore, in the absence of hypermomentum\footnote{If there is no hypermomentum, this is trivially true. Second, we can assume that the hypermomentum is antisymmetric, in which case our assertion is identically true ($H_{\,k}^{\,\,\,\mu\nu}= 0$). Finally, if $H_{\,k}^{\,\,\,\mu\nu}\neq 0$, we consider the assertion to be the hypermomentum conservation law.}, the connection field equations read as
\begin{equation}
\label{8}
\nabla_{\mu}\nabla_{\nu} \left(f_{Q}\sqrt{-g}\,P_{\,k}^{\,\,\,\mu\nu}\right)=0.
\end{equation}

We can also extend the $f(Q)$ gravity theory by adding the Lagrangian scalar field to the action \eqref{f(Q)action}. The action of $f(Q)$ gravity theory in the presence of a scalar field is defined as\footnote{In this action, we take \( Q + f(Q) \) instead of just \( f(Q) \) because we aim to retain the standard Einstein-Hilbert term in the limit where \( f(Q) \to 0 \), ensuring consistency with General Relativity in the appropriate regime.}  
\begin{equation}
\label{f(Q)scalarfield}
S=\int \left\{-\frac{1}{2\kappa^2}\left[Q+f(Q)\right]+\mathcal{L}_{\phi}+\mathcal{L}_m\right\}\sqrt{-g}\,d^4x,
\end{equation}
where $\mathcal{L}_{\phi}$ is the scalar field Lagrangian.\\
Varying action \eqref{f(Q)scalarfield} with respect to the metric, the gravitational field equation for $f(Q)$ is obtained as
\begin{multline}
\label{7}
\frac{2}{\sqrt{-g}}\nabla_{\sigma}\left((1+f_Q)\sqrt{-g}\,P^{\sigma}_{\,\,\mu\nu}\right)+\frac{1}{2}(Q+f(Q))\,g_{\mu\nu}
+\\(1+f_Q)\left(P_{\mu\sigma\lambda}Q_{\nu}^{\,\,\,\sigma\lambda}-2Q_{\sigma\lambda\mu}P^{\sigma\lambda}_{\,\,\,\,\,\,\nu}\right)= T_{\mu\nu}^{\phi}+T_{\mu\nu}.
\end{multline}
 The energy-momentum tensor for the scalar field is defined as
\begin{eqnarray}
    T_{\mu\nu}^{\phi}&\equiv & \partial_{\mu}\phi\,\partial_{\nu}\phi-\frac{1}{2}g_{\mu\nu}\,g_{\alpha\beta}\,\partial^{\alpha}\phi\,\partial^{\beta}\phi-g_{\mu\nu}V(\phi),
\end{eqnarray}
and for matter sector is defined in Eq.\eqref{EMT1}.

\subsection{Literature review}
Modified gravity theories have emerged as promising alternatives to GR in addressing unresolved issues in cosmology, such as the late-time accelerated expansion of the Universe and the nature of DE. Among these, STG or $f(Q)$ gravity, introduced by Jimenez, Heisenberg, and Koivisto \cite{Jimenez/2018}, where the nonmetricity $Q$ is responsible for the gravitational interaction. Investigations on $f(Q)$ gravity have been rapidly developed as well as observational constraints to
confront it against standard GR formulation. An interesting set of constraints on $f(Q)$ gravity were done by Lazkoz et al. \cite{Lazkoz/2019}, where the $f(Q)$ Lagrangian is written as a polynomial function of the redshift $z$. It has also been successfully confronted with a variety of background and perturbation observational data, including the Supernovae type Ia (SNIa), BAO, CMB, Redshift Space Distortion (RSD), growth data \cite{Lazkoz/2019,ob1,ob3,ob4,ob5,ob6,ob7,ob8}. This confrontation reveals that $f(Q)$ gravity may challenge the conventional $\Lambda$CDM scenario. Moreover, the Big Bang Nucleosynthesis (BBN) restrictions are easily overcome by $f(Q)$ gravity \cite{Anagnostopoulos/2023}. For the first time, Sokoliuk et al. \cite{simranlss} used N-body simulations of $f(Q)$ gravitation to study large-scale structure creation observables in order to evaluate the applicability of the theory to cosmological context. Mandal et al. \cite{Mandal/2021a} analyzed energy conditions to assess the stability of their cosmological models and constrained model parameters using current cosmological data in \( f(Q) \) gravity. In another study, Mandal et al. \cite{Mandal/2021b} employed cosmography to express \( f(z) \) and its derivatives up to fourth order in terms of observable parameters. By fitting these functions to the Pantheon Supernovae Ia dataset, they constrained the \( H_0 \) and cosmographic parameters, highlighting the dominance of the first two terms in the Taylor expansion of the luminosity distance in \( f(Q) \) gravity. Atayde and Frusciante \cite{Atayde/2021} investigated observational constraints on nonmetricity $f(Q)$-gravity, which reproduces the $\Lambda$CDM background expansion while modifying linear perturbations. Using CMB, BAO, RSD, SNIa, GC, and WL data, they found the model statistically favored over $\Lambda$CDM due to its improved fit to the low-$l$ tail of CMB temperature anisotropies. Capozziello and D'Agostino \cite{Capozziello/2022} reconstructed \( f(Q) \) gravity model-independently using Padé approximations, which ensured stability at high redshifts. Their numerical analysis identified \( f(Q) = \alpha + \beta Q^n \) as the best fit for accelerated expansion and considered potential deviations from \(\Lambda\)CDM. Furthermore, various methodologies exist for discussing the cosmophysical properties, including assuming specific forms of $f(Q)$, studying the dynamical behavior of backgrounds and perturbations, and validating outcomes through observational testing, including comparisons with the solar system \cite{Jimenez/2020,Khyllep/2021,Khyllep/2023}. Furthermore, several studies have investigated the existence and polarization of gravitational waves within the framework of STG and their modifications \cite{ob1,Hohmann/2019,Capozziello/2024,Capozziello/2024b,Capozziello/2024c}. These works aim to understand how such modifications affect the properties and dynamics of gravitational waves. The study by Gomes et al. \cite{Gomes/2024} highlights the presence of ghosts and strong coupling issues in $f(Q)$ theories, particularly in cosmological settings, raising concerns about their physical viability. The ghost mode can be eliminated by imposing specific constraints on it \cite{Hu/2023,Nojiri/2024,Nojiri/2024b}. Several other significant works have also been conducted in \( f(Q) \) theory \cite{Q1,Q2,Q3,Q4,Q5,Q6,Q7,Q8}. These studies have explored various aspects of the theory, including its extensions, cosmological applications, and potential to explain phenomena such as DE and cosmic acceleration. Such contributions have enriched the understanding of \( f(Q) \) gravity and its place within the broader framework of modified gravity theories. Ghosh et al. \cite{Ghosh1,Ghosh2} investigated the dynamics of scalar field cosmology within the framework of coincident $f(Q)$ gravity by employing a dynamical systems approach. They analyzed the cosmological implications of the theory under different scalar field potentials, aiming to understand the evolution of the Universe in this modified gravity context. For a more rigorous and detailed overview of \( f(Q) \) theory, readers can refer to the review article \cite{Heisenberg/2024}.

In addition, several modifications and extensions of \( f(Q) \) gravity have been proposed. Notable examples include \( f(Q,T) \) gravity \cite{Xu/2019}, where \( T \) represents the trace of the energy-momentum tensor, Weyl-type \( f(Q,T) \) gravity \cite{Xu/2020}, and \( f(Q,B) \) or \( f(Q,C) \) gravity \cite{Capozziello/2023,Avik/2024}, where \( B \) or \( C \) denotes a boundary term.

\section{Gaussian Process}\label{section 1.7}
This section introduces a fundamental statistical concept known as the Gaussian process (GP). The GP is widely used in cosmology to reconstruct smooth functions from observational datasets without requiring any specific parameterization of the underlying function. In the context of modified gravity theories, such as \(f(Q)\) gravity, the GP facilitates the reconstruction of the Hubble parameter \(H(z)\) and its derivatives directly from observational datasets \footnote{In this process, we particularly used the observational Hubble dataset.}. By providing a model-independent framework, the GP approach effectively bridges theoretical predictions with observational cosmology.
 
Gaussian processes (GPs) are highly adaptable and effective tools in machine learning and statistics, particularly for tasks such as regression, classification, and probabilistic modeling. GP regression, in particular, is widely used for reconstructing functions and their derivatives from observed data without relying on predefined functional forms. A comprehensive discussion of GPs can be found in the book by Rasmussen and Williams \cite{Rasmussen/2006}. Basically, a GP is defined as  
\[
f(x) \sim \mathcal{GP}(\mu(x), k(x, \Tilde{x})),
\]  
where \(k(x, \Tilde{x}) = \mathbb{E}[(f(x) - \mu(x))(f(\Tilde{x}) - \mu(\Tilde{x}))]\) is the kernel function that represents the covariance between the function values in \(x\) and \(\Tilde{x}\), and \(\mu(x) = \mathbb{E}[f(x)]\) is the mean of the function at each point \(x\).\\
This process provides a nonparametric approach to modeling functions \(f(x)\), assuming that the \(n\) observations of a dataset \(y = \{y_1, y_2, \dots, y_n\}\) are sampled from a multivariate Gaussian distribution. These data points are associated with a GP, often with a mean function set to zero for simplicity. Despite the straightforward nature of GP-based modeling, there are critical aspects to consider, particularly regarding the choice of components. For example, the values of the function at different points \(x_1\) and \(x_2\) are correlated, necessitating the use of a covariance function \(k(x_1, x_2)\). However, the specific form of this function is not unique. A commonly used covariance function is the squared exponential kernel, given by  
\[
k(x_1, x_2) = \sigma_f^2 \exp\left(-\frac{(x_1 - x_2)^2}{2l^2}\right),
\]  
where \(\sigma_f\) and \(l\) are hyperparameters. Here, \(l\) defines the length of the correlation, while \(\sigma_f\) governs the variation of \(f(x)\) relative to the mean. This kernel is widely favored for its smoothness and differentiability, making it suitable for reconstructing both functions and their derivatives.

The covariance matrix for \(n\) observation points \(\{x_1, x_2, \dots, x_n\}\) is expressed as  
\[
K =
\begin{bmatrix}
k(x_1, x_1) & k(x_1, x_2) & \cdots & k(x_1, x_n) \\
k(x_2, x_1) & k(x_2, x_2) & \cdots & k(x_2, x_n) \\
\vdots      & \vdots      & \ddots & \vdots      \\
k(x_n, x_1) & k(x_n, x_2) & \cdots & k(x_n, x_n)
\end{bmatrix}.
\]  
For a new point \(x_*\), the vector of covariances is  
\[
K_* = 
\begin{bmatrix}
k(x_*, x_1) & k(x_*, x_2) & \cdots & k(x_*, x_n)
\end{bmatrix},
\]  
and the covariance at the new point is \(K_{**} = k(x_*, x_*)\).

The joint distribution of observed data \(y\) and the prediction \(y_*\) at \(x_*\) is  
\[
\begin{bmatrix}
y \\ y_*
\end{bmatrix}
\sim \mathcal{N} \left( 0, 
\begin{bmatrix}
K & K_*^T \\
K_* & K_{**}
\end{bmatrix}
\right).
\]  
From this, the conditional probability  
\[
p(y_*|y) \sim \mathcal{N}(K_* K^{-1} y, K_{**} - K_* K^{-1} K_*^T),
\]  
provides the predictive mean and variance. The mean, which serves as an estimate of \(y_*\), is  
\[
\mu(y_*) = K_* K^{-1} y,
\]  
while the variance, which represents the uncertainty in the estimate, is  
\[
{\rm Var}(y_*) = K_{**} - K_* K^{-1} K_*^T.
\]
A detailed account of this method is provided in Ref. \cite{Seikel/2012}, and the associated algorithms can be downloaded from the author's website \url{http://www.acgc.uct.ac.za/~seikel/GAPP/index.html}.

\subsection{Observational Hubble data (OHD)}
The expansion rate of the Universe, \( H(z) \), is defined as \( H(z) = -\frac{1}{1+z} \frac{dz}{dt} \), where \( dz \) is derived from spectroscopic surveys. Measurements of the Hubble parameter are obtained from early-type galaxies undergoing passive evolution by estimating their differential ages along the line of sight. In the following chapters, we utilize the latest dataset of 58 Hubble parameter measurements, along with their associated error bars, for the Gaussian reconstruction process. Of these 58 data points \cite{Ratra/2018}, 32 are derived from cosmic chronometer (CC) observations \cite{Moresco/2012,Moresco/2016,Bruzual/2003,Maraston/2011,Zhang/2014,Moresco/2015,Stern/2010}, which estimate \( H(z) \) in a model-independent manner based on the age evolution of passively evolving galaxies. The remaining 26 points are obtained from radial Baryon Acoustic Oscillation (BAO) observations, which measure galaxy clustering using the BAO peak position as a standard ruler \cite{Sharov/2018}. The position of the BAO peak is sensitive to the sound horizon scale. The overall OHD dataset spans a redshift range of \( 0.07 < z < 2.42 \), providing valuable insight into the expansion history of the Universe.
\begin{table*}[hb]
\begin{center}

        \caption{\justifying \textit{Here, table contains the $58$ points of Hubble parameter values $H(z)$ with errors $\sigma _{H}$ from differential age }(\textit{$32$ points})\textit{, and BAO and other} (\textit{$26$ points})\textit{ approaches, along with references.}}
        \label{Table 1}     
\begin{tabular}{|c c c c c c c c|}\hline
\multicolumn{8}{|c|}{Table-1: $H(z)$ datasets consisting of 58 data points} \\ \hline
\multicolumn{8}{|c|}{CC data (32 points)}  \\ \hline
$z$ & $H(z)$ & $\sigma _{H}$ & Ref. & $z$ & $H(z)$ & $\sigma _{H}$ & Ref. \\ \hline
$0.070$ & $69$ & $19.6$ & \cite{h1} & $0.4783$ & $80$ & $99$ & \cite{h5} \\ \hline
$0.09$ & $69$ & $12$ & \cite{h2} & $0.480$ & $97$ & $62$ & \cite{h1} \\ \hline
$0.120$ & $68.6$ & $26.2$ & \cite{h1} & $0.593$ & $104$ & $13$ & \cite{h3} \\ \hline
$0.170$ & $83$ & $8$ & \cite{h2} & $0.6797$ & $92$ & $8$ & \cite{h3} \\ \hline
$0.1791$ & $75$ & $4$ & \cite{h3} & $0.75$ & $98.8$ & $33.6$ & \cite{h} \\ \hline
$0.1993$ & $75$ & $5$ & \cite{h3} & $0.7812$ & $105$ & $12$ & \cite{h3} \\ \hline
$0.200$ & $72.9$ & $29.6$ & \cite{h4} & $0.8754$ & $125$ & $17$ & \cite{h3} \\ \hline
$0.270$ & $77$ & $14$ & \cite{h2} & $0.880$ & $90$ & $40$ & \cite{h1} \\ \hline
$0.280$ & $88.8$ & $36.6$ & \cite{h4} & $0.900$ & $117$ & $23$ & \cite{h2} \\ \hline  
$0.3519$ & $83$ & $14$ & \cite{h3} & $1.037$ & $154$ & $20$ & \cite{h3} \\ \hline 
$0.3802$ & $83$ & $13.5$ & \cite{h5} & $1.300$ & $168$ & $17$ & \cite{h2} \\ \hline 
$0.400$ & $95$ & $17$ & \cite{h2} & $1.363$ & $160$ & $33.6$ & \cite{h7} \\ \hline 
$0.4004$ & $77$ & $10.2$ & \cite{h5} & $1.430$ & $177$ & $18$ & \cite{h2} \\ \hline 
$0.4247$ & $87.1$ & $11.2$ & \cite{h5} & $1.530$ & $140$ & $14$ & \cite{h2} \\ \hline
$0.4497$ & $92.8$ & $12.9$ & \cite{h5} & $1.750$ & $202$ & $40$ & \cite{h2} \\ \hline
$0.470$ & $89$ & $34$ & \cite{h6} & $1.965$ & $186.5$ & $50.4$ & \cite{h7}  \\ \hline
\multicolumn{8}{|c|}{From BAO \& other method (26 points)} \\ \hline
$z$ & $H(z)$ & $\sigma _{H}$ & Ref. & $z$ & $H(z)$ & $\sigma _{H}$ & Ref. \\ \hline
$0.24$ & $79.69$ & $2.99$ & \cite{h8} & $0.52$ & $94.35$ & $2.64$ & \cite{h10} \\ \hline
$0.30$& $81.7$ & $6.22$ & \cite{h9} & $0.56$ & $93.34$ & $2.3$ & \cite{h10} \\ \hline
$0.31$ & $78.18$ & $4.74$ & \cite{h10} & $0.57$ & $87.6$ & $7.8$ & \cite{h14} \\ \hline
$0.34$ & $83.8$ & $3.66$ & \cite{h8} & $0.57$ & $96.8$ & $3.4$ & \cite{h15} \\ \hline
$0.35$ & $82.7$ & $9.1$ & \cite{h11} & $0.59$ & $98.48$ & $3.18$ & \cite{h10} \\ \hline
$0.36$ & $79.94$ & $3.38$ & \cite{h10} & $0.60$ & $87.9$ & $6.1$ & \cite{h13} \\ \hline
$0.38$ & $81.5$ & $1.9$ & \cite{h12} & $0.61$ & $97.3$ & $2.1$ & \cite{h12} \\ \hline
$ 0.40$ & $82.04$ & $2.03$ & \cite{h10} & $0.64$ & $98.82$ & $2.98$ & \cite{h10}  \\ \hline
$0.43$ & $86.45$ & $3.97$ & \cite{h8} & $0.73$ & $97.3$ & $7.0$ & \cite{h13} \\ \hline
$0.44$ & $82.6$ & $7.8$ & \cite{h13} & $2.30$ & $224$ & $8.6$ & \cite{h16} \\ \hline
$0.44$ & $84.81$ & $1.83$ & \cite{h10} & $2.33$ & $224$ & $8$ & \cite{h17} \\ \hline
$0.48$ & $87.79$ & $2.03$ & \cite{h10} & $2.34$ & $222$ & $8.5$ & \cite{h18} \\ \hline
$0.51$ & $90.4$ & $1.9$ & \cite{h12} & $2.36$ & $226$ & $9.3$ & \cite{h19} \\ \hline
\end{tabular}
 \end{center}
 \end{table*}

\section{Conclusion} \label{section 1.8}
In this chapter, we provide a comprehensive introduction to the foundational concepts and challenges in modern cosmology that motivate the investigation of alternative gravity theories. The chapter began with an overview of the mathematical framework underpinning GR, which serves as the backbone for understanding gravitational phenomena. We then introduced the mathematical and theoretical foundations for exploring nonmetricity-based modified gravity theories, particularly \( f(Q) \) gravity, as alternatives to the \( \Lambda \)CDM model. Although GR and \( \Lambda \) CDM have achieved significant success in explaining cosmic phenomena, they face critical challenges, such as the cosmological constant problem, the coincidence problem, and tensions such as the Hubble and \( \sigma_8 \) discrepancies. These issues highlight the need for modified gravity frameworks that can address these limitations.

We discussed nonmetricity-based gravity as a promising extension of GR, focusing on \( f(Q) \) gravity, which reformulates gravity using nonmetricity rather than curvature. Additionally, we introduced GPs as a model-independent method for reconstructing cosmological functions directly from OHD data. This chapter lays the groundwork for the analytical and numerical studies in subsequent chapters, aimed at understanding the potential of \( f(Q) \) gravity to explain late-time cosmic acceleration and resolve outstanding issues in modern cosmology.




\chapter{Reconstruction of $\Lambda$CDM Universe in $f(Q)$ Gravity} 
\epigraph{\justifying \textit{``My goal is simple. It is a complete understanding of the Universe: why it is as it is and why it exists at all."}}{\textit{Stephen Hawking}}
\label{Chapter2} 

\lhead{Chapter 2. \emph{Reconstruction of $\Lambda$CDM Universe in $f(Q)$ Gravity}} 


\blfootnote{*The work in this chapter is covered by the following publication:\\
\textit{Reconstruction of $\Lambda$CDM Universe in $f(Q)$ Gravity}, Physics Letters B, \textbf{835}, 137509 (2022).}

The current chapter focuses on the reconstruction of the $\Lambda$CDM Universe within the framework of $f(Q)$ gravity. The detailed outline of the study is as follows:
\begin{itemize}
\item In the present study, we present a number of fascinating explicit reconstructions of $f(Q)$ gravity from the background of the FLRW evolution history.
\item The study identifies general functions of the nonmetricity scalar \( Q \) that replicate the exact \(\Lambda\)CDM expansion history. However, for more generic \( Q \) functions, the inclusion of additional degrees of freedom in the matter sector is required.
\item In addition, a cosmological reconstruction for modified $f(Q)$ gravity is constructed in terms of e-folding for two specific examples of $\Lambda$CDM cosmology. It is shown how any FLRW cosmology can arise from a specific $f(Q)$ theory.
\end{itemize}

\section{Introduction} 
The modified theory of gravity is well known for its successful representation of cosmic acceleration and can also reproduce the entire cosmological history, as well as the behavior of $\Lambda$ \cite{Nojiri/2006,Elizalde/2009,Dombriz/2006}. To recover the properties of \(\Lambda\)CDM through modified theories of gravity, various methodologies have been proposed in the literature. Among these, cosmological reconstruction schemes hold a particularly significant place. For instance, an interesting scheme for the cosmological reconstruction of $f(R)$ gravity in terms of e-folding was developed by Nojiri et al. \cite{Nojiri/2009}. For an accurate reconstruction of $\Lambda$CDM evolution in $f (R)$ gravity, Dunsby et al. \cite{Dunsby/2010} determined that the matter components must have extra degrees of freedom. Carloni et al. \cite{Carloni/2012} reconstructed $f (R)$ gravity utilizing cosmic parameters instead of any form of scale factors. Goheer et al. \cite{Goheer/2009a} demonstrated that exact power-law solutions in $f(G)$ gravity exist exclusively for a specific class of models because many popular $f(G)$ models (all examples addressed in \cite{Felice/2009}) do not allow for exact power-law solutions. This result is also found for the gravity models $f(R)$ and $f(R,T)$ \cite{Goheer/2009b,Sharif/2013}. In the context of $f(R, T)$ gravity, Jamil et al. \cite{Jamil/2012} have reconstructed cosmological models and demonstrated that the dust fluid reproduces the $\Lambda$CDM,  de Sitter Universe, Einstein static Universe, phantom and non-phantom eras, and phantom cosmology. In \cite{Sharif/2014}, the authors represented the cosmological reconstruction of the $f(R,T)$ models and their stability. An analysis of the stability of the $f (R, G)$ models is presented in \cite{Dombriz/2012} for power law and $\Lambda$CDM cosmology. Other than these curvature-based modified theories of gravity, there are two types of modified theories of gravity by which one can explain the Universe: teleparallel gravity and symmetric teleparallel gravity theories.\\
Inspired by the fascinating characteristics of $f(Q)$ gravity, this chapter aims to develop a class of $f(Q)$ functions, which will be able to mimic the properties of the $\Lambda$CDM model. For this purpose, we adopt a number of explicit reconstructions, which lead to a number of interesting results categorized into two schemes. In the first scheme, we obtain some real values $f(Q)$ functions that are able to retrieve the $\Lambda$CDM expansion history of the Universe filled with various matter components, respectively. In the second scheme, in terms of e-folding, a cosmological reconstruction is constructed for modified $f(Q)$ gravity. It is shown how any FLRW cosmology can arise from a specific $f(Q)$ theory. We further show that a theory can be constructed that can mimic the $\Lambda$CDM expansion history, which can be impossible to distinguish from GR at the fundamental level of the FLRW Universe.\\
This chapter is divided into different sections. The Friedmann equations for FLRW cosmology in \(f(Q)\) gravity are presented in section \ref{2a}. In Section \ref{2b}, we reconstruct the cosmological models for the various types of fluid. In section \ref{2c}, in terms of e-folding, we reconstruct the cosmological models for modified $f(Q)$ gravity. Lastly, the discussion and conclusions are presented in section \ref{2d}.

\section{Field Equations}\label{2a}

Throughout the study, we will assume a spatially flat FLRW Universe, whose metric is given by 
\begin{equation}
\label{FLRW}
ds^2=-dt^{2} + a^{2}(t) \left[dr^{2} + r^{2} d\theta^{2} + r^{2} sin^{2}\theta\,d\phi^{2}\right].
\end{equation}
The nonmetricity scalar in the FLRW metric is obtained as $Q=6H^2$, where $H=\frac{\dot{a}}{a}$ is the Hubble function, and the overhead point indicates the derivative with respect to $t$. In this case, the energy-momentum tensor for a perfect fluid is specified in Eq.\eqref{EMT}.\\
For the metric \eqref{FLRW}, the corresponding Friedmann equations are\footnote{In the Friedmann equations, an additional minus sign appears in the matter contribution because the minus sign in the action \eqref{f(Q)action} was not accounted for when deriving the gravitational field equations. If the minus sign in the action \eqref{f(Q)action} is properly accounted for (i.e., if the action is taken as it is), this extra minus sign disappears in the Friedmann equations, as discussed in Chapter \ref{Chapter5}.}\cite{Lazkoz/2019}: 
\begin{equation}
\label{FE1_fQ}
3 H^2=\frac{1}{2f_Q}\left(-\rho+\frac{1}{2}f\right),
\end{equation} 
\begin{equation}
\label{FE2_fQ}
\dot{H}+3H^2+\frac{\dot{f}_Q}{f_Q}H=\frac{1}{2f_Q}\left(p+\frac{1}{2}f\right).
\end{equation}

Now, one can investigate various cosmological applications using the above Friedmann equations in the background of $f(Q)$ gravity. Further, we would like to mention here that the $f(Q)$ gravity satisfies the conservation equation
\begin{equation*}
\dot{\rho}+3H(\rho+p)=0.
\end{equation*}

\section{Reconstruction of $f(Q)$ Gravity Theory that Admits an Exact $\Lambda$CDM Model}\label{2b}
In this section, we reconstruct the $f(Q)$ gravity model for various epochs such that it can precisely mimic the $\Lambda$CDM model. We can develop a real-valued function for the nonmetricity scalar that can provide a particular cosmological evolution of the $\Lambda$CDM model. The observational cosmology suggests that the Hubble rate in terms of redshift, which is described in the $\Lambda$CDM model, is given by
\begin{equation}
\label{11}
H(z)=\sqrt{\frac{\rho_0}{3}(1+z)^3+\frac{\Lambda}{3}},
\end{equation}
where $\rho_0\geq 0$ is the matter density. Now, we aim to construct the $f(Q)$ gravity theories that exactly mimic the $\Lambda$CDM expansion.\\
Using the scale factor $a$ and the redshift relationship $z$ as in $\frac{1}{a}=1+z$, the above equation can be demonstrated as
 
\begin{equation}
\label{2.5}
\frac{\dot{a}}{a}=\sqrt{\frac{\rho_0}{3a^3}+\frac{\Lambda}{3}}.
\end{equation}
From the above equation, we get the derivative of the scale factor $a(t)$ with respect to time ($t$) as
\begin{equation}
\label{13}
\dot{a}=\sqrt{\frac{\rho_0}{3a}+\frac{\Lambda}{3}a^2}.
\end{equation}
Now, we can rewrite the nonmetricity scalar $Q$ in terms of the scale factor by plugging Eq.\eqref{2.5},
\begin{equation}
\label{15}
Q(a)=\frac{2(\rho_0+\Lambda\, a^3)}{a^3}.
\end{equation}
With the help of above equation, we write the scale factor in terms of nonmetricity scalar as 
\begin{equation}
\label{2.8}
a(Q)=\left(\frac{2\,\rho_0}{(Q-2\,\Lambda)}\right)^{\frac{1}{3}}.
\end{equation}
We would like to note here that so far we have not presumed \textit{a priori} and a specific theory of gravity. The expression for Hubble parameter is obtained by observations, and the nonmetricity scalar is completely geometric for the FLRW spacetime, independent of gravitational theory. In addition, the scale factor in Eq.\eqref{2.8} has one real and two complex roots. The scale factor must be real. Therefore, with the real value of the scale factor, the nonmetricity scaler reaches the value $2\Lambda$ in an infinite time. From Eq.\eqref{2.8}, we can calculate the Hubble parameter in terms of nonmetricity scalar as 
\begin{equation}
\label{17}
H(Q)=\frac{1}{a(Q)}\sqrt{\frac{\rho_0}{3a(Q)}+\frac{\Lambda}{3}\,a^2(Q)}.
\end{equation}
In order to find a class of $f(Q)$ functions, which mimic the $\Lambda$CDM expansion, we plug all of the above quantities represented as functions of the nonmetricity scalar into the Friedmann equation \eqref{FE1_fQ}, yielding a first-order inhomogeneous differential equation for the function $f(Q)$ in $Q$-space,
\begin{equation}
\label{2.10}
Q\,\frac{df(Q)}{dQ}+\rho(Q)-\frac{f(Q)}{2}=0.
\end{equation}
Using the energy conservation equation, the inhomogeneous term $\rho(Q)$ may be derived in terms of nonmetricity $Q$. The inhomogeneous term disappears in the vacuum case, yielding the homogeneous first-order differential equation, and its solution is $f(Q)=c_1\sqrt{Q}$, where $c_1$ is an integration constant.\\ 
In further study, we focus only on the inhomogeneous term $\rho(Q)$ to construct more theories for the $\Lambda$CDM expansion. Let us explicitly reconstruct the theories that would lead to a particular matter field obeying a $\Lambda$CDM expansion history.  

\subsection{Reconstruction for dust-like matter}
First, we reconstruct the $f(Q)$ model, which may represent the $\Lambda$CDM era, taking dust into account as a matter content. For the dust-like case, the value of the EoS $w=0$. Imposing the dust-like case in the conservation equation, we obtained the energy density in terms of $a(t)$ as
\begin{equation}
\label{19}
\rho(a)=\frac{\rho_0}{a^3}.
\end{equation}
With the help of Eq.\eqref{2.8}, rewrite the above equation as
\begin{equation}
\label{20}
\rho(Q)=\frac{Q-2\Lambda}{2}.
\end{equation}
Substituting of $\rho(Q)$ in differential Eq.\eqref{2.10}, we get the general solution, which is given by
\begin{equation}
\label{2.13}
f(Q)=-Q-2\,\Lambda+c_1\sqrt{Q},
\end{equation}
This finding is intriguing because it demonstrates that the only real-valued Lagrangian $f(Q)$ is capable of accurately simulating the $\Lambda$CDM expansion history for a Universe composed primarily of a dust-like stuff. Also, if we put $\Lambda=0$ in Eq.\eqref{2.13}, the function for $f(Q)$ recovers the dust-dominated Universe other than GR.

\subsection{Reconstruction for perfect fluid with EoS $p=-\frac{1}{3}\rho$}
In this scenario, we reconstruct the $f(Q)$ model in which the Universe is accelerating, and the EoS value $w=-\frac{1}{3}$ is physically interesting because it appears near the limit of the set of matter fields that obey the strong energy condition.
Imposing the value of EoS $p=-\frac{1}{3}\rho$ in the conservation equation, we obtained the energy density in terms of $a(t)$
\begin{equation}
\label{22}
\rho(a)=\frac{\rho_0}{a^2}.
\end{equation}
With the help of Eq.\eqref{2.8}, the above equation can be written as
\begin{equation}
\label{23}
\rho(Q)=\left(\frac{\rho_0^2(Q-2\Lambda)}{2}\right)^{\frac{2}{3}}.
\end{equation}

Substituting the above equation in differential Eq.\eqref{2.10}, we get the general solution, which read as
\begin{equation}
\label{24}
f(Q)=c_1\sqrt{Q}+\mu_0\left(\frac{Q-2\Lambda}{2\Lambda-Q}\right)^{\frac{2}{3}}\times _2 F_1\left(-\frac{2}{3},-\frac{1}{2},\frac{1}{2};\frac{Q}{2\Lambda}\right),
\end{equation}
where $_2 F_1$ is a Hypergeometric function and $\mu_0=2\left(\frac{\rho_0^2}{\Lambda}\right)^{\frac{2}{3}}$.

\subsection{Reconstruction for multifluids}
Our present knowledge of the Universe is based on the interactions of multiple ``matter" components, which primarily interact via gravity and electromagnetic radiation. The nature of the various components and possible interactions is often based on the concept of coupled ideal fluids. To shed some light on this type of scenario, we start from the same place as in the previous sections, but with two fluid species instead of one. We assume that the Universe contains both dust-like matter and a non-interacting stiff fluid, with current densities of $\rho_0$ and $\rho_s$, respectively. In this case, total matter density is calculated using the conservation equation
\begin{equation}
\label{25}
\rho(a)=\frac{\rho_0}{a^3}+\frac{\rho_s}{a^6}.
\end{equation} 
With the help of Eq.\eqref{2.8}, the above equation can be written as
\begin{equation}
\label{26}
\rho(Q)=\frac{Q-2\Lambda}{2}+\frac{\rho_s}{\rho_0^2}\frac{(Q-2\Lambda)^2}{4}.
\end{equation}
Substituting the above equation in differential Eq.\eqref{2.10}, we get the general solution as
\begin{equation}
\label{27}
f(Q)=-Q-2\Lambda+\mu_1 Q-\mu_2 Q^2+\mu_3+c_1\sqrt{Q},
\end{equation}
where $\mu_1=\frac{2\rho_s\Lambda}{\rho_0^2}$, $\mu_2=\frac{\rho_s}{6\rho_0^2}$, $\mu_3=\frac{2\Lambda^2\rho_s}{\rho_0^2}$.\\
As a result, we conclude that the above-described theory of gravity would perfectly reproduce the history of the $\Lambda$CDM expansion if the Universe were composed of minimally connected noninteracting massless scalar fields with dust-like stuff.

\subsection{Reconstruction for nonisentropic perfect fluids}

The equation of state is used to characterise non-isentropic perfect fluids as
\begin{equation}
\label{28}
p=h(\rho,a).
\end{equation}
We may obtain the appropriate differential equation for $\rho$ by using the energy conservation relation,
\begin{equation}
\label{2.21}
\frac{d\rho}{da}=-\frac{3}{a}(\rho+h(\rho,a)).
\end{equation}
For simplicity of the above equation, we assume the separable function of $h(\rho,a)$ as $h(\rho,a)=w(a)\rho$. In this scenario, we can simply integrate Eq.\eqref{2.21} to obtain
\begin{equation}
\label{2.22}
\rho(a)=Exp\left[-3\int\frac{1+w(a)}{a}\,da\right].
\end{equation}
Let us use an example where the time-dependent barotropic index is represented by
\begin{equation}
\label{2.23}
w(a)=\frac{2\alpha-\beta a^3}{\alpha+\beta a^3},
\end{equation}
where $\alpha$ and $\beta$ are constants.
Using Eq.\eqref{2.23} in Eq.\eqref{2.22}, we get
\begin{equation}
\label{32}
\rho(a)=\frac{(\alpha+\beta a^3)^3}{a^9}.
\end{equation}
With the help of Eq.\eqref{2.8}, the above equation can be written as
\begin{equation}
\label{33}
\rho(Q)=\frac{(2\rho_0\beta+\alpha Q-2\Lambda \alpha)^3}{8\rho_0^3}.
\end{equation}
Substitute the above value in differential Eq.\eqref{2.10}, we get the general solution, which is given by
\begin{equation}
\label{34}
f(Q)=-\nu_1Q+\nu_2Q^2-\nu_3Q^3-\nu_4+c_1\sqrt{Q},
\end{equation}
where, $\nu_1=\frac{3\alpha(\Lambda\alpha-\beta \rho_0)^2}{\rho_0^3}$, $\nu_2=\frac{\alpha^2(\Lambda\alpha-\beta \rho_0)}{2\rho_0^3}$, $\nu_3=\frac{\alpha^3}{20\rho_0^3}$ and $\nu_4=\frac{2(\Lambda\alpha-\beta\rho_0)^3}{\rho_0^3}$.\\
Let us take a look at another type of non-isentropic perfect fluid whose equation of state is
\begin{equation}
\label{35}
p=w\rho+h(a).
\end{equation}
Integrating energy conservation Eq.\eqref{2.21}, we obtain 
\begin{equation}
\label{2.28}
\rho(a)=\left(-\int 3a^{(2+3w)}h(a)\,da+c\right)a^{-3(1+w)}.
\end{equation}
Let us use the particular cases of $h(a)=a^{-12}$ and $w=0$. This suggests that the Universe's matter field is made up of dust and a time-dependent cosmological component that diverges at the Big Bang singularity and quickly decays to zero at subsequent epochs.

After solving Eq.\eqref{2.28}, we obtained matter density, which is given by 
\begin{equation}
\label{37}
\rho(a)=\frac{\rho_0}{a^3}+\frac{\rho_1}{a^{12}}.
\end{equation}

With the help of Eq.\eqref{2.8}, the above equation can be written as
\begin{equation}
\label{38}
\rho(Q)=\frac{Q-2\Lambda}{2}+\frac{\rho_1}{\rho_0^4}\frac{(Q-2\Lambda)^4}{16}.
\end{equation}

Substituting the above equation in differential Eq.\eqref{2.10}, we get the general solution, which is given by
\begin{equation}
\label{39}
f(Q)=-Q+\gamma_1Q-\gamma_2Q^2+\gamma_3Q^3-\gamma_4Q^4+\gamma_5+c_1\sqrt{Q},
\end{equation}
where $\gamma_1=\frac{4\Lambda^3\rho_1}{\rho_0^4}$, $\gamma_2=\frac{\Lambda^2\rho_1}{\rho_0^4}$, $\gamma_3=\frac{\Lambda\rho_1}{5\rho_0^4}$, $\gamma_4=\frac{\rho_1}{56\rho_0^4}$ and $\gamma_5=-2\Lambda+\frac{2\Lambda^4\rho_1}{\rho_0^4}$.

\section{Cosmological Reconstruction of $f(Q)$ Gravity Theory}\label{2c}
The first Friedmann equation corresponding to the FLRW equation can be rewritten as
\begin{equation}
\label{2.32}
0=\frac{f(Q)}{2}-6H^2f'(Q)-\rho,
\end{equation}
where $Q=6H^2$ and $f_Q=f'$.
The above Friedmann equation is written as functions of the number of e-foldings instead of the time $t$, $N=log\frac{a}{a_0}$. The variable $N$ is related with the redshift $z$ by $e^{-N}=\frac{a_0}{a}=(1+z)$.  By adding the fluid densities with a constant EoS parameter $w_i$, we may get the matter energy density $\rho$ as
\begin{equation}
\label{2.33}
\rho=\sum_i \rho_{i0}\, a^{-3(1+w_i)}=\sum_i \rho_{i0}\,a_0^{-3(1+w_i)} e^{-3(1+w_i)N}.
\end{equation}
Let us write the Hubble parameter in terms of $N$ via the function $g(N)$ as 
\begin{equation}
\label{2.34}
H=g(N)=g(-ln(1+z)).
\end{equation}
Then the nonmetricity scalar is written as $Q=6H^2=6g(N)^2$, where $N=N(Q)$. 
By using Eq.\eqref{2.33} and Eq.\eqref{2.34}, Eq.\eqref{2.32} can be written as
\begin{equation}
\label{2.35}
0=\frac{f(Q)}{2}-6G(N)\frac{df(Q)}{dQ}-\sum_i \rho_{i0}\,a_0^{-3(1+w_i)} e^{-3(1+w_i)N}.
\end{equation}
Here, we denote $G(N)=g(N)^2=H^2$.

As an example, consider the CDM model. As will be demonstrated in the following, for the gravity theory represented by action \eqref{f(Q)action}, such an evolution may be reconstructed without the need of a cosmological constant factor, and the shift from a decelerated to an accelerated period is achieved. The FLRW equation for $\Lambda$CDM cosmology in Einstein gravity is given by
\begin{equation}
\label{44}
\frac{3}{\kappa^2}H^2=\frac{3}{\kappa^2}H_0^2+\frac{\rho_0}{a^3}=\frac{3}{\kappa^2}H_0^2+\rho_0\,a_0^{-3}e^{-3N},
\end{equation}
where $H_0$ and $\rho_0$ are constants in this case. The cosmological constant is represented by the first component in the RHS, while the CDM is represented by the second term. In the present Universe, the (effective) cosmological constant $\Lambda$ is given by $\Lambda=12H_0^2$. Then comes the
\begin{equation}
\label{45}
G(N)=H_0^2+\frac{\kappa^2}{3}\rho_0\,a_0^{-3}e^{-3N},
\end{equation}
and $Q=6G(N)=6H_0^2+2\kappa^2\rho_0\,a_0^{-3}e^{-3N}$, which can be solved for $N$ as follows
\begin{equation}
\label{46}
N=-\frac{1}{3}\,ln\left(\frac{Q-6H_0^2}{2\kappa^2\rho_0\,a_0^{-3}}\right).
\end{equation}
With the help of the above equation, Eq.\eqref{2.35} can be written as
\begin{equation}
\label{47}
0=\frac{f(Q)}{2}-Q\frac{df(Q)}{dQ}-K (Q-6H_0^2)^{(1+w)},
\end{equation}
where $K=\rho_0\left( \frac{1}{2\kappa^2\rho_0}\right)^{(1+w)}$.
The solution of the above differential equation is obtained as
\begin{equation}
\label{48}
f(Q)=c_1\sqrt{Q}-2K(6H_0^2)^{(1+w)}\left(\frac{Q-6H_0^2}{6H_0^2-Q}\right)^w 
\times _2 F_1\left(\frac{-1}{2},-1-w,\frac{1}{2};\frac{Q}{6H_0^2}\right),
\end{equation}
where $_2 F_1$ is a Hypergeometric function.\\
As a result, we have shown that modified $f(Q)$ gravity can characterize the $\Lambda$CDM era without introducing the effective cosmological constant.\\
As another example, in Einstein gravity, we reconstruct $f(Q)$ gravity by replicating the system with non-phantom matter and phantom matter, whose FLRW equation is provided by

\begin{equation}
\label{2.41}
\frac{3}{\kappa^2}H^2=\rho_p\,a^{-m}+\rho_q\,a^m,
\end{equation}
where $\rho_p$, $\rho_q$ and $m$ are positive constants. We can verify that the first component in the RHS in this solution relates to a non-phantom fluid with an EoS $w=-1+\frac{m}{3}>-1$, whereas the second term has an EoS $w=-1-\frac{m}{3}<-1$, which relates to a phantom fluid.
Then since $G(N)=g(N)^2=H^2$, we find 
\begin{equation}
\label{50}
G(N)=G_p\,e^{-mN}+G_q\,e^{mN},
\end{equation}
where $G_p$ and $G_q$ are constants. Since $Q=6G(N)$,
\begin{equation}
\label{51}
e^{mN}=\frac{Q\pm \sqrt{Q^2-144G_pG_q}}{12G_q}.
\end{equation}
The above equation can be solved for $N$ as follows
\begin{equation}
\label{52}
N=\frac{1}{m}ln\left(\frac{Q\pm \sqrt{Q^2-144G_pG_q}}{12G_q}\right).
\end{equation}

We consider the case $m = 4$. In this case, the non-phantom matter corresponding to the first term in the RHS of Eq.\eqref{2.41} could be radiation with $w = \frac{1}{3}$. Then Eq.\eqref{2.35} in this case is given by
\begin{equation}
\label{53}
0=\frac{f(Q)}{2}-Q\frac{df(Q)}{dQ}-\frac{\rho_{r0}}{a_0^4}\left(\frac{Q\pm \sqrt{Q^2-144G_pG_q}}{12G_q}\right).
\end{equation}
The solution of the above differential equation is 
\begin{equation}
\label{54}
f(Q)=-\alpha_1Q+c_1\sqrt{Q}-\alpha_2\sqrt{\frac{144G_pG_q-Q^2}{Q^2-144G_pG_q}}\times _2 F_1\left(-\frac{1}{2},-\frac{1}{4},\frac{3}{4};\frac{Q^2}{144G_pG_q}\right),
\end{equation}
where $\alpha_1=\frac{\rho_{r0}}{6a_0^4G_q}$, $\alpha_2=\frac{2\rho_{r0}}{a_0^4}\sqrt{\frac{G_p}{G_q}}$ are constants.

\section{Conclusions}\label{2d}

In this chapter, we have looked at the $f(Q)$ gravity theory that can mimic the exact $\Lambda$CDM expansion history of the Universe. We have adopted several cosmological reconstruction techniques for $f(Q)$ gravity to carry out this task. In our first approach, we have presumed different fluid components and obtained a class of $f(Q)$ theories. We have also reconstructed the $\Lambda$CDM Universe in terms of e-folding without using an auxiliary scalar in intermediate calculations. The great advantage of working on e-folding is that one can reconstruct any type of requested FLRW cosmology, such as the oscillating Universe, transition from deceleration to phantom phase without future singularity, the subsequent transition from deceleration to phantom super-acceleration leading to Big Rip singularity, and $\Lambda$CDM epoch.

In our study, we have obtained a class of $f(Q)$ theories that exactly mimic the $\Lambda$ CDM expansion history, even if it is impossible to distinguish it from the GR using measurements of the background cosmological parameters. Then, it is an exciting problem to see how the perturbation studies (such as growth factor, structure formations, or cosmological gravitational waves from GR) in these $f(Q)$ theories can be experimentally verified \cite{Jimenez/2020,Khyllep/2020,Anagnostopoulos/2021}. In fact, the $f(Q)$ theories presented in this study can be tested through the solar system test, which rules out the Lagrangians to have a modified theory that works for both local and cosmological scales. Furthermore, we have restricted the reconstruction scheme to the flat FLRW cases, whereas in non-flat FLRW cases, one additional term for curvature will appear in the Hubble parameter, nonmetricity scalar $Q$, and the Friedmann equations. As a result, a highly nonlinear differential equation will arise for $f(Q)$, which is an open problem for the readers. Hence, developing these types of modified theories adds a strong agreement in favor of inflation, DM, and DE in the context of a unified gravitational alternative theory.

In the present chapter, we reconstruct \(f(Q)\) gravity models corresponding to different phases of the Universe based on a theoretical study. In the next chapter, we extend this reconstruction process by aiming to reconstruct the \(f(Q)\) gravity model in a model-independent manner using observational datasets.



\chapter{Gaussian Process Approach for Model-Independent Reconstruction of $f(Q)$ Gravity with Direct Hubble Measurements} 
\label{Chapter3} 
\epigraph{\justifying \textit{``Observations always involve theory."
}}{\textit{Edwin Powell Hubble}}
\lhead{Chapter 3. \emph{Gaussian Process Approach for Model-Independent Reconstruction of $f(Q)$ Gravity with Direct Hubble Measurements}} 

\blfootnote{*The work in this chapter is covered by the following publication:\\
\textit{Gaussian Process Approach for Model-Independent Reconstruction of $f(Q)$ Gravity with Direct Hubble Measurements}, The Astrophysical Journal, \textbf{972}, 174 (2024).}

This chapter presents the GP approach for model-independent reconstruction of $f(Q)$ gravity with direct Hubble measurements. The detailed study of the work is described as follows:
\begin{itemize}
\item The increase in discrepancy in the standard procedure to choose the arbitrary functional form of the Lagrangian $f(Q)$ motivates us to solve this issue in modified theories of gravity.
\item In this regard, we investigate the GP, which allows us to eliminate this issue in a model-independent way $f(Q)$. 
\item In particular, we used the 58 Hubble measurements from CC and the radial BAO to reconstruct $H(z)$ and its derivatives $H'(z)$, $H''(z)$, resulting in us reconstructing the region of $f(Q)$, without any assumptions.
\item In addition, we probe the widely studied power-law and exponential $f(Q)$ models against the reconstructed region and can improve the parameter spaces significantly compared to observational analysis. In addition, direct Hubble measurements along with the reconstructed $f(Q)$ function allow $H_0$ tension to be alleviated.
\end{itemize}

\section{Introduction}\label{Ic}
 Advances in learning have allowed the construction of the function \(f(Q)\) directly from observational measurements, without the need to assume a specific functional form in advance. This reconstruction process, known as the GP method, has been successfully developed and used in various studies \cite{Rasmussen/2006,Holsclaw/2010,Seikel/2013}. This method has been studied and explored in various DE scenarios, such as (see the references herein \cite{Melia/2018a,Pinho/2018, zhang/2018, Yin/2019, Elizalde/2019, Rau/2020, Gomez-Valent/2018}) and the expansion history of the Universe \cite{Busti/2014, Verde/2014, Li/2016, wang/2017, Melia/2018b,Cai/2020}. This procedure allows for the reconstruction of $f(Q)$ in a model-independent manner, thereby enhancing the robustness of cosmological analyses. Consequently, the GP is poised to play a pivotal role in modern cosmological investigations, enabling the representation of reconstructions in terms of uncertainty and offering a means to reconstruct $f(Q)$ without assuming specific conditions.

The investigation in this study unfolds in several sequential stages. Initially, we provide a succinct overview of the symmetric teleparallel gravity framework for the FLRW spacetime metric in Section \ref{3a}, followed by an exploration of GPs with a focus on reconstructing the Hubble parameter and its derivative in Section \ref{3b}. Moving to Section \ref{3c}, we meticulously outline the step-by-step procedures used in the reconstruction process of \( f(Q) \), and we also confronted particular selections for \( f(Q) \) against it. Additionally, we delve into cosmological applications to corroborate the prevailing state of the Universe. Finally, in Section \ref{3d}, we encapsulate our findings and contemplate future perspectives.

\section{$f(Q)$ Gravity Theory}\label{3a}
Applying the FLRW metric and substituting \( f(Q) \) with \( Q + f(Q) \) in the general field equation \eqref{fQ_FE}, the relevant Friedmann equations of $f(Q)$ cosmology, namely
 
\begin{equation}
\label{3.1}
H^2+2H^2\,f_Q-\frac{f}{6}=\frac{\rho_m}{3},
\end{equation} 
\begin{equation}
\label{3.2}
\left(12H^2\,f_{QQ}+f_Q+1\right)\dot{H}=-\frac{1}{2}(p_m+\rho_m),
\end{equation}
where $f_Q=\frac{df}{dQ}$, and $f_{QQ}=\frac{d^2f}{dQ^2}$. Furthermore, in the equations provided, $\rho_m$ represents the energy density and $p_m$ denotes the pressure of the matter fluid. It can be easily derived that they accomplish the conservation equation $\dot{\rho}_m+3H(\rho_m+p_m)=0$. \\
We can rewrite Eqs.\eqref{3.1} and \eqref{3.2} as the standard form,
\begin{equation}
\label{13}
    3H^2=\rho_m+\rho_{DE},
\end{equation}
\begin{equation}
\label{14}
    2\dot{H}+3H^2=-(p_m+p_{DE}),
\end{equation}
where 
\begin{equation}
\label{rhode}
    \rho_{DE}=\frac{f}{2}-Q\,f_Q,
\end{equation}
\begin{equation}
\label{pde}
    p_{DE}=2\dot{H}\left(2Q\,f_{QQ}+f_Q\right)-\rho_{DE},
\end{equation}
are the DE density and pressure contributed by the modified part of geometry. Then, by using Eqs.\eqref{rhode} and \eqref{pde}, we can define the effective DE EoS as
\begin{equation}
\label{w}
    \omega_{DE}=\frac{p_{DE}}{\rho_{DE}}=-1+\frac{4\dot{H}\left(2Q\,f_{QQ}+f_Q\right)}{f-2Q\,f_Q}.
\end{equation}
Additionally, the conservation equation of the effective DE is
\begin{equation}
\dot{\rho}_{DE}+3H(\rho_{DE}+p_{DE})=0.
\label{c}
\end{equation}
In our analysis, we focus on the late-time evolution of the cosmic fluid, so that we can neglect radiation and consider the entire contribution due to pressureless matter. This implies $p_m=0$ and $\rho_m=3H_0^2\,\Omega_{0m}(1+z)^3$, where the zero subscript refers to the quantities evaluated at the present time, and $z$ is the redshift defined as $z=\frac{1}{a}-1$.

\section{GPs Using Observational Hubble Data}\label{3b}
In the present work, we employed GP with a squared exponential kernel, as discussed earlier in Section \ref{section 1.7}, to reconstruct the evolution of the Hubble function \(H(z)\) and its derivatives from OHD. For this reconstruction, we used the latest 58 observational Hubble data points (OHDs), along with their associated error bars.  

Although the BAO data are model dependent, relying solely on the CC data does not provide sufficient constraints for \(f(z)\). Therefore, we used a combination of both Hubble data samples to enhance statistical reliability, producing improved results from the GP reconstruction. The OHD spans a redshift range of \(0.07 < z < 2.42\).  

\begin{figure}[H]
\centering
\includegraphics[scale=0.5]{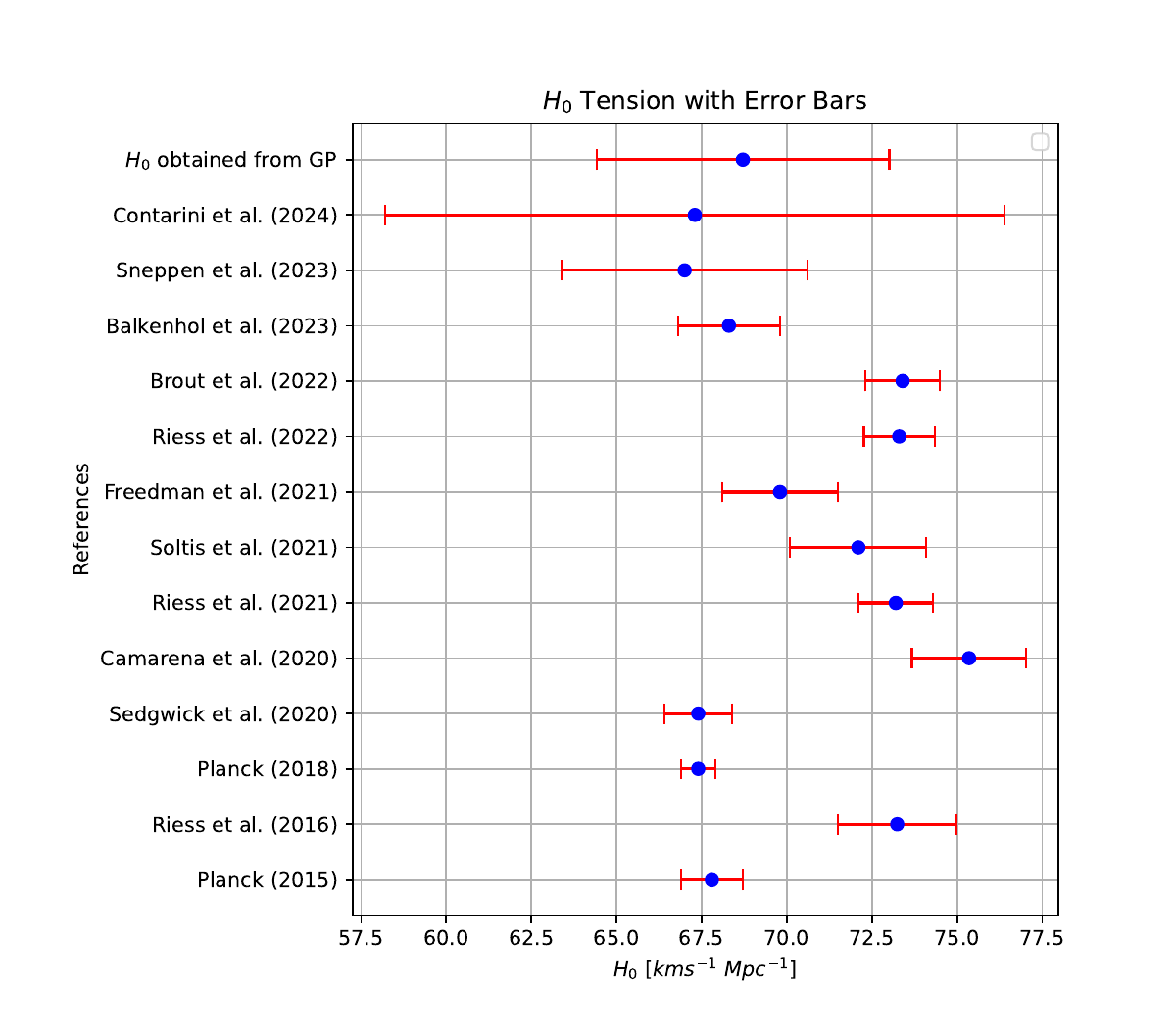}
\caption{\justifying Recent estimations of the Hubble constant $H_0$. The reconstructed $H(z)$ from GPs using the OHD dataset gives the value $H_0=68.74\pm 4.3$ $km\,s^{-1}\,\,Mpc^{-1}$.}
\label{H0}
\end{figure}

From this analysis, we determined the Hubble constant as \(H_0 = 68.71 \pm 4.3 \, \text{km}\,\text{s}^{-1}\,\text{Mpc}^{-1}\), which is consistent with the WMAP observations \cite{Bennett/2013} and the Planck 2015 collaboration value \cite{Ade/2016}, within the margin of error. However, we do not need to worry about the $H_0$ tension in this method, as it is model-independent and obtained directly from the dataset. A comparison of our results with recent \(H_0\) measurements is presented in Figure \ref{H0}. Additionally, the observational data points, along with their references, are summarized in Table \ref{Table 1}.  

\begin{figure}[H]
\includegraphics[scale=0.36]{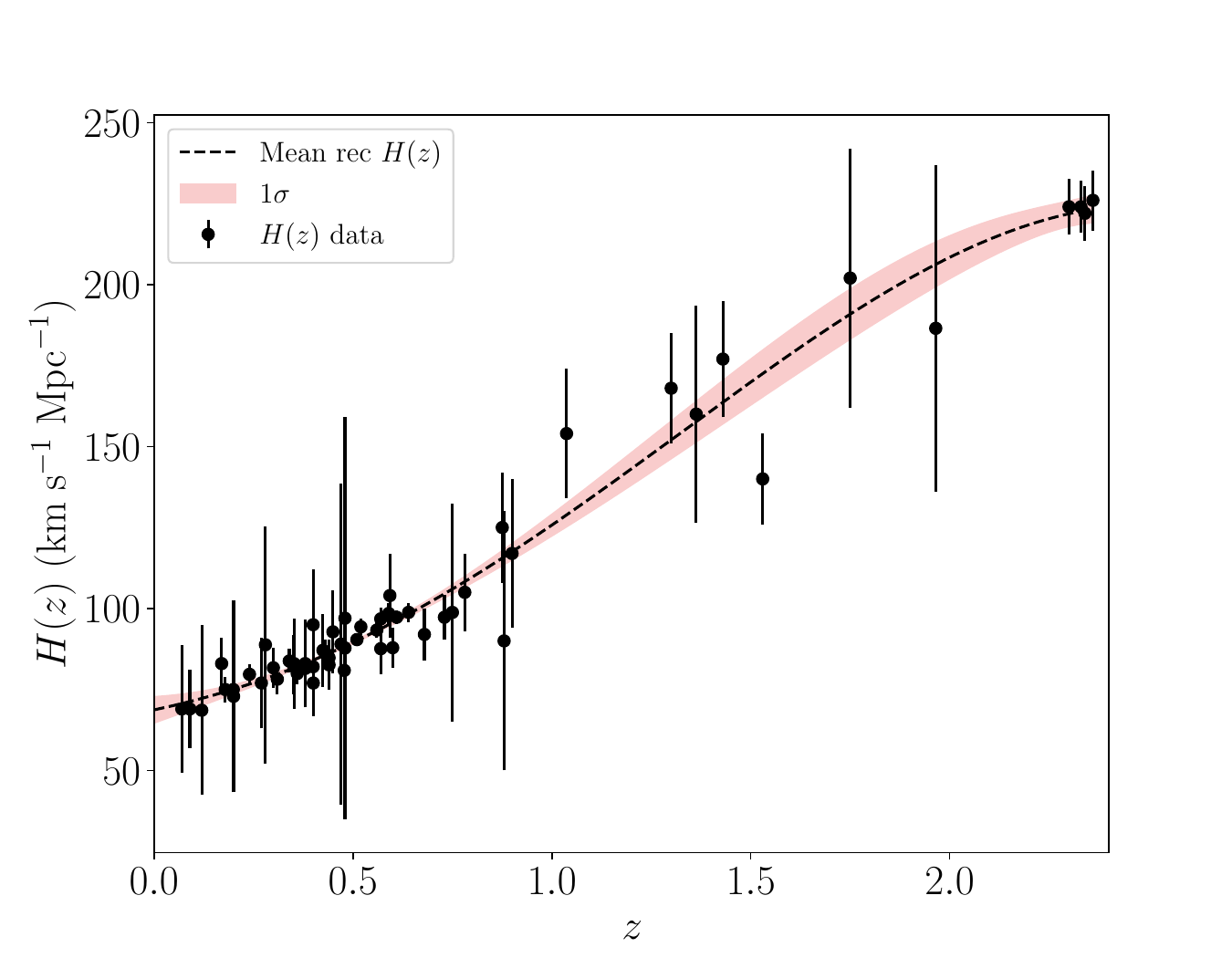}
\includegraphics[scale=0.36]{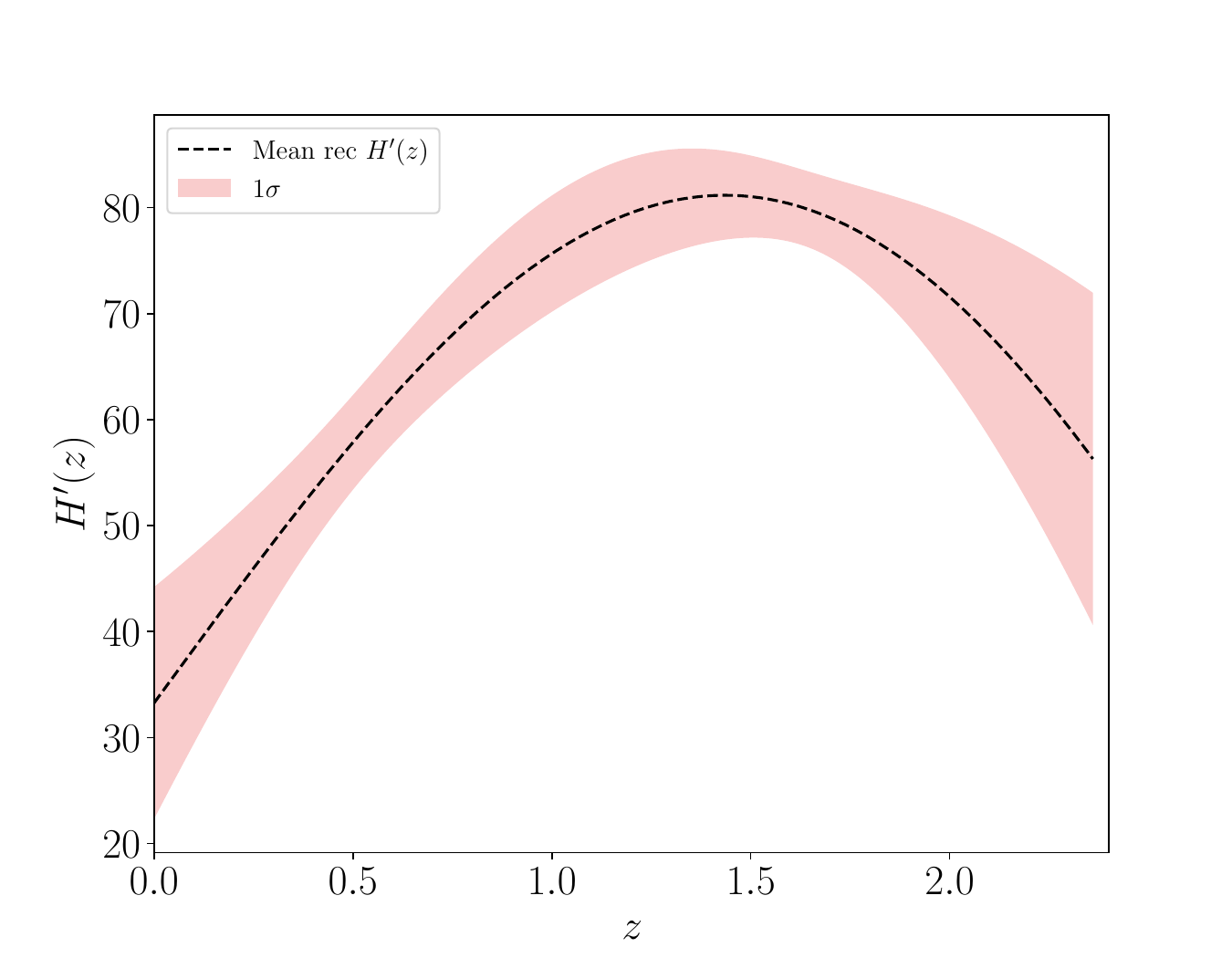}
\caption{\justifying In the left panel, we can see the reconstructed behavior of $H(z)$, which is derived from the 32 CC data points and the 26 BAO data points of the radial method. In the right panel, we can see the reconstructed behavior of the derivative of $H(z)$ with respect to $z$. The black dashed line in each graph represents the mean reconstructed curve, while the colored region indicates $1\sigma$ errors arising from the GP errors.}
\label{Hz}
\end{figure}

The Hubble parameter \(H(z)\) and its derivative \(H'(z)\) (prime denotes the derivative with respect to $z$), successfully reconstructed in a model-independent manner, are shown in Figure \ref{Hz}.

\section{Reconstructing the $f(Q)$ Function from GPs Utilizing OHD Data}\label{3c}
In this section, we will attempt to derive the functional form of $f(Q)$ by using the reconstructed Hubble function and its derivative, which we obtained in the previous section by applying the GP to OHD data. The reconstruction process is simpler in FLRW cosmology in $f(Q)$ gravity, since it only depends on the Hubble function and its first-order derivative. Our goal is to establish the relationship between the redshift $z$ and $f$, or, in other words, to find $f(z)$.\\
To use the model-independent reconstruction approach, we need first to extract the expressions for the involved derivatives $f_Q$ as
\begin{equation}
    f_Q\equiv \frac{df}{dQ}=\frac{df/dz}{dQ/dz}=\frac{f'}{12HH'}\,,
\end{equation}
where primes represent the derivative with respect to redshift $z$. The following step in the application of the GP is to take the approximation of $f'$ as
\begin{equation}
\label{3.10}
    f'(z)\approx \frac{f(z+\Delta z)-f(z)}{\Delta z},
\end{equation}
for small $\Delta z$.
Let us compute the approximation error. We write a Taylor expansion of $f(z + \Delta z)$ about $z$, and then we obtain
\begin{equation}
\label{3.11}
    f'(z)=\frac{f(z + \Delta z)-f(z)}{\Delta z}-\frac{\Delta z}{2}f''(\zeta),\,\,\,\,\,\zeta\in (z,z+\Delta z).
\end{equation}
The second term on the right-hand side of \eqref{3.11} is the error term. Since the approximation \eqref{3.10} can be thought of as being obtained by truncating this term from the exact formula \eqref{3.11}, this error is called the truncation error.
The small parameter $\Delta z$ denotes the distance between the two points $z$ and $z+\Delta z$. As this distance tends to zero, i.e., $\Delta z\to 0$, the two points approach each other, and we expect the approximation \eqref{3.10} to improve. This is indeed the case if the truncation error goes to zero, which in turn is the case if $f''(\zeta)$ is well defined in the interval $(z,z+\Delta z)$. The “speed” in which the error goes to zero as $\Delta z \to 0$ is called the rate of convergence. When $\Delta z\to 0$, it could increase the compatibility of the reconstructed model with the $\Lambda$CDM model.\\
Using the modified Friedmann equation \eqref{3.1} and the approximation above for $f'(z)$, we can extract a recursive relation between consecutive redshifts ($z_i$ and $z_{i+1}$). This involves writing $f(z_i+1)$ as a function of $f(z_i)$, and $H(z_i)$ and $H'(z_i)$ as
\begin{equation}
    f(z_{i+1})-f(z_i)\\=-6(z_{i+1}-z_i)\frac{H'(z_i)}{H(z_i)}\left[H^2(z_i)-\frac{f(z_i)}{6}-\frac{\rho_m(z_i)}{3}\right].
\end{equation}
Ultimately, we arrived at the final phrase as follows by using the EoS parameter for the matter sector:
\begin{equation}
    f(z_{i+1})=f(z_i)-6(z_{i+1}-z_i)\frac{H'(z_i)}{H(z_i)}\\\left[H^2(z_i)-\frac{f(z_i)}{6}-H_0^2\,\Omega_{m0}(1+z_i)^3\right].
\end{equation}
Using the expression provided, we can compute the value of $f$ in the redshift $z_{i+1}$, given that we have information about the parameters in the redshift $ z_i$. Furthermore, through an analysis of the connection between $Q$ and $H$, and by observing the evolution of $H(z)$, we can derive the expression of $f$ in relation to the redshift $z$.
\begin{figure}[H]

\begin{minipage}{.5\linewidth}
\centering
\subfloat[]{\label{fa}\includegraphics[scale=.35]{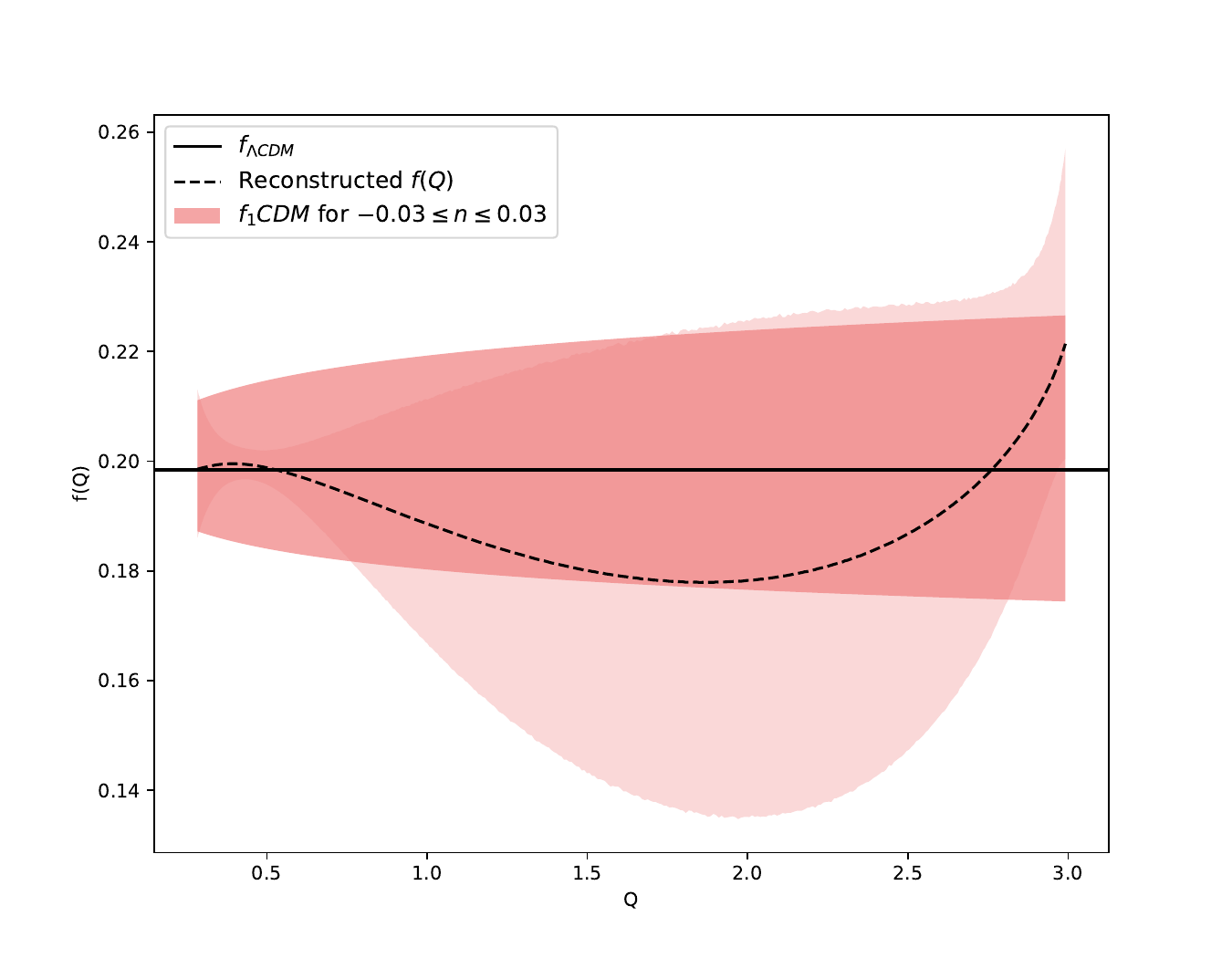}}
\end{minipage}%
\begin{minipage}{.5\linewidth}
\centering
\subfloat[]{\label{fb}\includegraphics[scale=.35]{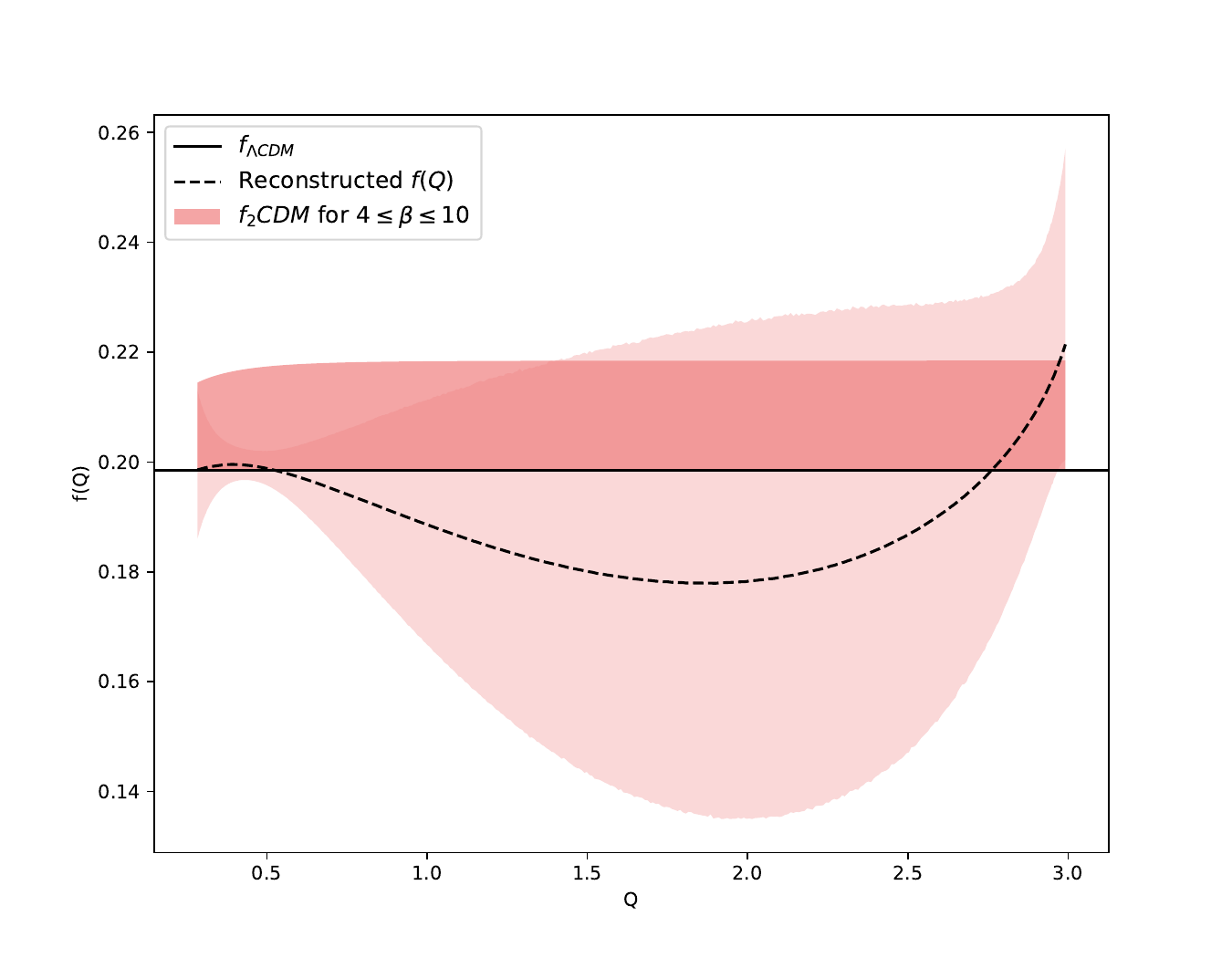}}
\end{minipage}\par\medskip

\caption{\justifying The reconstructed behaviour for $f(Q)$ as a function of $Q$, resulting from data-driven reconstructions of $H(z)$ and $H'(z)$. The black dashed line in the graph represents the mean reconstructed curve, while the light pink colored regions indicate $1\sigma$ errors arising from the GP errors. Moreover, the black solid line marks the scenario for the cosmological constant $f_{\Lambda CDM}=6H_0^2(1 -\Omega_{m0})$. $f(Q)$ and $Q$ are both expressed in $H^2(z)$ units of $(km\,\,s^{-1}\,\,\,\,Mpc^{-1})^2$, and we display them normalized by $10^5$. In the left panel (a), the dark pink region displays the $f_1CDM$ model for $-0.03\le n \le 0.03$. In the right panel (b), the dark pink region displays the $f_2CDM$ model for $4\le \beta \le 10$. To present the plots in a simplified manner, we divided them by a factor of $10^5$. }
\label{Recf}
\end{figure}

In Figure \ref{Recf}, we present the reconstructed function $f(Q)$ using the GP against $Q$. Now, we aim to find the appropriate functional form of $f(Q)$ from our results, which will be able to mimic the reconstructed $f(Q)$.

In the reconstruction profile, we presented the mean reconstruction curve along with the $\Lambda$ CDM model depicted by the straight line. This line maintains a constant value of $2\Lambda=-19267$, derived from the analysis. Observably, the reconstructed mean curve does not have a constant value like $\Lambda$, but embodies the best-fit curve from the Gaussian analysis. It adopts a second-order polynomial form as $f(Q)= -2\Lambda+ \eta Q+\epsilon Q^2$, with the parameter values $\eta\simeq-1.45\times 10^{-3}$ and $\epsilon\simeq5.05\times 10^{-9}$, constrained by the reconstructed $f(Q)$ data. In particular, the functional form of $f(Q)$ simplifies to $f(Q)= -2\Lambda+ \eta Q+\epsilon Q^2$. Consequently, the reconstructed functional form now relies solely on one parameter $\epsilon$, as the linear term merges with the standard linear form in the action. Although one could introduce additional parameters into the reconstructed function, a model with fewer parameters typically represents a better model than one with more parameters. Therefore, we adhere to the one free parameter form of $f(Q)$ denoted as
\begin{equation}
\label{3.14}
f(Q)= -2\Lambda+\epsilon\,Q^2,
\end{equation}
where $\epsilon$ is the sole free parameter with units of $km\,s^{-1}\,\,Mpc^{-1}$. Note that if we used a dimensionless parameter, then we may rewrite Eq.\eqref{3.14} as $f(Q)= -2\Lambda+\epsilon'\,\frac{Q^2}{Q_0^2}$, where the dimensionless parameter $\epsilon'$ is defined as $\epsilon'=36\,H_0^4\,\epsilon$.\\
Subsequently, the reconstructed curve $f(Q)$, along with its shaded error regions, helps to discern the true form of some widely studied functions of $f(Q)$. To this end, we compare two $f(Q)$ models: a power law type and an exponential type, in search of suitable functions.

\subsection{$f_1$CDM: $f(Q)=\alpha \left(\frac{Q}{Q_0}\right)^{n}$}
\begin{figure}[]

\begin{minipage}{.5\linewidth}
\centering
\subfloat[]{\label{a}\includegraphics[scale=.35]{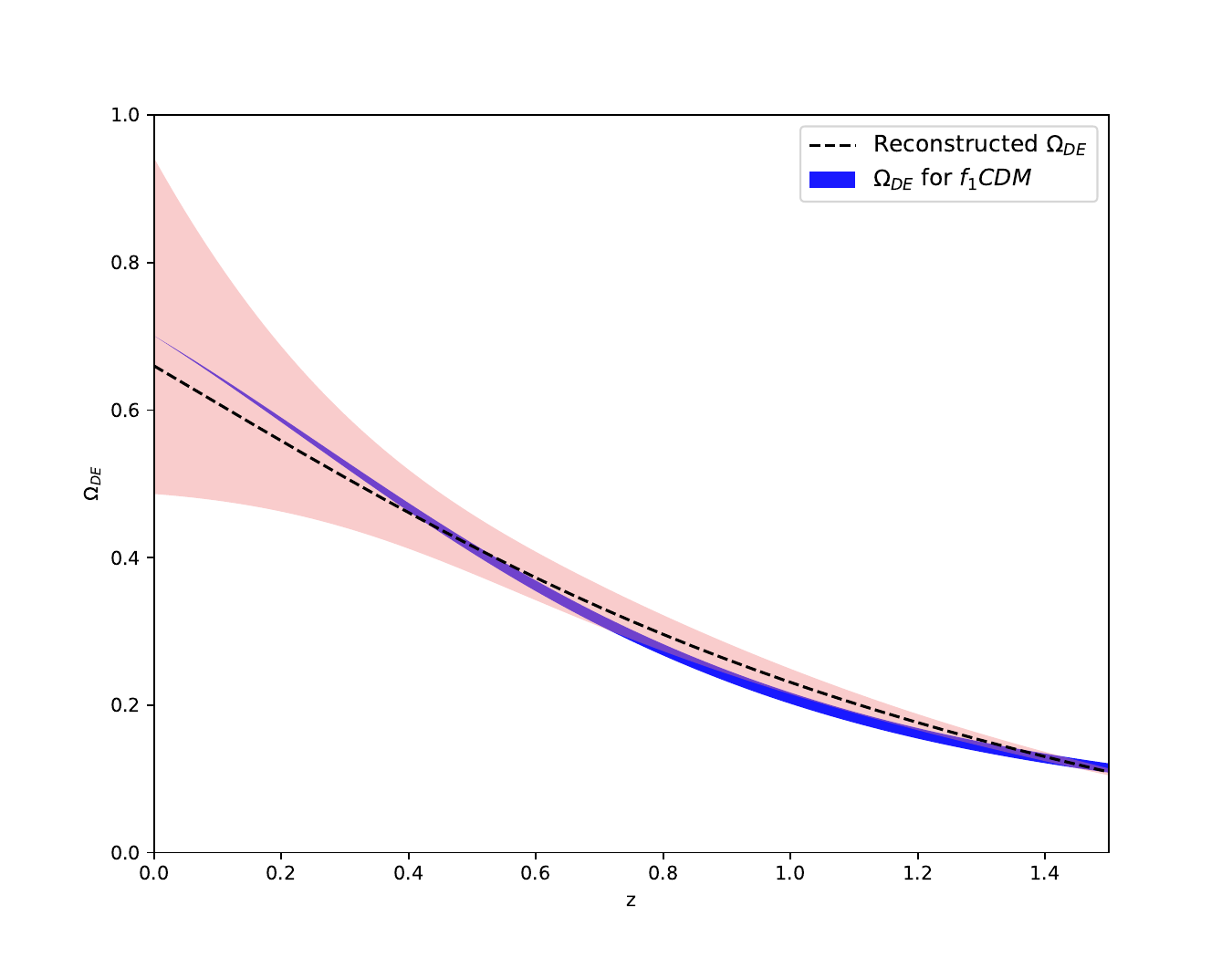}}
\end{minipage}%
\begin{minipage}{.5\linewidth}
\centering
\subfloat[]{\label{b}\includegraphics[scale=.35]{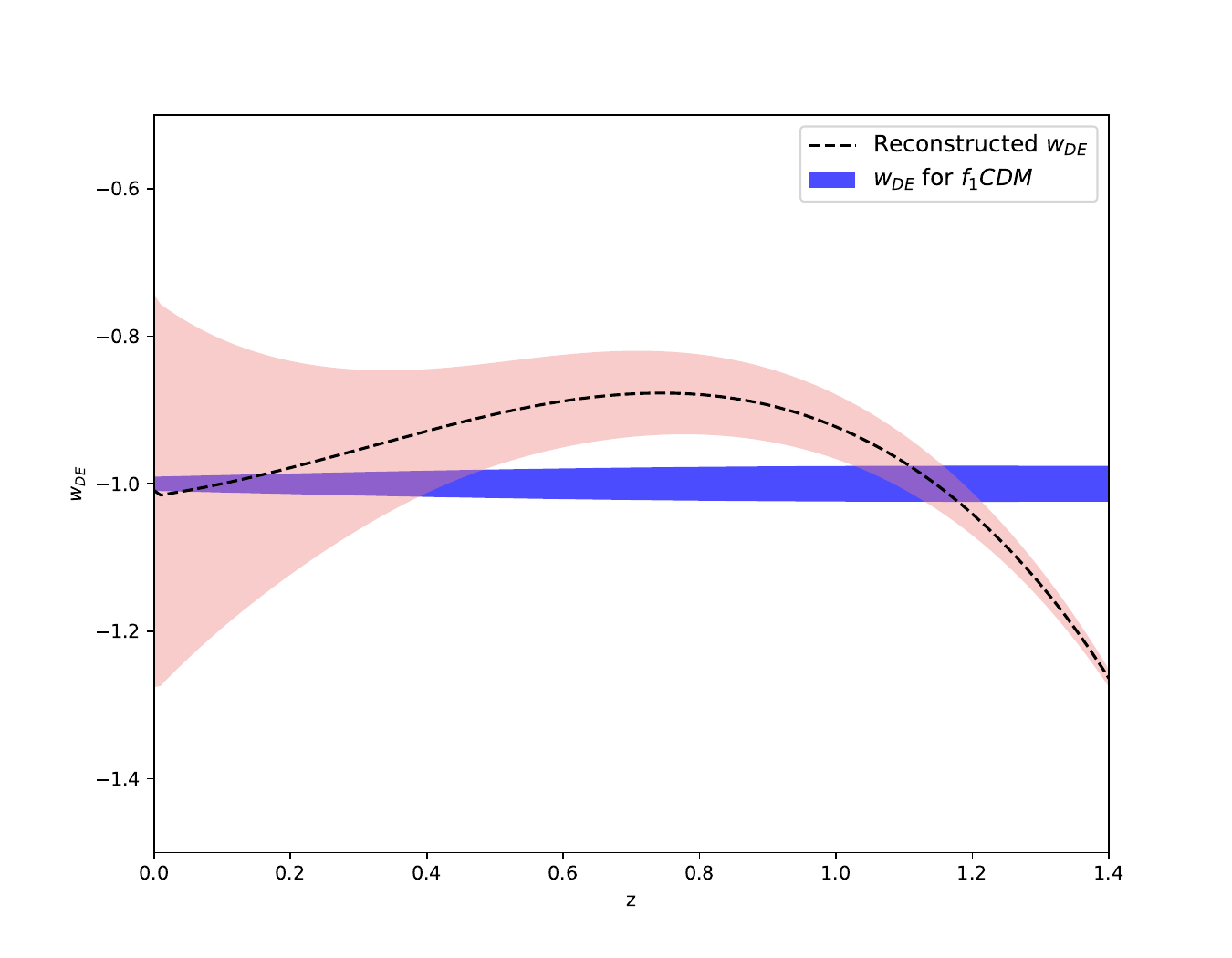}}
\end{minipage}\par\medskip
\centering
\subfloat[]{\label{c1}\includegraphics[scale=.35]{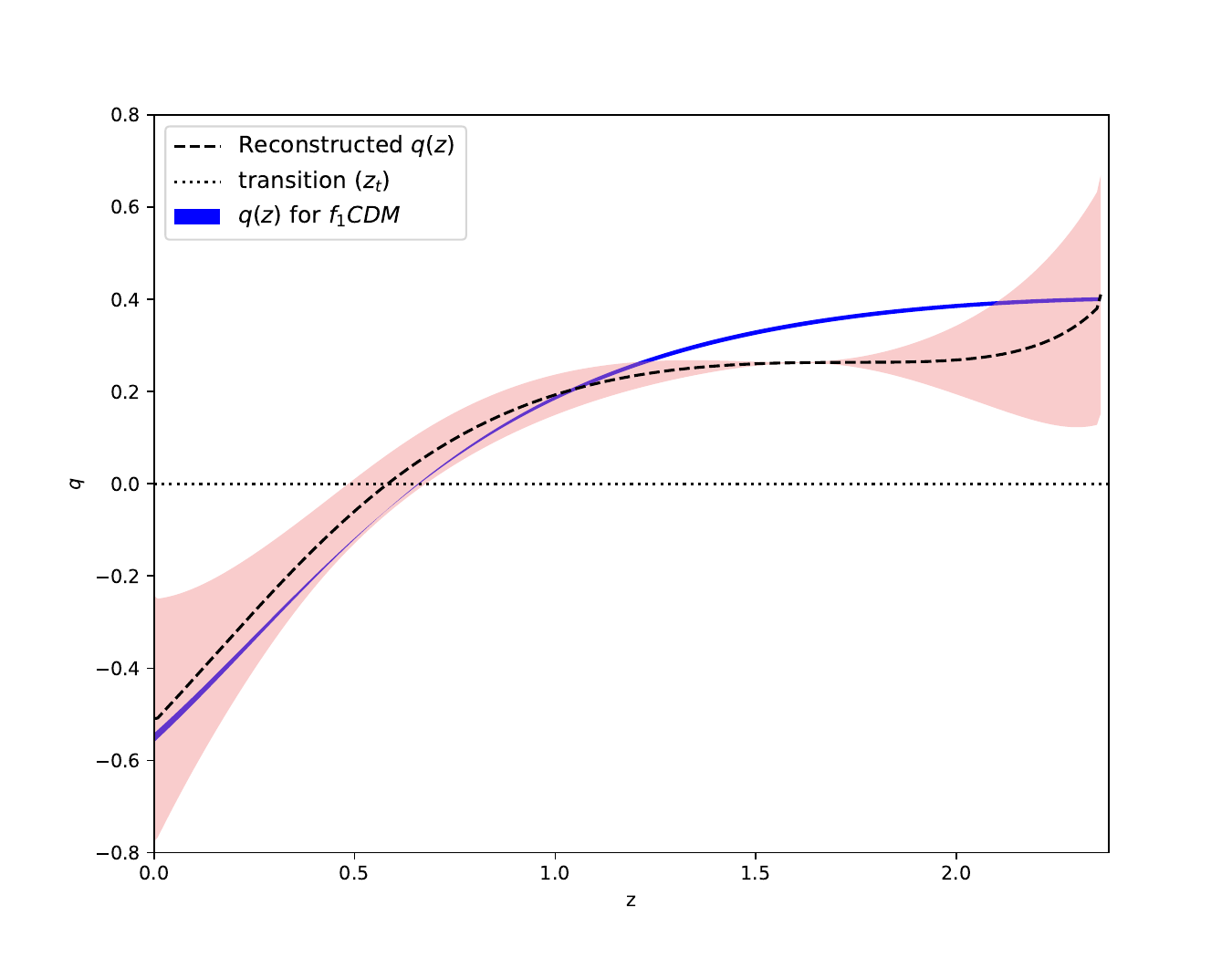}}

\caption{\justifying The reconstructed forms of the DE density parameter $\Omega_{DE}$ (panel a), the DE EoS parameter $w_{DE}$ (panel b), and the deceleration parameter $q$ (panel c) are derived using the reconstructed functions of $H(z)$, $H'(z)$, and the $f(Q)$, with obtained $H_0=68.71\pm 4.3$ $km\,s^{-1}\,\,Mpc^{-1}$ from the GP. In each graph, the black dashed line represents the mean reconstructed curve, while the shaded regions in different colors indicate the $1\sigma$ error resulting from the GP uncertainties. Furthermore, we have incorporated the projection of a plausible $f_1$CDM model by using reconstructed $H(z)$ and $H'(z)$ from GP and considering the range of the free parameter $-0.03 \leq n \leq 0.03$, depicted by a dark blue-shaded area in each graph. }
\label{f1}
\end{figure}

First, we consider the power-law $f(Q)$ model ($f_1$CDM) \cite{Jimenez/2020,Lazkoz/2019,Khyllep/2023}, which is of the form $f(Q)=\alpha \left(\frac{Q}{Q_0}\right)^{n}$, with $\alpha=\frac{(\Omega_{m0}-1)\,6H_0^2}{2n-1}$. 
When $n=0$, the model reduces to $f_{\Lambda CDM}=-2\Lambda=6H_0^2(1-\Omega_{m0})$, which recover the $\Lambda$CDM expansion history of the Universe.\\
It should be noted that any curve that falls within the shaded area in Figure \ref{Recf} can be considered the true form of $f(Q)$, in addition to the mean reconstruction curve. Hence, we constrain the free parameter $n$ to determine which values of $n$ allow the $f_1$CDM model to fit within the reconstructed area. As shown in Figure \ref{fa}, the constraint value indicates that $n$ might fall into the range of $-0.03\le n \le 0.03$.\\
The DE EoS parameter corresponding to $f_1$CDM is
\begin{equation}
    w_{DE}(z)=-1+\frac{2n}{3}\frac{(1+z)}{H(z)}\frac{dH(z)}{dz},
\end{equation}
and the deceleration parameter is 
\begin{equation}
    q(z)=-1+\frac{3}{2}\left[ \frac{H^2(z)+(\Omega_{m0}-1)H_0^2\left(\frac{H(z)}{H_0}\right)^{2n}}{H^2(z)+n\,(\Omega_{m0}-1)H_0^2\left(\frac{H(z)}{H_0}\right)^{2n}}\right].
\end{equation}
In Figures \ref{a}, \ref{b}, and \ref{c1}, we present the reconstructed profiles of $\Omega_{DE}$, $\omega_{DE}$, and $q(z)$ for $f_{1}$CDM, respectively. The constraint values are shown in Table \ref{table 2}.

\subsection{$f_2$CDM: $f(Q)=\alpha Q_0\left(1-e^{-\beta\sqrt{\frac{Q}{Q_0}}}\right)$ }

\begin{figure}

\begin{minipage}{.5\linewidth}
\centering
\subfloat[]{\label{main:a}\includegraphics[scale=.35]{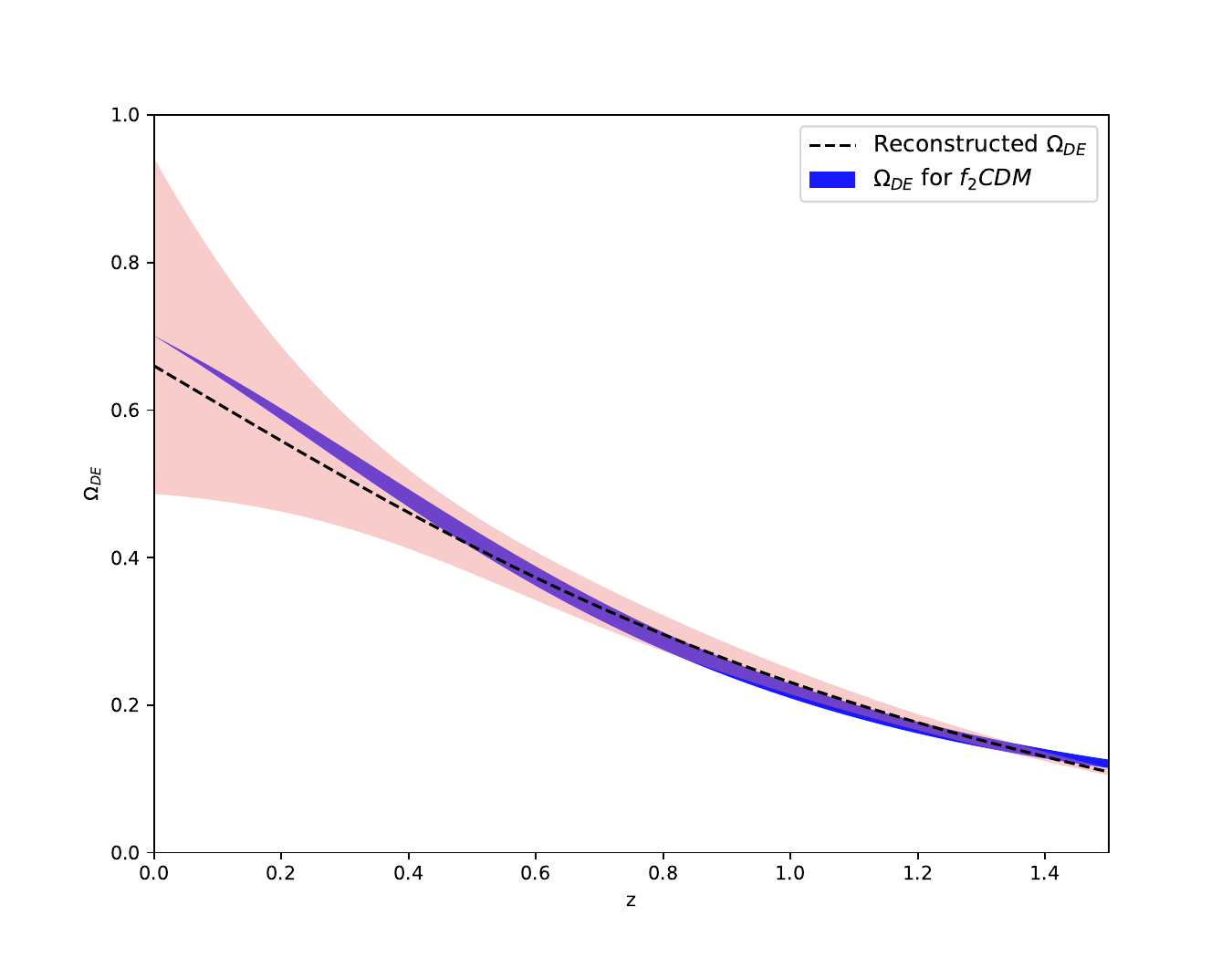}}
\end{minipage}%
\begin{minipage}{.5\linewidth}
\centering
\subfloat[]{\label{main:b}\includegraphics[scale=.35]{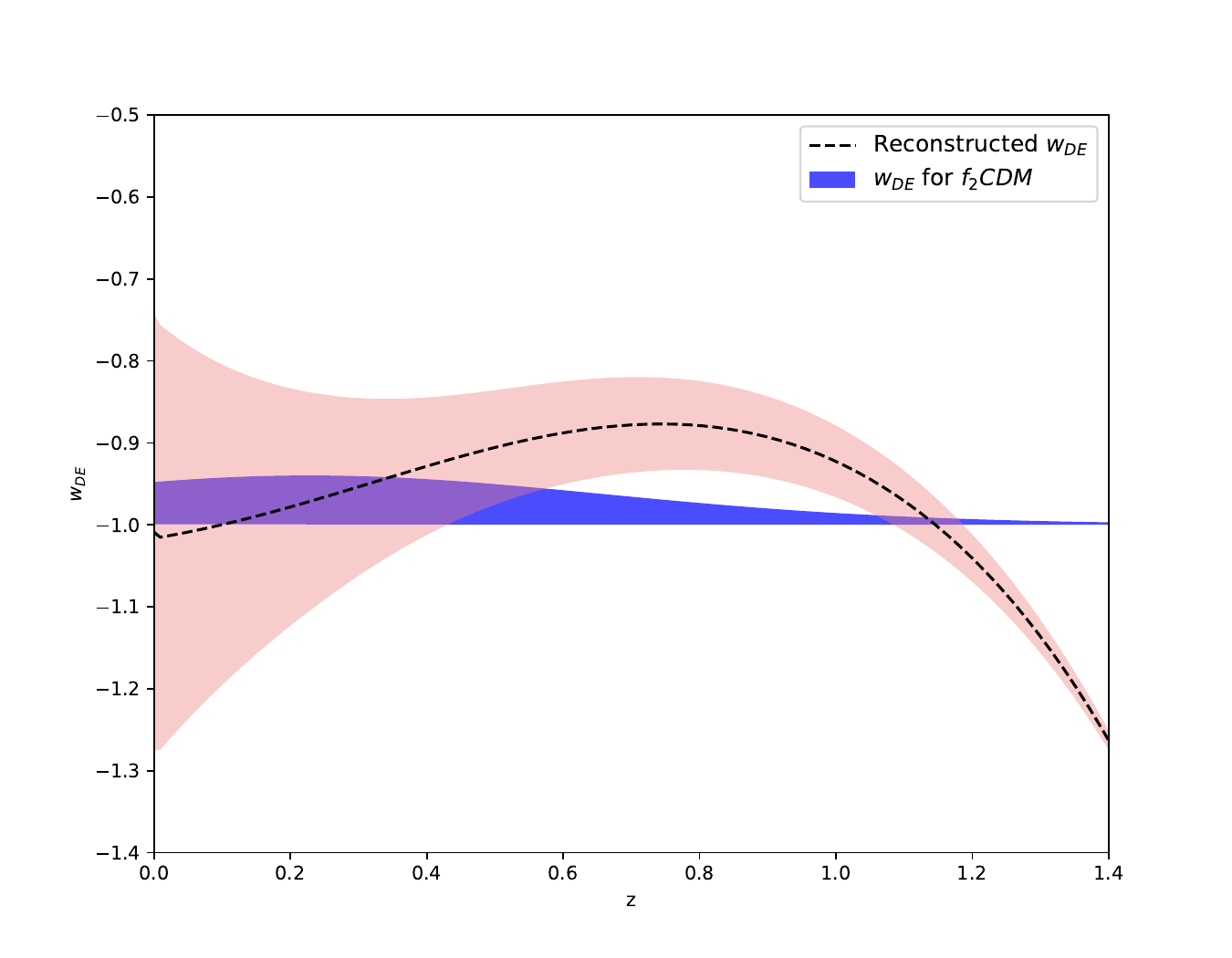}}
\end{minipage}\par\medskip
\centering
\subfloat[]{\label{main:c}\includegraphics[scale=.35]{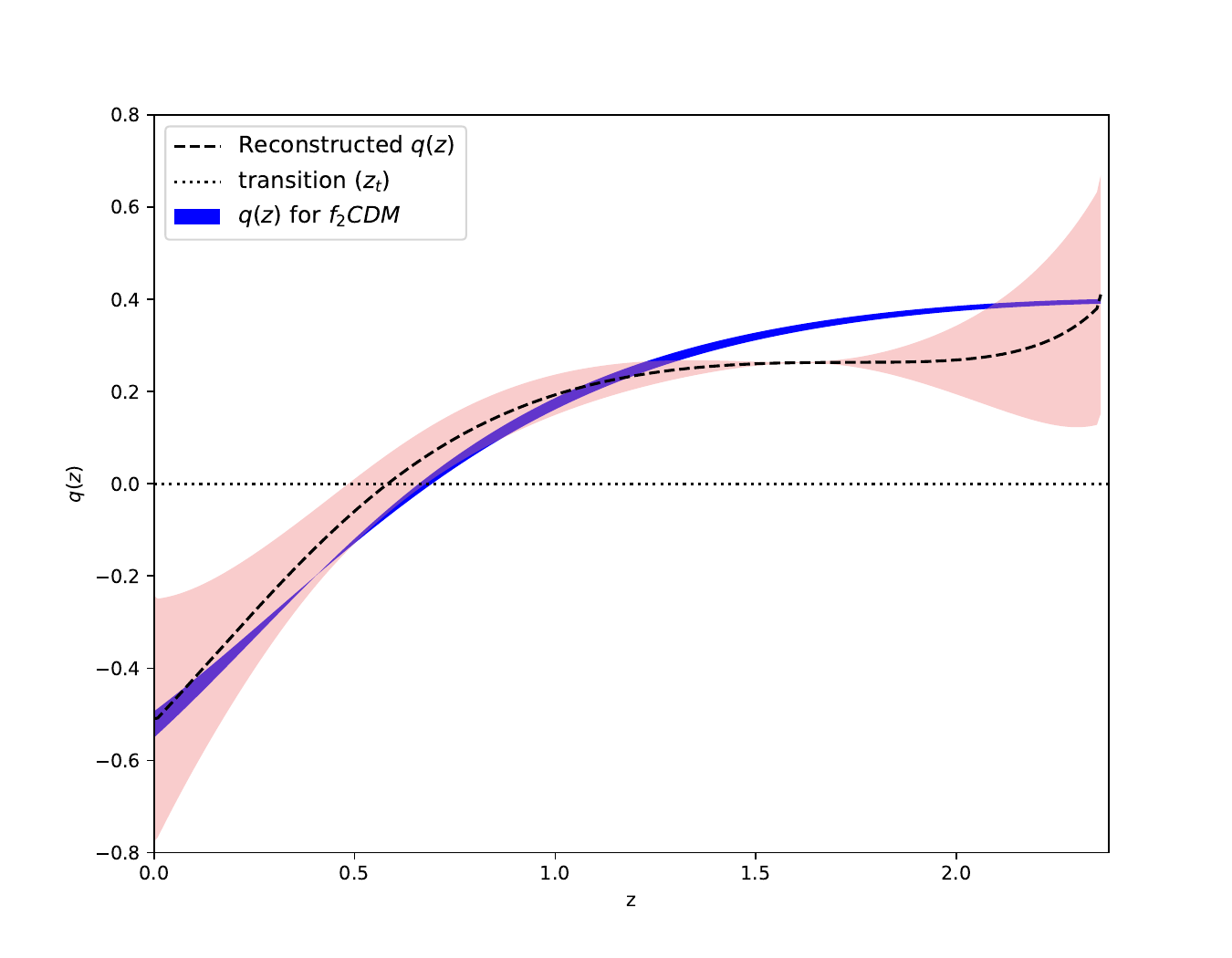}}

\caption{\justifying Here, we have incorporated the projection of a plausible $f_2$CDM model by using reconstructed $H(z)$ and $H'(z)$ from GP and considering the range of the free parameter $4 \leq \beta \leq 10$, depicted by a dark blue-shaded area in each graph. }
\label{f2}
\end{figure}

Next, we consider the exponential $f(Q)$ model ($f_2$CDM) \citep{simranlss,Anagnostopoulos/2023,Khyllep/2023}, which is of the form $f(Q)=\alpha Q_0\left(1-e^{-\beta\sqrt{\frac{Q}{Q_0}}}\right)$, with $\alpha=\frac{1-\Omega_{m0}}{1-(1+\beta)e^{-\beta}}$. For $\beta=0$ the model reduces to the symmetric teleparallel theory equivalent to GR without a cosmological constant. When $\beta \to +\infty$, the model reduces to $f_{\Lambda CDM}=-2\Lambda=6H_0^2(1-\Omega_{m0})$, which recovers the $\Lambda$CDM expansion history of the Universe.\\
Here, we also constrain the free parameter $\beta$ to determine which values of $\beta$ allow the $f_2$CDM model to fit within the reconstructed area. As shown in Figure \ref{fb}, the constraint value indicates that $\beta$ might fall into the range of $4\le \beta \le 10$.\\
The DE EoS parameter corresponding to $f_2$CDM is
\begin{equation}
    w_{DE}(z)=-1+\frac{\beta^2(1+z)H(z)}{3H_0\left(-H_0-\beta\,H(z)+H_0\,e^{\beta\frac{H(z)}{H_0}}\right)}\frac{dH(z)}{dz},
\end{equation}
and the deceleration parameter is 
\begin{equation}
    q(z)=-1+\frac{3}{H^2(z)}\left[\frac{H^2(z)+\frac{1-\Omega_{m0}}{1-(1+\beta)e^{-\beta}}H_0^2\left(-1+\left(1+\beta \frac{H(z)}{H_0}\right)e^{-\beta\frac{H(z)}{H_0}}\right)}{2-\frac{1-\Omega_{m0}}{1-(1+\beta)e^{-\beta}}\beta^2\,e^{-\beta\frac{H(z)}{H_0}}}\right].
\end{equation}
In Figures \ref{main:a}, \ref{main:b}, and \ref{main:c}, we present the reconstructed profiles of $\Omega_{DE}$,  $\omega_{DE}$, and $q(z)$ for  $f_{2}$CDM, respectively. The constraint values are shown in Table \ref{table 2}.

\begin{table}[]
\begin{center}
  \caption{Cosmographic parameter values have been acquired for the reconstructed $f(Q)$ model, as well as for the $f_1$CDM and $f_2$CDM models.}
    \label{table 2}
    \begin{tabular}{l c c c }
\hline\hline 
Model      & $\Omega_{DE}$ & $w_{DE}$ & $q_{0}$     \\ \hline\hline
    Reconstructed $f(Q)$       &  $0.661^{+0.279}_{-0.176}$ & $-1.0089^{+0.2689}_{-0.2681}$ & $-0.5091^{+0.2641}_{-0.2659}$\\[1ex]
    $f_1$CDM   &  $0.7$ &  $-1.0113\le w_{DE}\le-0.9887$ & $-0.5593\le q_0\le-0.5403$  \\[1ex] 
$f_2$CDM &  $0.7$ &  $-1\le w_{DE}\le -0.9462$ & $-0.55\le q_0\le -0.494$  \\[1ex] 
\hline
\end{tabular}
\end{center}
\end{table}


\section{Conclusions}\label{3d}

In the present chapter, we have independently reconstructed the \( f(Q) \) function using observational measurements. To achieve this objective, we have incorporated local Hubble measurements, including CC and BAO, and employed GPs for statistical analysis. Recent inquiries into modified gravities have spurred the search for a function derivable from observational data. Typically, researchers assume specific functional forms for \( f(Q) \) and then constrain the free parameters using observational measurements, often arbitrary assumptions. However, the GP methodology allows us to discern the functional form of \( f(Q) \) independently, without imposing specific conditions based on observational measurements.

Our analysis encompasses Hubble measurements, including CC and BAO measurements, for GP analysis. From this analysis, we determined the value of \( H_0 \), which not only resolves the tension issue \( H_0 \) in a model-independent manner, but also aligns closely with recent precise studies on the subject. To advance our investigation, we first reconstructed \( H(z) \) and its first derivative ${H'(z)}$ from observational samples. Given that the non-metric scalar \( Q \) is a function of \( H \), all Friedmann equations can be expressed in terms of \( H(z) \) and its first derivative ${H'(z)}$. Using the reconstructed functions of \( H \) and its derivative $H'$, we reconstructed \( f(Q) \) without resorting to any assumptions. This reconstructed \( f(Q) \) addresses the \( H_0 \) problem by employing local Hubble measurements. We have presented recent estimates of the Hubble constant $H_0$ along with its constraint value from our GP analysis in Figure \ref{H0}.

Figure \ref{Recf} displays the profile of the reconstructed function \( f(Q) \) concerning the nonmetricity scalar \( Q \), where the dark dotted line represents the mean reconstructed function, the shaded region denotes the error, and the black line indicates the $\Lambda$CDM model. The deviation of the reconstructed function from $\Lambda$CDM suggests a quadratic behavior, leading us to propose a quadratic functional form of \( f(Q) \) with a single free parameter $\epsilon$, quantifying the deviation from $\Lambda$CDM as $f(Q)= -2\Lambda+\epsilon Q^2$. Furthermore, we constrain the range of the free parameter $\epsilon$ within which the one-parameter \( f(Q) \) function lies in the reconstructed \( f(Q) \) region given by $-4.809 \times 10^{-9}< \epsilon < 5.658 \times 10^{-10}$.

Additionally, we scrutinized two widely studied forms of \( f(Q) \) against the reconstructed \(f(Q) \) function, enhancing the constraint on the free parameters compared to traditional observational constraints, and presented the improved parameter range. Moreover, we explore cosmological implications, investigating deceleration parameters, dimensionless DE, and the DE EoS parameters for specific models. As anticipated, our findings corroborate the current accelerated expansion of the Universe, consistent with recent studies.

In conclusion, our study presents a model-independent reconstruction and a functional form proposition \( f(Q) \), based solely on observational measurements through GP analysis. This approach not only enhances constraints on the free parameters of specific models, but also circumvents arbitrary choices for gravitational Lagrangian functions. Although our study focuses on Hubble measurements, future endeavors could extend this analysis to other observational measurements, such as supernovae, which we aspire to explore in forthcoming research.

Hence, \(f(Q)\) gravity is an effective framework for describing the accelerated expansion of the Universe, apart from its role in explaining inflation. In the subsequent chapter, we will explore the present dynamical DE sector and the early Universe's exponential acceleration (inflation) within the \(f(Q)\) gravity framework. This will be achieved by incorporating a scalar field Lagrangian into the action of \(f(Q)\) gravity.


\chapter{Reconstruction of the Scalar Field Potential in Nonmetricity Gravity through GPs} 

\label{Chapter4} 
\epigraph{\justifying \textit{``Inflation hasn’t won the race, but so far it’s the only horse."}}{\textit{Andrei Linde}}
\lhead{Chapter 4. \emph{Reconstruction of the Scalar Field Potential in Nonmetricity Gravity through GPs}} 

\blfootnote{*The work in this chapter is covered by the following publication:\\
\textit{Reconstruction of the scalar field potential in nonmetricity gravity through Gaussian processes}, Physics Letter B, \textbf{860}, 139232 (2025).}

This chapter focuses on reconstructing the scalar field potential within the framework of \(f(Q)\) gravity using GPs. Additionally, it examines whether this potential contributes to the late-time acceleration behavior of the Universe. The detailed study of the work follows as follows:
\begin{itemize}
\item In the present chapter, we incorporate a quintessence scalar field into the nonmetricity framework to model both inflation and late-time acceleration.
\item Using the GP method with a squared exponential kernel, we reconstruct the scalar field potential \(V(\phi)\) in a model-independent manner from OHD. This enables the development of a quintessence scalar field model consistent with observational data within the framework of power-law \(f(Q)\) gravity.
\item Additionally, we compare our reconstructed potential with some standard scalar field potentials and also examine whether this potential contributes to the late-time acceleration of the Universe.
\end{itemize}

\section{Introduction}

In order to acquire a dynamical DE sector and early Universe exponential acceleration (known as inflation) within the framework of nonmetricity gravity, we need to include the quintessence scalar field Lagrangian in the $f(Q)$ action, given by \citep{Caldwell/1998,Bahamonde/2018}
\begin{equation}
\mathcal{L}_{\phi} = -\frac{1}{2} g^{\mu\nu} \partial_{\mu}\phi\, \partial_{\nu}\phi - V(\phi).
\end{equation}

Choosing a quintessence scalar field model requires selecting a suitable potential \( V(\phi) \) to drive the various dynamical phases of the Universe. Various scalar field potentials have been extensively studied in the literature, including quadratic free field potentials \citep{FF1,FF2}, power law potentials \citep{PP1,PP2}, exponential potentials \cite{exp}, and hyperbolic potentials \citep{Obs1}. Recently, Heisenberg et al. \citep{exp} investigated the observational implications of future surveys on quintessence models with \( V(\phi) \sim e^{-\lambda\phi} \), placing constraints on the parameter \( \lambda \). Yang et al. \citep{Obs1} also explored a variety of general potentials, such as exponential, hyperbolic, and power-law types, using data from BAO, CC, CMB, joint light curve analysis (JLA), and redshift space distortions (RSD) to constrain cosmological models. More recently, the power-law potential has been applied to determine the cosmological parameters through HII starburst galaxy magnitude data and other observational measurements \citep{Obs2}. So far in the literature, researchers have considered various scalar field potentials to explore cosmological scenarios in theoretical and observational studies in the context of gravitational theories, as discussed previously \cite{Hussain/2024,Motohashi/2019}. However, it will be interesting to reconstruct the analytical form of scalar field potential without prior assumptions and based on observational measurements instead of assuming any arbitrary forms of scalar field potential. 

In this work, we select a quintessence scalar field model and attempt to reconstruct the scalar field potential $V(\phi)$ directly from the OHD in the context of the power law nonmetricity model. This approach allows us to obtain a suitable quintessence scalar field model that aligns with the OHD under the framework of power-law nonmetricity gravity. The reconstruction is performed using the GP method, originally developed by Seikel et al. \citep{Seikel/2012}. Recently, Jesus et al. \citep{RP1} used data from Type Ia supernovae $H(z)$, while Elizalde et al. \cite{RP2} used 40 Hubble data points to reconstruct the DE potential via the GP in a model-independent manner. Similarly, Niu et al. \citep{RP3} employed three data sets-CC, BAO, and a combination of CC+BAO-along with two priors from Planck 2018 and the Nine-Year WMAP. They demonstrated how different priors and data sets affect the reconstruction results of the DE potential using the GP. Apart from these, some other works have been explored for various DE scenarios using the same approach; for instance, see \cite{Cai/2020,Yang/2024,jls1,jls2,jls3}.

The structure of this chapter is as follows. In Section \ref{4b}, we provide a brief overview of $f(Q)$ cosmology in the presence of a scalar field. In Section \ref{4c}, we introduce OHD and employ GP methods to reconstruct the Hubble function and its first-order derivative. Section \ref{4e} details the step-by-step reconstruction of the scalar field potential using the OHD dataset within the framework of the power law nonmetricity model, and we compare our reconstructed model with other scalar field potentials. Finally, in Section \ref{4d}, we discuss our results and their implications.\\

\section{$f(Q)$ Cosmology in the Presence of a Scalar Field}\label{4b}
Substituting the FLRW metric into the general field equation \eqref{7}, we obtain the relevant Friedmann equations for \( f(Q) \) cosmology in the presence of a scalar field as
 
\begin{equation}
\label{4.2}
3H^2=\rho_m+\rho_{f}+\rho_{\phi},
\end{equation} 
\begin{equation}
\label{4.3}
2\dot{H}+3H^2=-(p_m+p_{f}+p_{\phi}).
\end{equation}
Here, 
\begin{eqnarray}
\label{4.4}
    \rho_{f}&=&\frac{f}{2}-Q\,f_Q,\\
    p_{f}&=&2\dot{H}(2Q\,f_{QQ}+f_Q)-\rho_{f},
\end{eqnarray}
are the energy density and pressure contributed by the modified geometry part of $f(Q)$ gravity, and
\begin{eqnarray}
\label{4.6}
    \rho_{\phi}&=&\frac{1}{2}\dot{\phi}^2+V(\phi),\\
    p_{\phi}&=&\frac{1}{2}\dot{\phi}^2-V(\phi),\label{4.7}
\end{eqnarray}
are the energy density and pressure corresponding to the scalar field.\\
Additionally, the conservation equations of matter fluid, DE, and scalar field are
\begin{eqnarray}
\label{4.8}
\dot{\rho}_{m}+3H(\rho_{m}+p_{m})&=&0,\\
\dot{\rho}_{f}+3H(\rho_{f}+p_{f})&=&0,\\ \text{and}\,\,\,\,\,\,
\dot{\rho}_{\phi}+3H(\rho_{\phi}+p_{\phi})&=&0,
\label{c}
\end{eqnarray}
respectively.
In this analysis, we concentrate on the late-time behavior of the cosmic fluid, allowing us to disregard radiation and focus solely on the contribution from pressureless matter. As a result, we set the matter pressure \( p_m = 0 \), and the matter density is given by \( \rho_m = 3H_0^2 \, \Omega_{0m}(1+z)^3 \), where the subscript zero refers to the values measured at the present epoch.\\ 
Using the equations from \eqref{4.2} to \eqref{4.7}, along with the expression for the matter density, we can determine the kinetic energy \( T \equiv \frac{1}{2}\dot{\phi}^2 \) and the scalar field potential \( V \) as follows
\begin{equation}
\label{KE}
    T=-\dot{H}(2Qf_{QQ}+f_Q+1)-\frac{3}{2}H_0^2\,\Omega_{0m}(1+z)^3,
\end{equation}
and
\begin{equation}
\label{Vz}
    V=\bigg(3H^2+Qf_Q-\frac{f}{2}\bigg)-\frac{3}{2}H_0^2\,\Omega_{0m}(1+z)^3\\+\dot{H}(2Qf_{QQ}+f_Q+1),
\end{equation}
respectively. In this study, we need to express the time dependence of the equation in terms of the redshift $z$ to investigate the observational study. This can be achieved using the relation
\begin{equation}
\label{tz}
    \frac{d}{dt} = -(1+z)H(z)\,\frac{d}{dz},
\end{equation}
where $H(z)$ is the Hubble parameter as a function of redshift.
\section{Dataset and GPs} \label{4c}
As in the previous chapter, this chapter also employs the GP with a squared exponential kernel, as described in Section \ref{section 1.7}, to reconstruct the evolution of the Hubble parameter \(H(z)\) and its derivatives from the OHD data. For this reconstruction, we used the most recent 58 OHD points along with their corresponding error bars. A summary of these data points and their references is provided in Table \ref{Table 1}. The reconstructed Hubble parameter \(H(z)\) and its derivative \(H'(z)\) (where the prime denotes differentiation with respect to \(z\)) are presented in Figure \ref{Hz}, demonstrating a model-independent reconstruction approach.  

\section{Reconstruction of Scalar Field Potential From GPs Utilizing OHD Data}\label{4d}
\begin{figure}[H]
\includegraphics[scale=0.36]{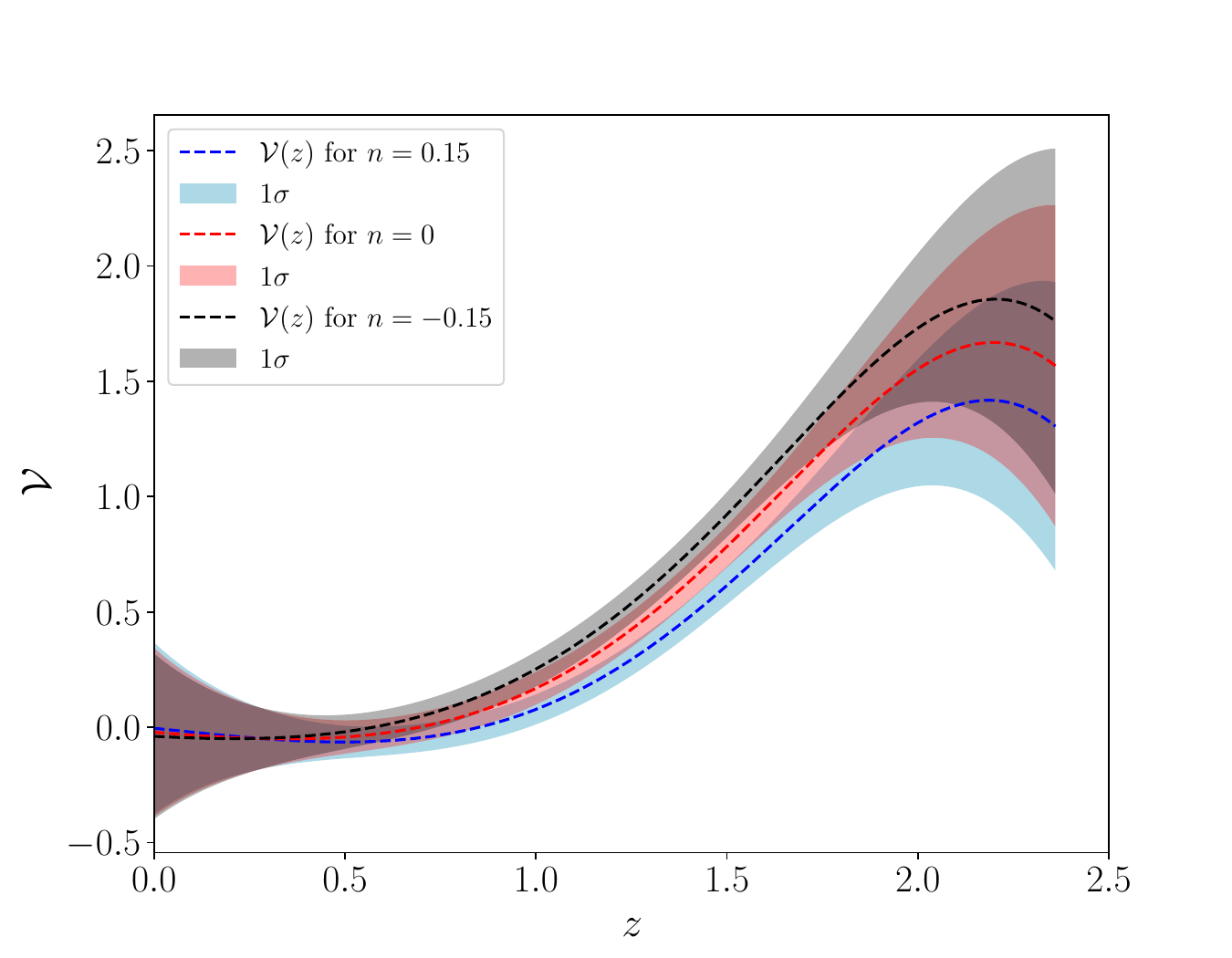}
\includegraphics[scale=0.36]{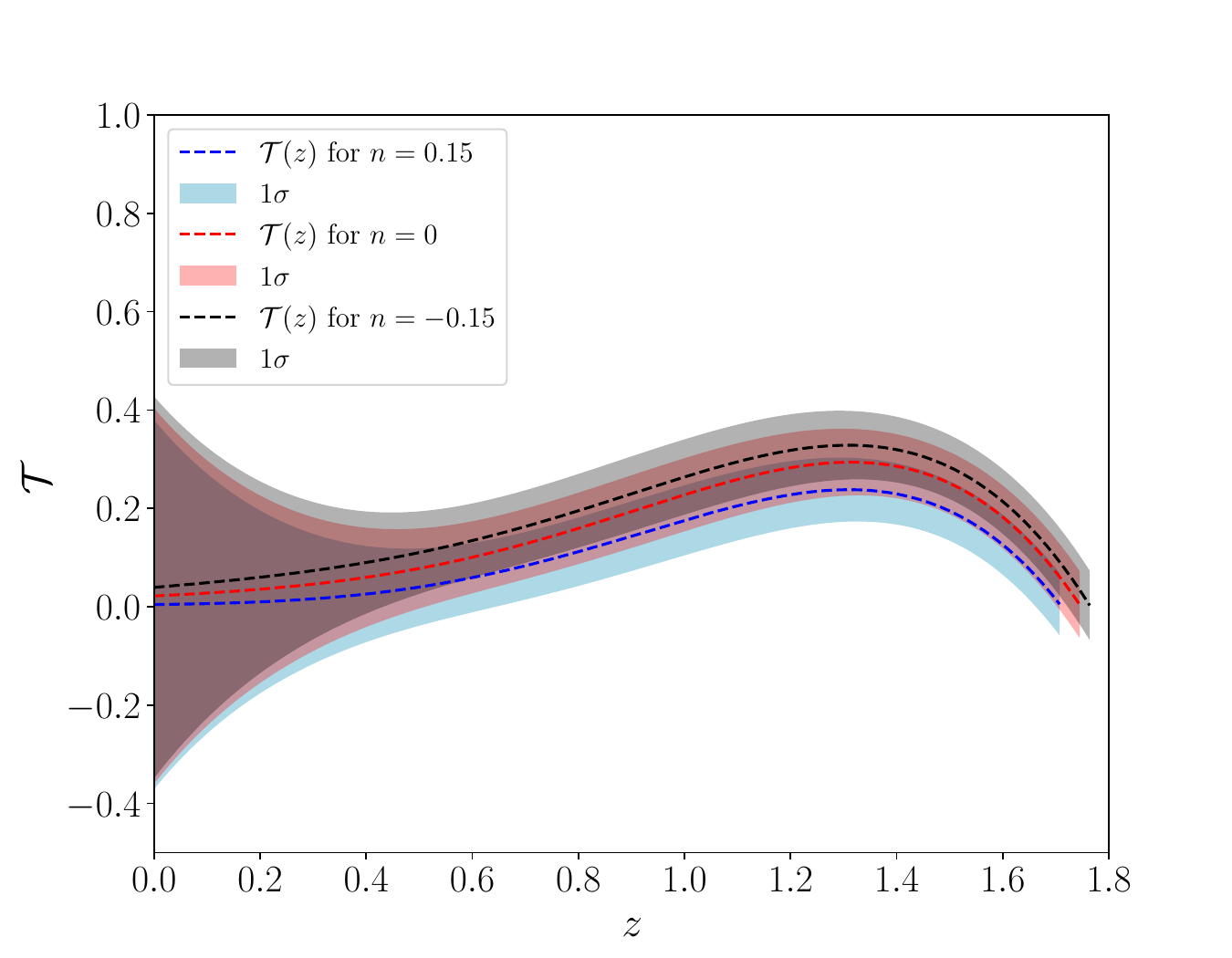}
\caption{\justifying GP reconstruction of \(\mathcal{V}(z)\) and \(\mathcal{T}(z)\) based on data-driven reconstructions of \(H(z)\) and \(H'(z)\). The dashed lines in both graphs represent the mean reconstructed curves for different values of \(n\), while the lightly shaded regions indicate the \(1\sigma\) uncertainty due to GP errors.}
\label{potentialz}
\end{figure}

\begin{figure}
\begin{minipage}{.5\linewidth}
\centering
\subfloat[]{\label{Va}\includegraphics[scale=.35]{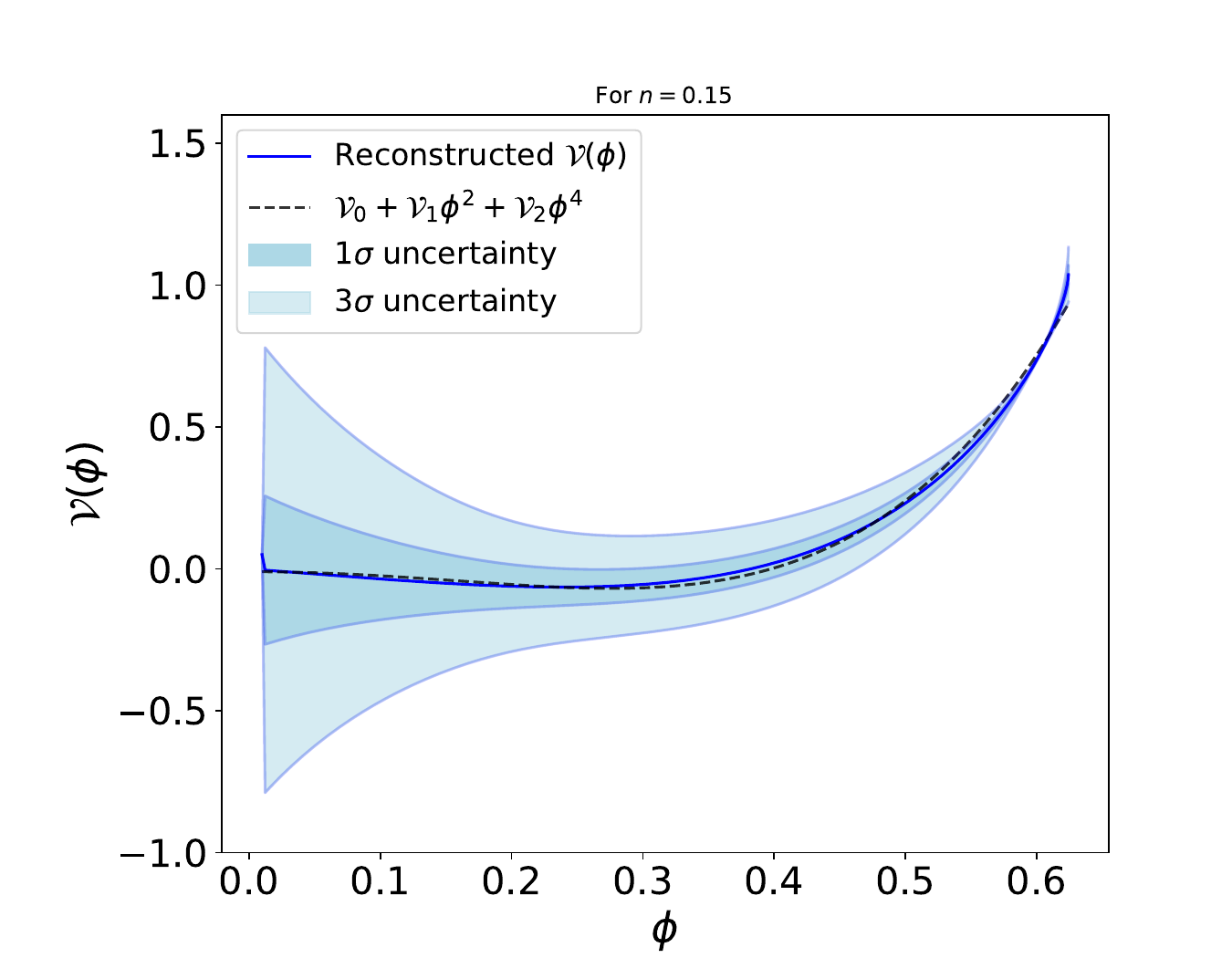}}
\end{minipage}%
\begin{minipage}{.5\linewidth}
\centering
\subfloat[]{\label{Vb}\includegraphics[scale=.35]{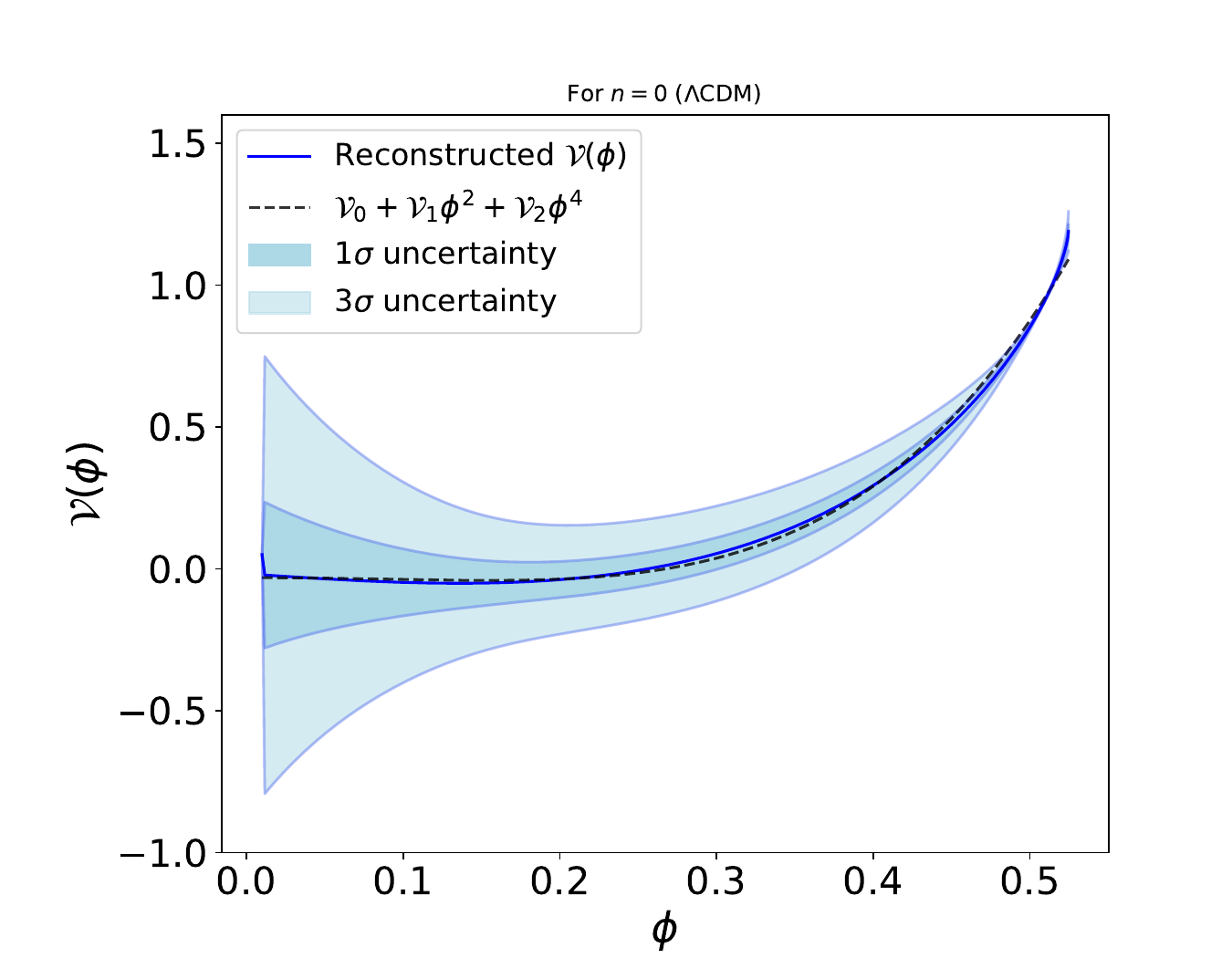}}
\end{minipage}\par\medskip
\centering
\subfloat[]{\label{Vc}\includegraphics[scale=.35]{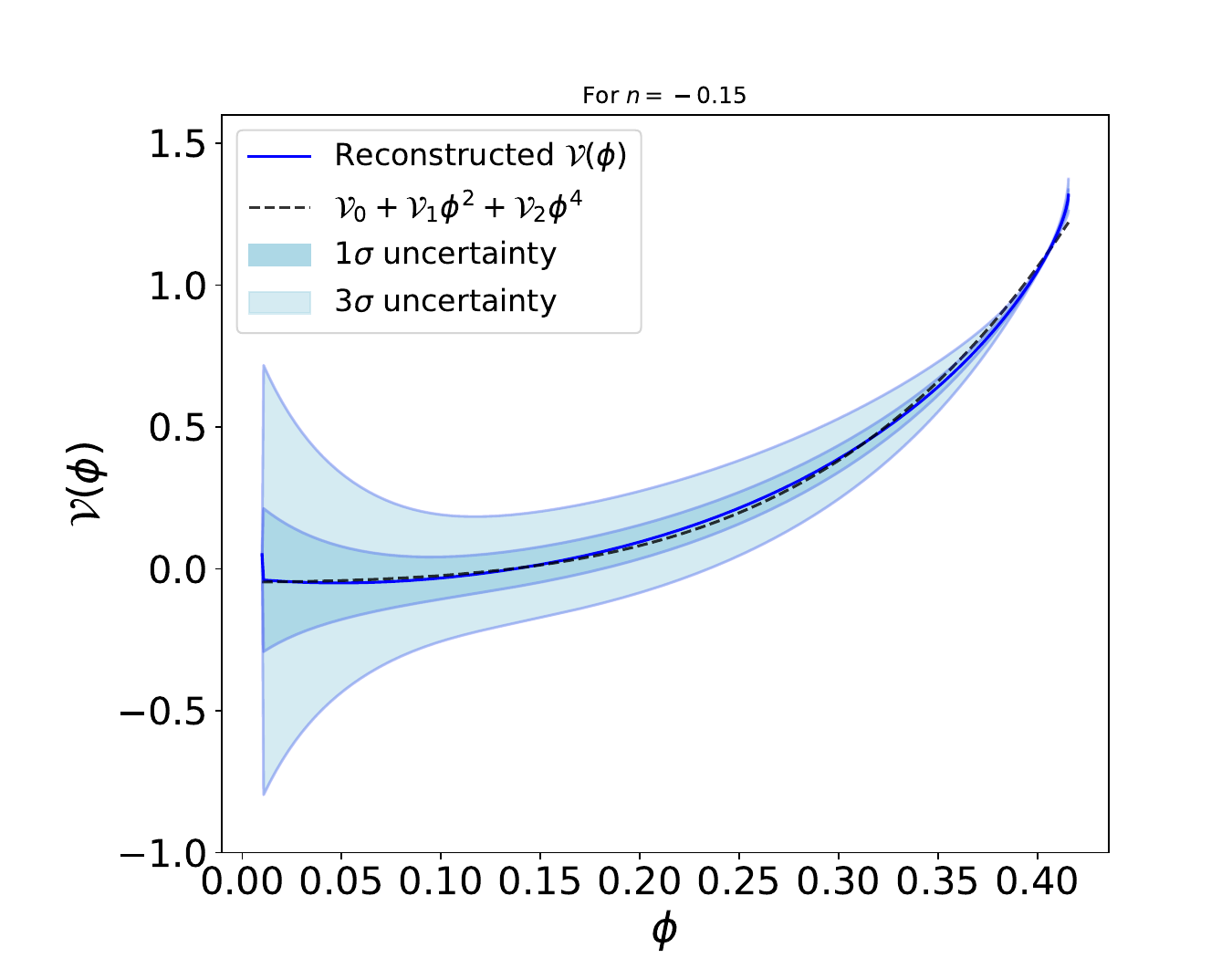}}
\caption{\justifying The reconstructed behavior of the potential \( \mathcal{V}(\phi) \) as a function of the scalar field \( \phi \), derived from the OHD dataset for three different values of the model parameter \( n \), are presented. The blue and light blue regions represent the \( 1\sigma \) and \( 3\sigma \) uncertainties in the reconstruction, respectively, while the black dotted line corresponds to the approximate explicit form of the reconstructed potential.}
\label{potvsphi}
\end{figure}

In this section, we reconstruct the scalar field potential corresponding to the power-law nonmetricity model. The functional form of this model is given by $f(Q) = \alpha \left(\frac{Q}{Q_0}\right)^n$, where the parameter $\alpha$ is defined as $\alpha = \frac{(\Omega_{m0}-1)\,6H_0^2}{2n-1}$ and $n$ is a free parameter \cite{Lazkoz/2019,Jimenez/2020,Khyllep/2023}.\\ 
Using this power law model along with Eq.\eqref{tz}, we find kinetic energy $T$ and scalar field potential $V$ in terms of the Hubble function and its derivative (with respect to $z$) as
\begin{equation}
    T(z)=-\frac{3}{2}H_0^2\,\Omega_{m0}(1+z)^3+(1+z)H'\left[H+n\,H_0(\Omega_{m0-1})\left(\frac{H}{H_0}\right)^{2n-1}\right],
\end{equation}
and
\begin{multline}
\label{V(z)}
    V(z)=3H^2+3H_0^2(\Omega_{m0}-1)\left(\frac{H}{H_0}\right)^{2n}-\frac{3}{2}H_0^2\,\Omega_{m0}(1+z)^3\\
    -\frac{(1+z)H'}{H}\left[H^2+n\,(\Omega_{m0}-1)H_0^2\left(\frac{H}{H_0}\right)^{2n}\right],
\end{multline}
respectively. Here, the dash represents the derivative with respect to $z$.\\  For numerical analysis, it is beneficial to work with dimensionless quantities. This approach eliminates the need to manage units throughout the calculations, reducing the likelihood of unit mismatches or conversion errors. The dimensionless forms of kinetic energy, $\mathcal{T}(z)$, and potential energy, $\mathcal{V}(z)$, are defined as
\begin{equation}
\label{dimensionlessV}
    \mathcal{T}=\frac{8\pi G}{3H_0^2}T,\,\,\,\,\,\mathcal{V}=\frac{8\pi G}{3H_0^2}V.
\end{equation}
To plot these quantities, we constrained the model parameter $n$ to the range $-0.15 \leq n \leq 0.15$. Extending the parameter beyond $n > 0.15$ causes the kinetic energy of the scalar field to become negative, introducing a ghost field. Ghost fields are characterized by unphysical behavior, such as negative energy densities, which lead to instabilities in the system and violate fundamental physical principles like energy conservation. Therefore, values of $n > 0.15$ are avoided to maintain the physical consistency of the model.\\ 
Similarly, for $n < -0.15$, the potential $\mathcal{V}(\phi)$ takes on imaginary values, which are nonphysical and signal the breakdown of the theoretical framework. Imaginary potentials would imply unphysical solutions and introduce mathematical complexities that render the model unsuitable for meaningful cosmological interpretation or further study. Consequently, this lower bound on $n$ ensures that the potential remains real and physically viable.\\
Compared to the power-law model, the dimensionless reconstructed scalar field potential and kinetic energy with respect to $z$ are plotted in Figure \ref{potentialz}.
To clarify, we need to reconstruct the scalar field potential in terms of $\phi$. For this, it is necessary to find the scalar field $\phi$. By applying the chain rule to Eq.\eqref{KE}, we obtain the following expression
\begin{equation}
\label{Dphi}
    [\phi'(z)]^2=-3\Omega_{m0}(1+z)\left(\frac{H_0}{H}\right)^{2}+\frac{2H'}{(1+z)H}\left[1+n\,(\Omega_{m0}-1)\left(\frac{H}{H_0}\right)^{2n-2}\right],
\end{equation}
where primes represent the derivative with respect to redshift $z$. The following step in the application of the GP is to take the approximation of $\phi'$ as
\begin{equation}
\label{4.18}
    \phi'(z)\approx \frac{\phi(z+\Delta z)-\phi(z)}{\Delta z},
\end{equation}
for small $\Delta z$.
Let us compute the approximate error. We write a Taylor expansion of $\phi(z + \Delta z)$ about $z$, and then we obtain
\begin{equation}
\label{4.19}
    \phi'(z)=\frac{\phi(z + \Delta z)-\phi(z)}{\Delta z}-\frac{\Delta z}{2}\phi''(\zeta),\,\,\,\,\,\zeta\in (z,z+\Delta z).
\end{equation}
The second term on the right-hand side of Eq.\eqref{4.19} represents the error term. Since the approximation in Eq.\eqref{4.18} is derived by omitting this term from the exact expression in Eq.\eqref{4.19}, this error is referred to as the truncation error. The small parameter \(\Delta z\) indicates the separation between two points, \(z\) and \(z + \Delta z\). As \(\Delta z\) approaches zero, meaning the two points are closer together, the approximation in Eq.\eqref{4.18} is expected to become more accurate. This holds true if the truncation error vanishes, which happens when \(\phi''(\zeta)\) is well defined within the interval \((z, z + \Delta z)\). The rate at which the error decreases as \(\Delta z \to 0\) is called the rate of convergence.\\
Using Eq.\eqref{Dphi} along with the approximation for \(\phi'(z)\), we can derive a recursive relation between consecutive redshifts \(z_i\) and \(z_{i+1}\). This allows us to express \(\phi(z_{i+1})\) as a function of \(\phi(z_i)\), along with \(H(z_i)\) and \(H'(z_i)\) as
\begin{multline}
\label{phii}
    \phi(z_{i+1})=\phi(z_i)+(z_{i+1}-z_i)\left[-3\Omega_{m0}(1+z)\left(\frac{H_0}{H}\right)^{2}\right.\\
    \left.+\frac{2H'}{(1+z)H}\left(1+n\,(\Omega_{m0}-1)\left(\frac{H}{H_0}\right)^{2n-2}\right)\right]^{1/2}.
\end{multline}
Using Eqs.\eqref{V(z)}, \eqref{dimensionlessV}, and \eqref{phii}, we numerically plotted the dimensionless scalar field potential \(\mathcal{V}(\phi)\) as a function of the scalar field \(\phi\), as shown in Figure \ref{potvsphi}.\\
Now, our objective is to determine an explicit functional form for $\mathcal{V}(\phi)$ that is closely aligned with the reconstructed $\mathcal{V}(\phi)$ from our analysis. We adopt a fourth-order polynomial form as
\begin{equation}
    \mathcal{V}(\phi)=\mathcal{V}_0+\mathcal{V}_1\,\phi^2+\mathcal{V}_2\,\phi^4,
\end{equation}
 where $\mathcal{V}_0$, $\mathcal{V}_1$, and $\mathcal{V}_2$ are polynomial coefficient. In the reconstruction profile, we presented the mean reconstruction curve along with the adopted polynomial function. Now, the coefficients are dependent on the model parameter $n$. Here, we take the interval for parameter $n$ as
 \begin{equation*}
     n\in[-0.15,0.15]
 \end{equation*}
 and the corresponding polynomial coefficient takes 
 \begin{eqnarray*}
     \mathcal{V}_0&\in &[-0.0463,-0.0069]\\
     \mathcal{V}_1&\in &[-1.5533,1.9601]\\
     \mathcal{V}_2&\in &[10.2105,31.1459].
 \end{eqnarray*}
When we set $n = 0$, our nonmetric gravity theory reduces to the standard $\Lambda$CDM cosmological model. This implies that the reconstruction of the potential for $n = 0$ corresponds to the $\Lambda$CDM cosmological model. We have presented the most precise values of the constant coefficients in Table \ref{Table 2}.

Our reconstructed potential is matched with the Higgs scalar field potential with $\mathcal{V}_0$ being a constant term, representing the vacuum energy, $\mathcal{V}_1$ is typically associated with the mass term of the scalar field (ie $\mathcal{V}_1=\mu^2/2$, where $\mu=m/H_0$) and $\mathcal{V}_2$ is related to the self-interaction of the Higgs field (i.e., $\mathcal{V}_2=\lambda/4$). This potential is commonly used in the Standard Model of particle physics to describe spontaneous symmetry breaking.\\
Our reconstructed potential can be divided into various distinct forms, as discussed below.\\
\textit{Massive free scalar field potential:} The first form is the massive free scalar field potential \(\mathcal{V}(\phi) = \mathcal{V}_1 \phi^2\), where the coefficient \(\mathcal{V}_1\) is defined as \(\mathcal{V}_1 = \mu^2/2\), with \(\mu = m/H_0\) and \(m\) representing the mass of the scalar field. This potential is valid only when \(n\) lies between \([-0.08, -0.15]\), where the mass \(m\) falls within the range \(0.5348 H_0\) eV to \(1.9799 H_0\) eV. Outside this range, the free-field potential does not align with the reconstructed potential.\\
\textit{Massless self-interacting scalar field potential:} The second form is the massless self-interacting scalar field potential \(\mathcal{V}(\phi) = \mathcal{V}_2 \phi^4\), where \(\mathcal{V}_2 = \lambda/4\) and \(\lambda > 0\) represent the self-interaction coupling constant. This ensures that the potential remains bounded from below, maintaining stability. For \(n \in [0.15, -0.15]\), \(\mathcal{V}_2\) is positive, keeping the potential positive definite and compatible with the \(3\sigma\) region of our reconstructed potential.\\
 \textit{$\phi^2$ potential with vacuum energy:} The third form is the \(\phi^2\) potential with vacuum energy, \(\mathcal{V}(\phi) = \mathcal{V}_0 + \mathcal{V}_1 \phi^2\), known as Linde's hybrid inflation model. Here, \(\mathcal{V}_0\) represents the constant vacuum energy, while \(\mathcal{V}_1\) corresponds to the mass term, as in the massive free scalar field potential. This form is compatible with the reconstructed potential for \(n \in [-0.08, -0.15]\), similar to the massive free field case.\\
 \textit{Massive self-interacting scalar field potential:} Lastly, we have the massive self-interacting scalar field potential is given by 
$\mathcal{V}(\phi) = \mathcal{V}_1 \phi^2 + \mathcal{V}_2 \phi^4$,
where \(\mathcal{V}_1 = -\frac{\mu^2}{2} < 0\) and \(\mathcal{V}_2 = \frac{\lambda}{4} > 0\). Based on these restrictions on \(\mathcal{V}_1\) and \(\mathcal{V}_2\), this potential is consistent with our reconstructed potential only for \(n \in (-0.08, 0.15]\).

\begin{table}
\begin{center}
  \caption{The table below specifies the polynomial coefficients with $3\sigma$ error for three different values of the model parameter $n$.}
    \label{Table 2}
    \begin{tabular}{l c c c }
\hline\hline 
Model parameter $n$      & $\mathcal{V}_0$ & $\mathcal{V}_1$ & $\mathcal{V}_2$    \\ \hline\hline 
    $n=0.15$       &  $-0.0069\pm 0.0071$ & $-1.5533\pm 0.0984$ & $10.2105\pm 0.2545$\\[1ex]
    $n=0$ ($\Lambda$CDM)       &  $-0.0307\pm 0.0064$ & $-0.8494\pm 0.1300$ & $17.8900\pm 0.4820$\\[1ex]
$n=-0.15$       &  $-0.0463\pm 0.0055$ & $1.9601\pm 0.1951$ & $31.1459\pm 1.1872$\\[1ex]
\hline
\end{tabular}
\end{center}
\end{table}

\section{Conclusions}\label{4e}
In this chapter, we have focused on reconstructing the early DE scalar potential, \( V(\phi) \), using the GP method within the framework of modified gravity theories, specifically nonmetricity gravity. Although previous studies, such as those by Niu et al. \citep{RP1}, Jesus et al. \citep{RP2}, and Elizabeth et al. \citep{RP3}, have explored scalar-field DE reconstructions, our approach differs in methodology. We build on our previous work, where the Lagrangian function of \( f(Q) \) gravity was reconstructed, as well as Niu's approach \citep{RP1}, which focused on reconstructing the scalar potential within the framework of GR. In our study, we have introduced both early- and late-time DE within a unified framework, filling a gap in the literature that typically focuses on early- or late-time DE. To perform the reconstruction, we utilized 32 Hubble parameter measurements and 26 BAO measurements. For priors, we applied constraints from Planck 2018 and used the square exponential kernel as the covariance function in the GP.

To carry out the reconstruction process, we first reconstructed the Hubble function and its derivatives using the GP. The presence of both early- and late-time DE components in our model has enabled us to proceed with reconstructing the scalar field potential. For this, we have considered the widely studied power-law form of the Lagrangian function $f(Q)$ in nonmetricity gravity. As noted in previous studies, the free parameter $n$ has played a crucial role in describing both the present and late-time acceleration of the Universe. With this in mind, we have reconstructed the scalar field potential and kinetic energy for different values of $n$. We then reconstructed the potential \(V(\phi)\) as a function of the scalar field \(\phi\) to compare it with the potentials explored in the literature. For better comparison, we have adopted an analytical form for the scalar field potential using a fourth-order polynomial function. Furthermore, we have discussed the parameter space of the coefficients in \(V(\phi)\) to determine the best fit for the reconstructed potential. Furthermore, we compared our results with various other potentials and their parameter values, aiming to achieve better agreement with observational datasets.

In the early phase of the Universe, the influence of the reconstructed scalar field potential is high (see Figure \ref{potentialz}), and it can properly study the early inflation of the Universe. Later, for redshifts $z < 0.15$, both the kinetic and potential energies of the scalar field converge to near zero. This implies that the influence of the scalar field becomes negligible in the present-time evolution of the Universe, particularly within the framework of the nonmetricity gravitational theory. As a result, the present-day accelerated expansion of the Universe is no longer driven by the scalar field but is instead attributed solely to the effects of the modified geometrical part of nonmetricity gravity. In this context, the modified geometric part of nonmetric gravity acts as the primary mechanism responsible for the current cosmic acceleration, and the present value of the EoS $w_{f}$ parameter lies in the range $-1.0484\lesssim w_{f}\lesssim-0.9516$ for $n\in[-0.15,0.15]$. In this scenario, the modified part of the geometry acts as a source of DE. 

This behavior suggests that in the early Universe, the dynamics of inflation and matter-dominated expansion were influenced by the scalar field, while in the current epoch, the late-time acceleration is purely governed by the geometrical modifications introduced by the nonmetricity gravity. In future work, we aim to conduct a more rigorous study of the inflationary scenario of the Universe by exploring the reconstructed scalar field potential in detail. This analysis will allow us to better understand the role of this potential in driving inflation, providing deeper insights into the early Universe's dynamics and the conditions that led to its rapid expansion. Further, it would be interesting to see the same analysis with more data points for the other observational samples, such as SNIa supernova and gamma-ray bursts, which may provide better results.

In the previous chapters, we reconstructed the \( f(Q) \) models using various methods, predominantly obtaining polynomial-type or power-law functional forms. In the subsequent chapter, we focus on studying the generalized form of these models by exploring scenarios involving interactions between DE and DM through the dynamical systems approach.

\chapter{Cosmological Dynamics of Interacting Dark Energy and Dark Matter in $f(Q)$ Gravity} 
\label{Chapter5} 
\epigraph{\justifying \textit{``Light brings us the news of the Universe.”}}{\textit{William Bragg}}
\lhead{Chapter 5. \emph{Cosmological Dynamics of Interacting DE and DM in $f(Q)$ Gravity}} 

\blfootnote{*The work in this chapter is covered by the following publication:\\
\textit{Cosmological Dynamics of Interacting Dark Energy and Dark Matter in $f(Q)$ Gravity}, Fortschritte der Physik \textbf{73}, 2400205 (2025).}


The current chapter presents the cosmological dynamics of interacting DE and DM in $f(Q)$ gravity. A detailed study of the work is given below.
\begin{itemize}
\item The present study aims to explore the behavior of interacting DE and DM within a model of $f(Q)$ gravity, employing a standard framework for dynamical system analysis.
\item  We consider the power-law model $f(Q)$ that incorporates two different forms of interacting DE and DM: $3\alpha H\rho_m$ and $\frac{\alpha}{3H}\rho_m \rho_{DE}$.
\item The evolution of \(\Omega_m\), \(\Omega_r\), \(\Omega_{DE}\), \(q\), and \(\omega\) for different values of the model parameter \(n\) and the interaction parameter \(\alpha\) has been examined.
\end{itemize}


\section{Introduction} \label{sec:3.1}

The interaction between DM and DE is a promising mechanism for addressing the cosmic coincidence problem, although its initial motivation was to resolve the discrepancy in the cosmological constant \cite{Campo/2009,Pavon/2005}. Early work by Wetterich demonstrated that an interaction between a scalar field and gravity could result in a dynamic effective cosmological constant that asymptotically approaches a small value, providing a plausible explanation for the mismatched cosmological constant \cite{Wetterich/1995}. These dual motivations laid the foundation for exploring interactions within the dark sector. Initially, the primary motivation for interacting dark energy (IDE) models was to address or mitigate the coincidence problem. However, more recently, the focus has shifted toward resolving the discrepancy between the Hubble constant values derived from CMB and local measurements. As the tension between high-redshift and low-redshift Hubble constant measurements has persisted with improving data, IDE models have emerged as promising candidates to reconcile these discrepancies with the $\Lambda$CDM model \cite{Vagnozzi/2017,Aljaf/2021,Cheng/2020,Yang/2017,Nong/2024}.

The interaction within the dark sector is largely a phenomenological concept, as no fundamental principle explicitly dictates its existence. However, from a theoretical perspective, particularly in particle physics, any two matter fields, such as the DM and DE fields, can interact. This idea has garnered significant attention within the cosmological community because of its potential implications. Notably, allowing such an interaction can transition the DE equation of state (EoS) from the quintessence regime to the phantom regime, effectively introducing a quintom-like behavior. Phenomenological models are typically formulated by incorporating an energy exchange between dark components into their continuity equations with an interacting kernel $\mathcal{U}$. Similarly to interactions in particle physics, the kernel is generally expected to depend on the energy densities involved, such as $\rho_{DE}$ and $\rho_{CDM}$, as well as on time, represented by $H^{-1}$. Moreover, within the framework of field theory, considering interactions between the dark sectors is both natural and unavoidable. Exploring these interactions could provide valuable insight into the fundamental nature of DM and DE. Various interaction models that have been proposed and tested in the literature are linear models, such as $\mathcal{U} = \alpha_{m} H \rho_{m}$, $\mathcal{U} = \alpha_{DE}H\rho_{DE}$ and $\mathcal{U} = H(\alpha_{m} \rho_{m} + \alpha_{DE}\rho_{DE})$ \cite{Amendola/2007,Quartin/2008,Pavon/2009,Wang/2016}. However, only a limited number of non-linear models have been proposed and thoroughly investigated in the literature \cite{Wang/2024}.

 For several decades, dynamical system analysis has been used in cosmology to qualitatively study these models, proving to be effective in identifying and classifying their asymptotic behaviors \cite{Wainwright/2005,Coley/1999}. Studying autonomous dynamical analysis and point stability is essential for understanding the evolution of the Universe and the behavior of cosmological models. Autonomous dynamical systems reformulate the equations governing cosmic evolution into first-order differential equations, simplifying the analysis and enabling a deeper exploration of the Universe's dynamics. Critical points within these systems represent key asymptotic behaviors of the Universe, such as matter domination, radiation domination, or accelerated expansion. Stability analysis of these points is critical for determining whether the Universe can evolve toward specific states, such as the current dark-energy-dominated epoch, thereby linking theoretical predictions with observational data. Moreover, dynamical system analysis provides insights into the influence of model parameters, such as coupling constants or power law indices, on the Universe's trajectory, helping identify ranges that lead to physically viable solutions, including late-time acceleration or scaling behaviors. This framework has been applied to investigate various DE models in various modified theories of gravity \cite{Jimenez/2020,Bahamonde/2018,Samart/2021,Khyllep/2022,Hussain/2024,An/2016,Carloni/2016,D’Ambrosio/2022,Arora/2022}. 

In this chapter, we investigate the interaction between DE and DM within a viable model of $f(Q)$ gravity. The structure of this chapter is organized as follows. In Section \ref{5a}, we present the fundamental cosmological equations of the general theory $f(Q)$ and derive the Friedmann equation corresponding to the FLRW metric. In Section \ref{5b}, we derive the autonomous dynamical system within the framework of gravity theory $f(Q)$, focusing on the interaction between DE and DM. In Section \ref{5c}, we apply the power law model $f(Q)$ to enclose the dynamical system and conduct a further study of $f(Q)$ gravity. Finally, we discuss our findings in Section \ref{5d}.

\section{Formulation of $f(Q)$ Gravity Theory}
\label{5a}
To apply $ f(Q)$ gravity in a cosmological context, we consider the spatially FLRW spacetime, characterized by the metric \eqref{FLRW}. Applying the FLRW metric into the general field equation \eqref{7}, the Friedmann equations of $f(Q)$ cosmology read as  
\begin{eqnarray}
\label{f1}
6H^2f_Q-\frac{f}{2}&=&\rho\,,\\
\label{f2}
\left(12H^2\,f_{QQ}+f_Q\right)\dot{H}&=&-\frac{1}{2}(p+\rho)\,.
\end{eqnarray} 
To study the interaction between DE and DM, we have to modify the conservation equations of the DE and the matter given above by adding some coupling term $\mathcal{U}$ as \cite{Samart/2021}
 \begin{eqnarray*}
  && \dot{\rho}_{DE} + 3H(p_{DE}+\rho_{DE})=\mathcal{U}\,, \\
  && \dot{\rho}_{m} + 3H\,\rho_m =-\mathcal{U}\,,\\
  && \dot{\rho}_{r}+ 4H\,\rho_r =0\,.
\end{eqnarray*}
In addition, the coupling term $\mathcal{U}$ between DE and DM can be interpreted as the exchange rate of the energy density between these two components. When \(\mathcal{U} > 0\), energy is transferred from DM to DE, while when \(\mathcal{U} < 0\), energy is transferred from DE to DM.

\section{Autonomous Dynamical System of Interacting DE
and DM}\label{5b}
In this section, we derive the autonomous dynamical system within the framework of $f(Q)$ gravity theory, focusing on the interaction between DE and DM. Our approach involves incorporating two interacting terms. The first term relies solely on the energy density of the matter sector, the Hubble parameter, and a coupling constant $\alpha$, structured multiplicatively as $\mathcal{U}=3\alpha\,H\,\rho_m$. The second term encompasses the multiplication of energy densities from both sectors, capturing the immediate effects of both DM and DE on the interaction term, with a dimensionless coupling constant $\alpha$, denoted as $\mathcal{U}=\frac{\alpha}{3H}\rho_m\,\rho_{DE}$.\\
According to the Friedmann equation \eqref{f1}, we can define the dimensionless variable as
\begin{equation}
\label{16}
    x=\frac{f}{12H^2\,f_Q},\,\,\,\,\,y=\frac{\rho_r}{6H^2\,f_Q}.
\end{equation}
Invoking the above dimensionless variables, it is straightforward to derive the autonomous equations which play a key role in studying the dynamical system for the interacting DE and DM.

\subsection{Case I: $\mathcal{U}=3\alpha\,H\,\rho_m$}

Using Eq.\eqref{f1}, the autonomous dynamical system is given by
\begin{equation}
    \frac{dx}{dN}=(1-x)\frac{\dot{H}}{H^2}+3x-3x^2+xy+3\alpha\,x(1-x-y),
     \end{equation}
\begin{equation}
    \frac{dy}{dN}=-\frac{\dot{H}}{H^2}\,y-y+y^2-3xy+3\alpha\,y(1-x-y),
\end{equation}
where the variable $N =ln\,a$ is e-folding number and leads to $\frac{d}{dN}=\frac{1}{H}\frac{d}{dt}$.

\subsection{Case II: $\mathcal{U}=\frac{\alpha}{3H}\rho_m\,\rho_{DE}$}
For this particular case, the autonomous dynamical system is given by
\begin{equation}
    \frac{dx}{dN}=(1-x)\frac{\dot{H}}{H^2}+3x-3x^2+xy+\alpha\,x^2(1-x-y),
 \end{equation}
\begin{equation}
    \frac{dy}{dN}=-\frac{\dot{H}}{H^2}\,y-y+y^2-3xy+\alpha\,xy(1-x-y).
\end{equation}
Furthermore, we obtain the generalized equation for $\frac{\dot{H}}{H^{2}}$ as 
\begin{equation}
\label{21}
    \frac{\dot{H}}{H^{2}} = \frac{-3f_{Q}(1-x+\frac{y}{3})}{2Qf_{QQ}+f_{Q}}.
\end{equation}
Consequently, the density parameters for individual matter species can be linked to the dimensionless variables outlined in Eq.\eqref{16}. These variables serve as a set of constraint equations
\begin{equation}
    \Omega_m=1-x-y,\,\,\,\,\,\,\,\Omega_r=y,\,\,\,\,\,\,\,\Omega_{DE}=x.
\end{equation}
Finally, we define the EoS and deceleration parameters corresponding to the dimensionless variables to check the appropriate acceleration expansion of the Universe, that is,
\begin{equation}
    \omega=-1+\frac{2f_{Q}(1-x+\frac{y}{3})}{2Qf_{QQ}+f_{Q}}
\end{equation}
and 
\begin{equation}
    q=-1+\frac{3f_{Q}(1-x+\frac{y}{3})}{2Qf_{QQ}+f_{Q}}.
\end{equation}
Both $\omega$ and $q$ represent different aspects of cosmic evolution, each uniquely influencing the large-scale structure of the Universe. Grasping the significance of these parameters is essential for studying the effects of DE across various stages of cosmic evolution. Moreover, these parameters aid in comparing and differentiating between various DE models, each employing distinct mechanisms to drive cosmic acceleration.

\section{Model: $f(Q)=6\gamma\,H_0^2\left(\frac{Q}{Q_0}\right)^n$}\label{5c}
In the previous section, we derived autonomous dynamical systems, and our primary objective is to investigate and analyze them. To accomplish this goal, we will focus on the designated power-law model $f(Q)$, which has the form $f(Q)=6\gamma\,H_0^2\left(\frac{Q}{Q_0}\right)^n$, where $\gamma$ and $n$ are free model parameters.\\
Corresponding to our power-law $f(Q)$ model, Eq.\eqref{21} can be simplified to obtain
\begin{equation}
    \frac{\dot{H}}{H^2}=-\frac{3(1-x+\frac{1}{3}y)}{2n-1}.
\end{equation}
We will now derive the critical points of the system by setting $dx/dN = dy/dN = 0$ and determine the corresponding eigenvalues for both dynamical systems. The critical points, which represent the solutions of the dynamical system, provide an initial qualitative understanding of the phase space. As discussed in the following, these points can be classified on the basis of their stability properties. In the absence of singularities or strange attractors, the trajectories of $x(N)$ and $y(N)$, which are generally obtained numerically, tend to evolve from unstable fixed points to stable fixed points, possibly passing through intermediate saddle points.\\
Analyzing the critical points and evaluating their stability is essential for a thorough understanding of the critical aspects of cosmic evolution driven by interactions between DE and DM, as explored in this study. Furthermore, the properties of the dynamical system depend significantly on the values of the constants $\alpha$ and $n$.

\subsection{For case I}
The fixed points $(x,y)$ of the general dynamical system are described in Table \ref{TABLE-I}.
Let us now focus on the dynamics of each critical point and its features in the following subsections.

\begin{table*}[!htb]
\centering 
\resizebox{\textwidth}{!}{%
\begin{tabular}{|*{6}{c|}}\hline 
    \parbox[c][0.8cm]{3cm}{Critical Points} & $\Omega_m$ & $\Omega_{DE}$ & $w$ & Eigenvalues & Stability\\ [0.5ex]\hline \hline 
    \parbox[c][1.2cm]{3cm}{$P_1:(1,0)$} & $0$ & $1$ & $-1$ & $\{ -4,\,\,\,-3 (\alpha +1) \}$ & \begin{tabular}{@{}c@{}}Stable node for $\alpha>-1$, \\ Saddle for $\alpha<-1$,\\  Non-hyperbolic for $\alpha=-1$.\end{tabular}\\
    \hline
    \parbox[c][1cm]{3cm}{$P_2:\left(\frac{1}{2n},\frac{2n-1}{2n}\right)$} & $0$ & $\frac{1}{2n}$ & $-1+\frac{4}{3n}$ & $\{4,1-3\alpha\}$  &  \begin{tabular}{@{}c@{}}Saddle node for $\alpha>1/3$,\\ Unstable for $\alpha<1/3$,\\ Non-hyperbolic for $\alpha=1/3$.
    \end{tabular}\\
    \hline
    \parbox[c][0.8cm]{3cm}{$P_3\text{:}\left(\frac{1}{2n+\alpha(2n-1)},0\right)$} & $\frac{1}{\alpha -2 (\alpha +1) n}+1$ & $\frac{1}{2n+\alpha(2n-1)}$ & $\frac{3 \alpha -2 (\alpha +1) n+2}{2 (\alpha +1) n-\alpha }$ & $\{3 (\alpha +1),3 \alpha -1\}$ & \begin{tabular}{@{}c@{}}Stable node for $\alpha<-1$,\\ Saddle for $-1<\alpha<1/3$,\\ Unstable node for $\alpha>1/3$.
    \end{tabular} \\
    \hline
\end{tabular}}
\caption{This table summarizes the analysis of the critical points for Case I. }
\label{TABLE-I}
\end{table*}

\subsubsection{$P_1$: The de Sitter fixed point}
We start the discussion with the first fixed point. Here, we have
\begin{equation*}
    P_1: (x,y)=(1,0).
\end{equation*}
This critical point is independent of both the model and the coupling parameters. Corresponding to this fixed point, the matter density, DE density, and effective EoS parameters are
\begin{equation}
   \Omega_{m}=0,\,\,\,\,\Omega_{DE}=1,\,\,\,\,\omega=-1.
\end{equation}
At this point, both DM and radiation are absent, and the Universe is dominated by DE. The EoS at this fixed point indicates that the Universe is experiencing an accelerated expansion. To study the stability of this fixed point, we can analyze the eigenvalues of the Jacobian matrix, which describe the behavior of the fixed point. The eigenvalues for the fixed point $P_1$ read $  \lbrace -4,\,\,\,-3 (\alpha +1) \rbrace$.
The stability behavior of this fixed point (which depends on $\alpha$) is obtained as
\begin{itemize}
    \item Stable node for $\alpha>-1$,
    \item Saddle for $\alpha<-1$,
    \item Non-hyperbolic for $\alpha=-1$.
\end{itemize}

\subsubsection{$P_2$: Nonmetricity-dominated fixed point}
Next, for the second fixed point, we have
\begin{equation*}
    P_2: (x,y)=\left(\frac{1}{2n},\frac{2n-1}{2n}\right).
\end{equation*}
The characteristics of this critical point are solely dependent on the model parameter $n$.
 Corresponding to this fixed point, the matter density, DE density, and effective EoS parameters are obtained as
\begin{equation}
   \Omega_{m}=0,\,\,\,\,\Omega_{DE}=\frac{1}{2n},\,\,\,\,\omega=-1+\frac{4}{3n}.
\end{equation}
The conditions for an accelerating Universe are $n<0$ or $n>2$. When $n<0$, the Universe exhibits phantom-like behavior, i.e., $\omega<-1$. When $n>2$, the Universe exhibits quintessence-like behavior, i.e., $-1<\omega<-1/3$. In this fixed-point solution, a matter-dominated era occurs for $n=4/3$ and a radiation-dominated era for $n=1$. The eigenvalues for the fixed point $P_2$ are as follows $ \lbrace 4,\,\,\,1-3\alpha \rbrace$. The stability behavior for this particular fixed point can be obtained for different $\alpha$, given by
\begin{itemize}
    \item Saddle node for $\alpha>1/3$,
    \item Unstable for $\alpha<1/3$,
    \item Non-hyperbolic for $\alpha=1/3$.
\end{itemize}

\subsubsection{$P_3$: Scaling solution fixed point}

\begin{figure}[]
\centering
\includegraphics[scale=0.5]{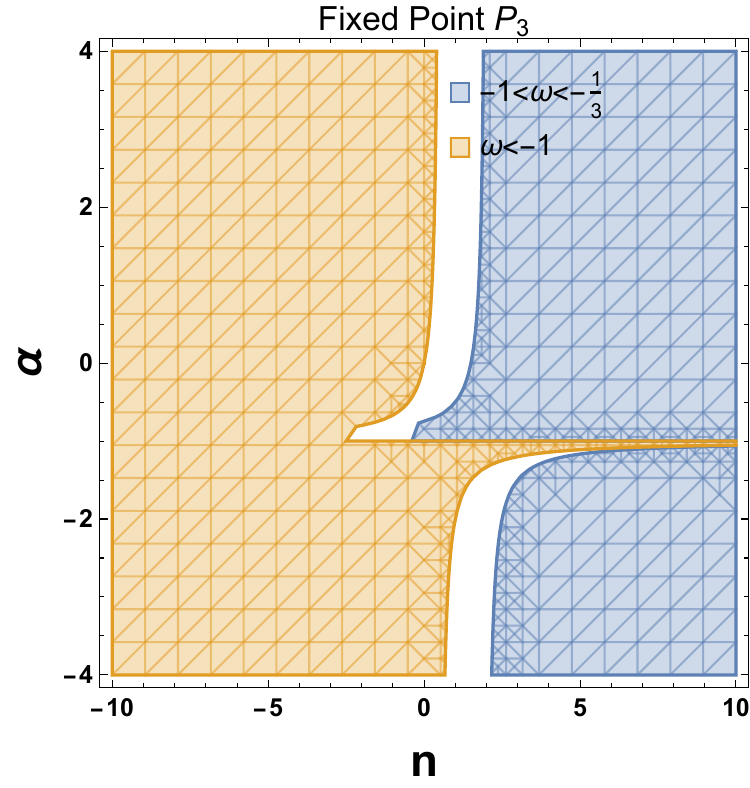}
\caption{\justifying The plot illustrates the relationship between the model parameter $n$ and the interaction parameter $\alpha$ for an accelerating Universe.  The light blue region represents the quintessence-like behavior of an accelerating Universe (i.e., $-1<\omega<-1/3$), while the golden region represents the phantom-like behavior of an accelerating Universe (i.e., $\omega<-1$).} 
\label{fig1}
\end{figure}

It should be noted that our result modifies the scaling solution of gravity $f(Q)$ with the interacting DE parameter $\alpha$. The fixed point $P_3$ reads
\begin{equation*}
     P_3: (x,y)=\left(\frac{1}{2n+\alpha(2n-1)},0\right).
\end{equation*}
The properties of this critical point are influenced both by the model parameter $n$ and the coupling parameter $\alpha$. Corresponding to this fixed point, the matter density, DE density, and effective EoS parameters read as
\begin{equation*}
   \Omega_{m}=\frac{1}{\alpha -2 (\alpha +1) n}+1,\quad \Omega_{DE}=\frac{1}{2n+\alpha(2n-1)}, \quad \omega=\frac{3 \alpha -2 (\alpha +1) n+2}{2 (\alpha +1) n-\alpha }.
\end{equation*}
The conditions for an accelerating Universe are shown in Figure \ref{fig1}.

In this fixed-point solution, we obtain a matter-dominated era for $n=\frac{3 \alpha +2}{2 (\alpha +1)}$, $1+\alpha\neq 0$, and a radiation-dominated era for $n=\frac{5 \alpha +3}{4 (\alpha +1)}$, $1+\alpha\neq 0$. The eigenvalues for the fixed point $P_3$ are $\lbrace 3 (\alpha +1),\,\,3 \alpha -1 \rbrace$.
We determine the stability of this fixed point for various ranges of $\alpha$. The ranges are as follows:
\begin{itemize}
    \item Stable node for $\alpha<-1$,
    \item Saddle for $-1<\alpha<1/3$,
    \item Unstable node for $\alpha>1/3$.
\end{itemize}

\begin{figure}[H]
\centering
\includegraphics[scale=0.55]{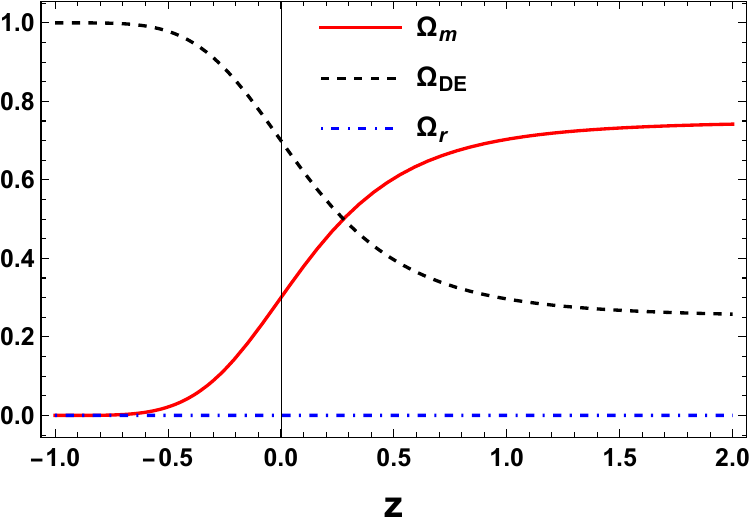} \hspace{0.2in}
\includegraphics[scale=0.55]{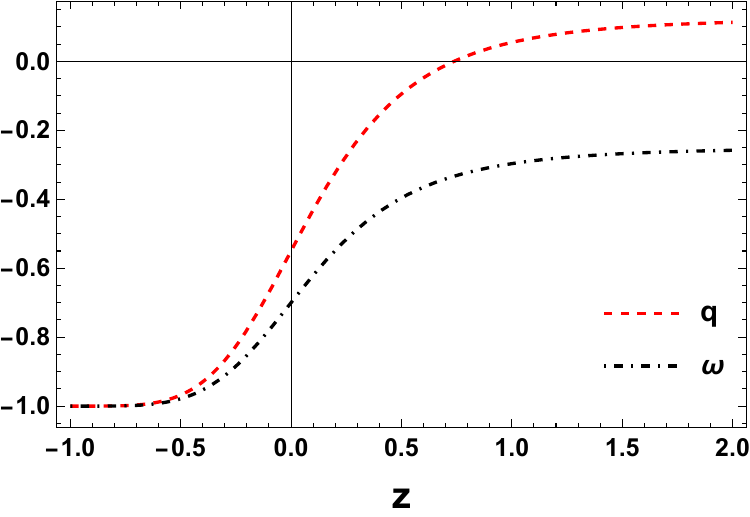}
\caption{\justifying The left panel shows the density parameters for matter ($\Omega_m$), DE ($\Omega_{DE}$), and radiation ($\Omega_r$) as functions of redshift ($z$). The right panel displays the deceleration parameter ($q$) and the equation of state parameter ($\omega$) as functions of redshift ($z$). Together, these two panels illustrate the trajectory for a positive coupling parameter, which can facilitate the transfer of energy from DM to DE.}
\label{fig2}
\end{figure}

\begin{figure}[H]
\centering
\includegraphics[scale=0.55]{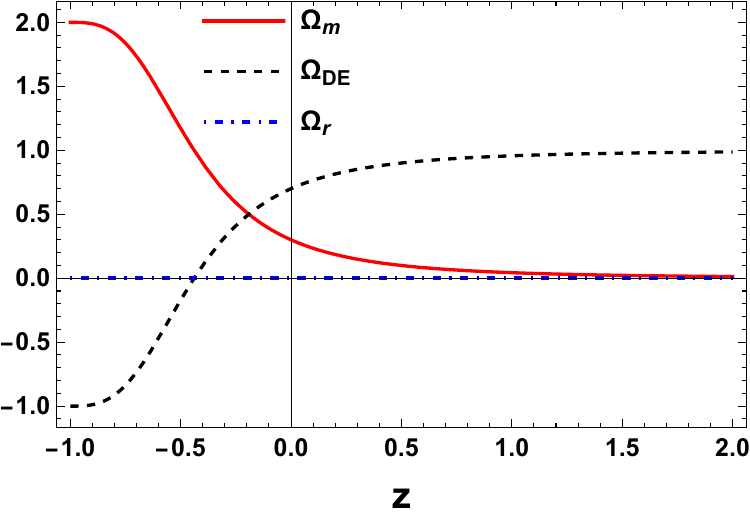} \hspace{0.2in}
\includegraphics[scale=0.55]{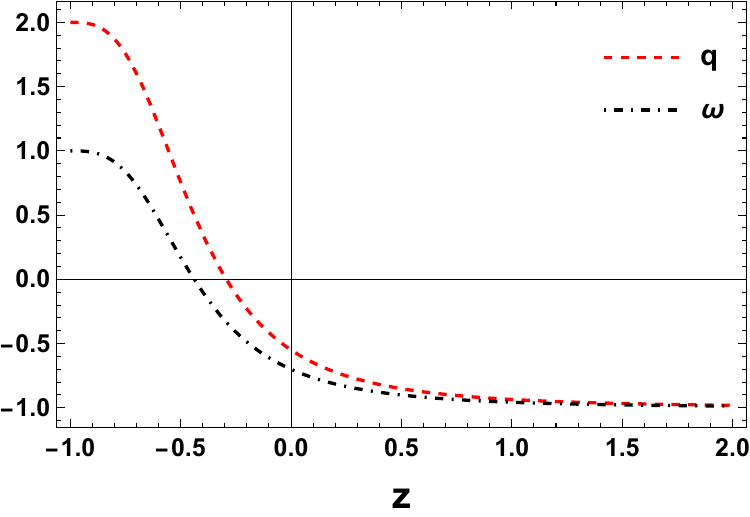}
\caption{\justifying The left panel displays density parameters for matter ($\Omega_m$), DE ($\Omega_{DE}$), and radiation ($\Omega_r$) as functions of redshift $z$. The right panel shows the deceleration parameter $q$ and the EoS parameter $\omega$ as functions of redshift $z$. These two panels illustrate the trajectory for the negative coupling parameter, which can transfer energy from DE to DM.} 
\label{fig3}
\end{figure}

In Figures \ref{fig2} and \ref{fig3}, we illustrate the evolution of \(\Omega_m\), \(\Omega_r\), \(\Omega_{DE}\), \(q\), and \(\omega\) for different values of the model parameter \(n\) and the interaction parameter \(\alpha\). This is done using the numerical solution of the dynamical system with the initial conditions $x(0)=0.7$ and $y(0)=0.00005$. The motivation behind the choice of initial conditions is as follows: \( x \) represents the DE density, which currently has a value of $0.7$. Similarly, \( y \) represents the radiation density, which currently has a value of $0.00005$. \\

\begin{figure}[]
\centering
\includegraphics[scale=0.5]{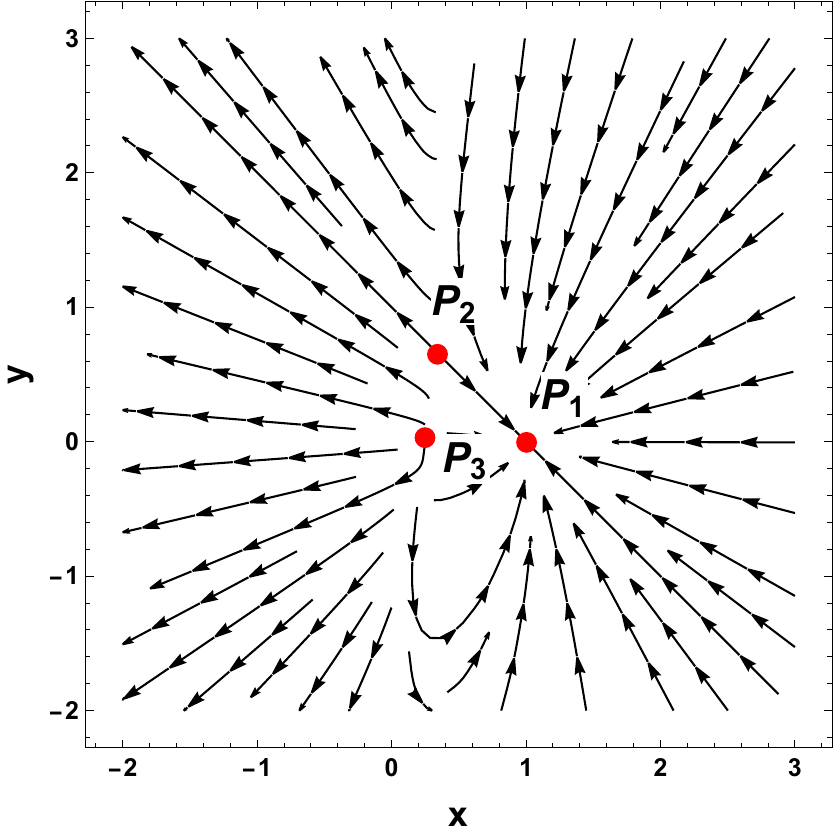} \hspace{0.2in} 
\includegraphics[scale=0.5]{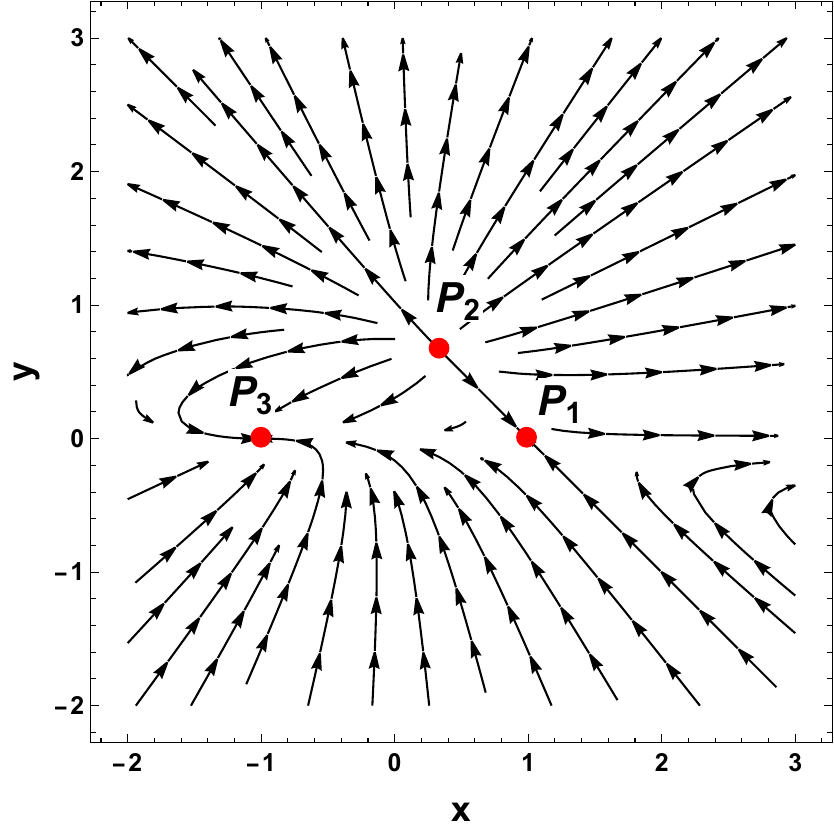}
\caption{\justifying Phase space of $x$ vs $y$. The left phase plot illustrates the stable behavior of $P_1$ when $\alpha>0$, while the right phase plot showcases the stable behavior of $P_3$ for $\alpha<-1$.} 
\label{fig4}
\end{figure}

For \(n = 3/2\) and \(\alpha = 1/2\) in Figure \ref{fig2}, \(\alpha > 0\) indicates the coupling term \(\mathcal{U} > 0\), which means the transfer of energy from DM to DE. This figure shows that the Universe was dominated by matter in the early stages and will be dominated by DE in later stages. Currently, the Universe is dominated by DE, with parameters \(\Omega_m = 0.3\), \(\Omega_r = 0.00005\), \(\Omega_{DE} = 0.7\), \(q_0 = -0.55\), and \(\omega_0 = -0.70\). For these values, the fixed point \(P_1\) is stable and represents the de Sitter acceleration solution, while the fixed point \(P_2\) is a saddle node, and \(P_3\) is an unstable node that cannot demonstrate universal acceleration.\\
In Figure \ref{fig3}, with \(n = 3/2\) and \(\alpha = -2\), \(\alpha < 0\) indicates the coupling term \(\mathcal{U} < 0\), which means energy transfers from DE to DM. This figure shows that the Universe was dominated by DE in the early stages and will be dominated by DM at later stages. Currently, the Universe remains dominated by DE, with parameters \(\Omega_m = 0.3\), \(\Omega_r = 0.00005\), \(\Omega_{DE} = 0.7\), \(q_0 = -0.563\), and \(\omega_0 = -0.71\). Here, fixed point \(P_3\) is stable and exhibits acceleration for the early and present Universe but fails to show acceleration for late times, while fixed point \(P_1\) is a saddle node, and \(P_2\) is an unstable node. Figure \ref{fig4} shows the phase space trajectories of the fixed points \(P_1\), \(P_2\), and \(P_3\).

\subsection{For case II}
Next, the second case in $f(Q)$ gravity is also investigated in detail for the interacting DE system. The models have been systematically constructed with possible ranges of their model parameters derived from dynamical system analysis within the interacting DE framework. It is important to examine how the interaction affects the cosmological viability of the $f(Q)$ gravity models.

The fixed points $(x,y)$ of the dynamical system are described in Table \ref{TABLE-II}. The expressions of $\mathcal{A}_1$ and $\mathcal{A}_2$ in the table are given by $\mathcal{A}_1=\frac{2 \sqrt{(2 n-1) \left(3 (\alpha -1) \alpha ^2+8 n^3-4 n^2-6 (\alpha -1) \alpha  n\right)}}{\alpha -2 \alpha  n}$ and $\mathcal{A}_2=\sqrt{9n^2-3\alpha+6n\alpha}$.

\begin{table}[!htb]
\centering 
\resizebox{\textwidth}{!}{%
\begin{tabular}{|*{6}{c|}}\hline 
    \parbox[c][0.8cm]{3.3cm}{Critical Points} & $\Omega_m$ & $\Omega_{DE}$ & $w$ & Eigenvalues & Stability\\ [0.5ex]\hline \hline 
    \parbox[c][1.2cm]{5cm}{$R_1:(1,0)$} & $0$ & $1$ & $-1$ & $\{-4,-3 -\alpha\}$ & \begin{tabular}{@{}c@{}}Stable node for $\alpha>-3$,\\ Saddle for $\alpha<-3$,\\ Non-hyperbolic for $\alpha=-3$.\end{tabular}\\
    \hline
    \parbox[c][1cm]{5cm}{$R_2:\left(\frac{1}{2n},\frac{2n-1}{2n}\right)$} & $0$ & $\frac{1}{2n}$ & $-1+\frac{4}{3n}$ & $\{4,1-\frac{\alpha}{2n}\}$  &  \begin{tabular}{@{}c@{}}Saddle node for $\alpha>2n$,\\ Unstable for $\alpha<2n$,\\ Non-hyperbolic for $\alpha=2n$.
    \end{tabular}\\
    \hline
    \parbox[c][0.8cm]{5cm}{$R_3:\left(\frac{1}{\alpha },\frac{-3 \alpha +8 n-1}{\alpha }\right)$} & $4-\frac{8 n}{\alpha }$ & $\frac{1}{\alpha }$ & $\frac{8}{3 \alpha }-1$ & $\{\mathcal{A}_1-\frac{4n}{\alpha}+4,-\mathcal{A}_1-\frac{4n}{\alpha}+4\}$ & \begin{tabular}{@{}c@{}}The stability conditions\\ are shown in Fig. \ref{fig5}
     \end{tabular} \\
     \hline
     \parbox[c][2cm]{5cm}{$R_4:\left(\frac{-\sqrt{-3\alpha +9 n^2+6 \alpha  n}-3 n}{2\alpha n-\alpha },0\right)$} & Sec.\ref{R4}    &  \ref{R4} &  \ref{R4} & $\left\lbrace  \frac{5n-1+\mathcal{A}_2}{1-2n},\frac{-2\mathcal{A}_2^2-2(3n-\alpha+2n\alpha)\mathcal{A}_2}{(2n-1)^2\alpha}\right\rbrace$ & \begin{tabular}{@{}c@{}}The stability conditions\\ are shown in Fig. \ref{fig7}
     \end{tabular} \\
     \hline
     \parbox[c][2cm]{5cm}{$R_5:\left(\frac{ \sqrt{-3\alpha +9 n^2+6 \alpha  n}-3 n}{2 \alpha  n-\alpha },0\right)$} & Sec.\ref{R5} & \ref{R5} & \ref{R5} & $\left\lbrace \frac{-2\mathcal{A}_2^2+2(3n-\alpha+2n\alpha)\mathcal{A}_2}{(2n-1)^2\alpha},\frac{1-5n+\mathcal{A}_2}{2n-1}\right\rbrace$ & \begin{tabular}{@{}c@{}}The stability conditions\\ are shown in Fig. \ref{fig8}
     \end{tabular} \\
     \hline
\end{tabular}}
\caption{\justifying This figure summarizes the results of the analysis of the critical points for Case II. The critical points \( R_4 \) and \( R_5 \), however, are difficult to present comprehensively in a table format. Detailed analyses of these points can be found directly in Sections \ref{R4} and \ref{R5}.}
\label{TABLE-II}
\end{table}

\subsubsection{$R_1$: The de Sitter fixed point}\label{sub2.1}

Corresponding to fixed point $R_1: (x,y)=(1,0)$, the matter density, DE density, and effective EoS parameters are
\begin{equation}
   \Omega_{m}=0,\,\,\,\,\Omega_{DE}=1,\,\,\,\,\omega=-1.
\end{equation}
At this point, both DM and radiation are absent, and the Universe is dominated by DE. The EoS at this fixed point indicates that the Universe is experiencing an accelerated expansion. To study the stability of this fixed point, we can analyze the eigenvalues of the Jacobian matrix, which describe the behavior of the fixed point. The eigenvalues for the fixed point $R_1$ are $\lbrace{ -4,\,\,\,-\alpha -3\rbrace}$.

The stability behavior of this fixed point is as follows: 
\begin{itemize}
    \item Stable node for $\alpha>-3$,
    \item Saddle for $\alpha<-3$,
    \item Non-hyperbolic for $\alpha=-3$.
\end{itemize}
\subsubsection{$R_2$: Nonmetricity-dominated fixed point}
The fixed point $R_2$ can be used to explain the late-time accelerating expansion of the Universe in $f(Q)$ gravity. The point is given by 
\begin{equation*}
    R_2: (x,y)=\left(\frac{1}{2n},\frac{2n-1}{2n}\right).
\end{equation*}
The characteristics of this critical point are solely dependent on the model parameter $n$. Corresponding to this fixed point, the matter density, DE density, and effective EoS parameters are obtained as
\begin{equation}
   \Omega_{m}=0,\,\,\,\,\Omega_{DE}=\frac{1}{2n},\,\,\,\,\omega=-1+\frac{4}{3n}.
\end{equation}
The conditions for an accelerating Universe are $n<0$ or $n>2$. When $n<0$, the Universe exhibits phantom-like behavior, i.e., $\omega<-1$. When $n>2$, the Universe exhibits quintessence-like behavior, i.e., $-1<\omega<-1/3$. In this fixed-point solution, we have obtained a matter-dominated era for $n=4/3$ and a radiation-dominated era for $n=1$.\\
The corresponding eigenvalues for the fixed point $R_2$ are $\lbrace{ 4,\,\,1-\frac{\alpha}{2n} \rbrace}$. We determine the stability of this fixed point for various ranges of $\alpha$. The ranges are as follows:
\begin{itemize}
    \item Saddle node for $\alpha>2n$,
    \item Unstable for $\alpha<2n$,
    \item Non-hyperbolic for $\alpha=2n$.
\end{itemize}

\subsubsection{$R_3$: Scaling solution point}\label{sub2.2}
\begin{figure}[]
\centering
\includegraphics[scale=0.7]{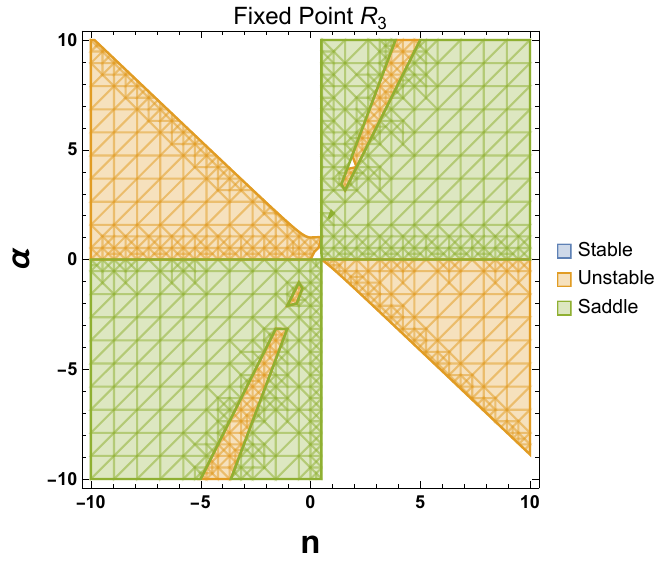}
\caption{\justifying The region plot illustrates the stable, unstable, and saddle behaviors associated with the fixed point $R_3$. In this plot, the stable region is very small $\left(\frac{1}{8}<n\leq 0.4393\,\, \text{and}\,\, 0<\alpha <\frac{1}{3} (8 n-1)\right)$. } 
\label{fig5}
\end{figure}

The fixed point $R_3$ represents a scaling solution for the Universe. This scaling solution makes the ratio $\Omega_{m}/\Omega_{DE}$ constant. The fixed point in this case is given by
\begin{equation*}
    R_3: (x,y)=\left(\frac{1}{\alpha },\frac{-3 \alpha +8 n-1}{\alpha }\right).
\end{equation*}
Corresponding to this fixed point, the matter density, DE density, and effective EoS parameters read as
\begin{equation}
    \Omega_m=4-\frac{8 n}{\alpha },\,\,\,\,\Omega_{DE}=\frac{1}{\alpha },\,\,\,\,\omega=\frac{8}{3 \alpha }-1.
\end{equation} 
The conditions for an accelerating Universe are $\alpha<0$ or $\alpha>4$. When $\alpha<0$, the Universe exhibits phantom-like behavior, i.e., $\omega<-1$. When $\alpha>4$, the Universe exhibits quintessence-like behavior, i.e., $-1<\omega<-1/3$. In this fixed-point solution, we have obtained a matter-dominated era for $\alpha=8/3$ and a radiation-dominated era for $\alpha=2$. The stability conditions are shown in Figure \ref{fig5}.



\subsubsection{$R_4$: Scaling solution point}\label{R4}

The fixed point $R_4$ represents the scaling solution point of the Universe. In addition, it is worth noting that our result modifies the scaling solution with the interacting DE parameter $\alpha$. The point $R_4$ is 
\begin{equation*}
    R_4: (x,y)=\left(\frac{-\sqrt{-3\alpha +9 n^2+6 \alpha  n}-3 n}{2 \alpha  n-\alpha },0\right).
\end{equation*}

\begin{figure}

\begin{minipage}{.5\linewidth}
\centering
\subfloat[]{\label{A1}\includegraphics[scale=.5]{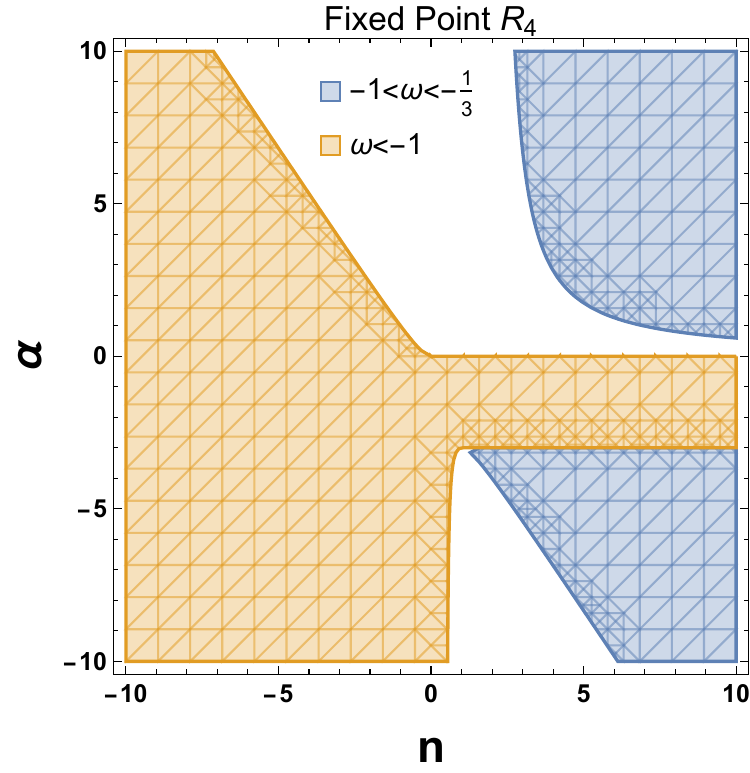}}
\end{minipage}%
\begin{minipage}{.5\linewidth}
\centering
\subfloat[]{\label{B1}\includegraphics[scale=.7]{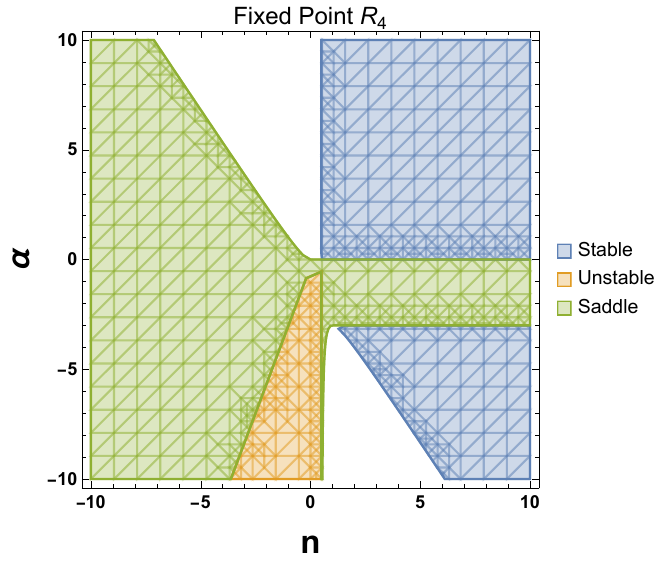}}
\end{minipage}\par\medskip

\caption{\justifying The region plot (A) shows the relationship between the model parameter $n$ and the interacting parameter $\alpha$ for the accelerating Universe. The light blue region represents the quintessence-like behavior of an accelerating Universe (i.e., $-1<\omega<-1/3$), while the golden region represents the phantom-like behavior of an accelerating Universe (i.e., $\omega<-1$). The region plot (B) illustrates the stable, unstable, and saddle behaviors associated with the fixed point $R_4$.}
\label{fig7}
\end{figure}

Corresponding to this fixed point, the matter density, DE density, and effective EoS parameters are obtained as 
 \begin{eqnarray*}
  &&  \Omega_m = \frac{-\alpha +\sqrt{-3 \alpha +9 n^2+6 \alpha  n}+(2 \alpha +3) n}{\alpha  (2 n-1)}\,, \\
  && \Omega_{DE} = \frac{-\sqrt{3} \sqrt{-\alpha +3 n^2+2 \alpha  n}-3 n}{2 \alpha  n-\alpha }\,,\\\text{and}\,
  && w = \frac{-3 \alpha -4 \alpha  n^2+2 \sqrt{-3 \alpha +9 n^2+6 \alpha  n}+(8 \alpha +6) n}{\alpha  (1-2 n)^2}\,.
\end{eqnarray*}
We note that all three parameters depend on the model parameter $n$ and the interacting parameter $\alpha$. The detailed calculation can be seen in the Appendix \ref{Appendix}. The conditions for an accelerating Universe and the stability are shown in Figures \ref{A1} and \ref{B1}.

\subsubsection{$R_5$: Additional scaling solution point}\label{R5}
\begin{figure}

\begin{minipage}{.5\linewidth}
\centering
\subfloat[]{\label{A2}\includegraphics[scale=.5]{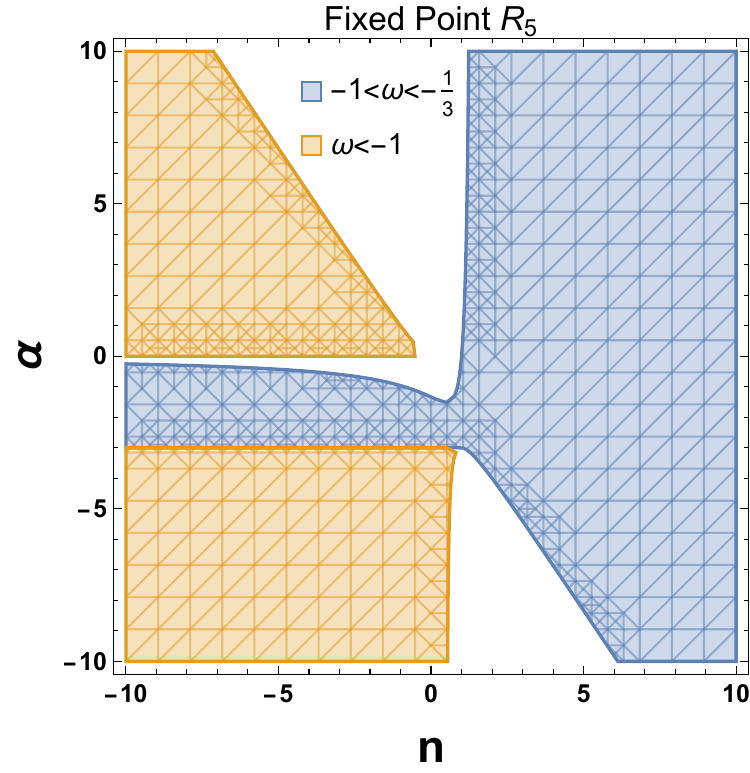}}
\end{minipage}%
\begin{minipage}{.5\linewidth}
\centering
\subfloat[]{\label{B2}\includegraphics[scale=.7]{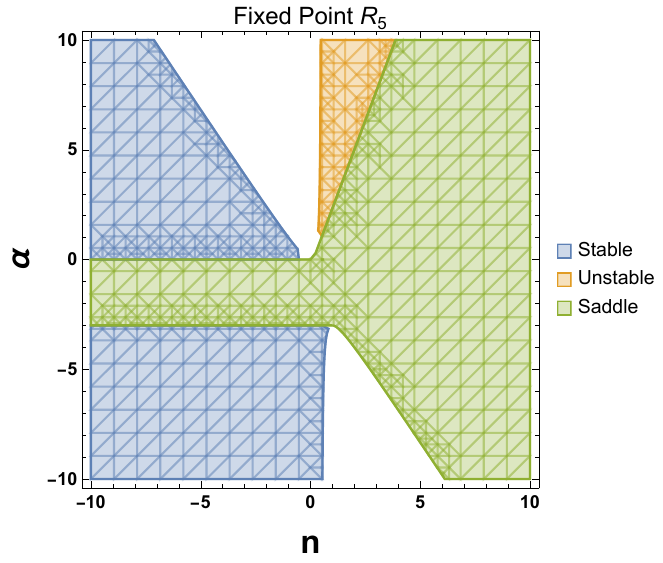}}
\end{minipage}\par\medskip

\caption{\justifying The region plot (A) shows the relationship between the model parameter $n$ and the interacting parameter $\alpha$ for the accelerating Universe. The light blue region represents the quintessence-like behavior of an accelerating Universe (i.e., $-1<\omega<-1/3$), while the golden region represents the phantom-like behavior of an accelerating Universe (i.e., $\omega<-1$). The region plot (B) illustrates the stable, unstable, and saddle behaviors associated with the fixed point $R_5$.}
\label{fig8}
\end{figure}

\begin{figure}[H]
\centering
\includegraphics[scale=0.55]{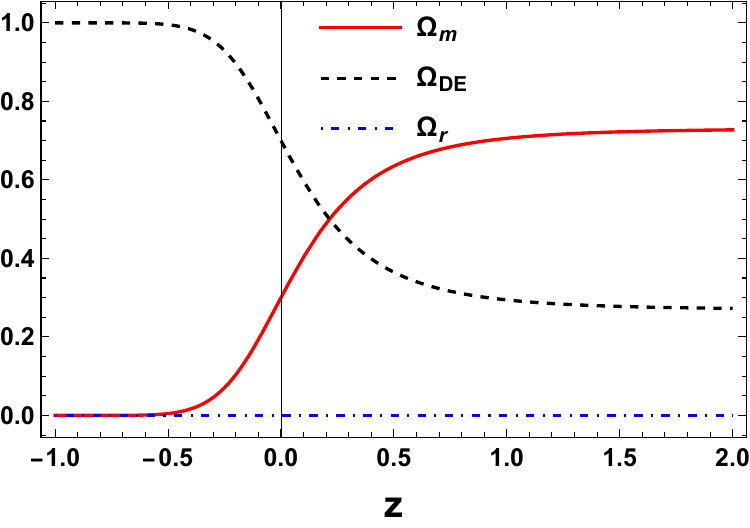} \hspace{0.2in} 
\includegraphics[scale=0.55]{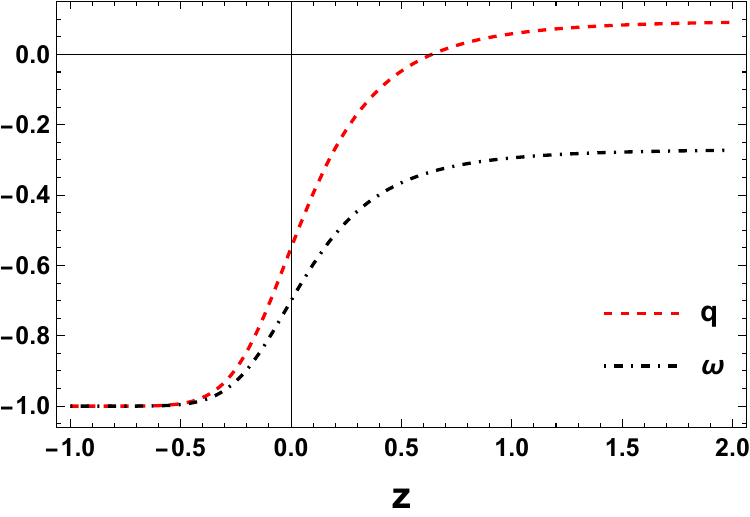}
\caption{\justifying The left panel displays density parameters for matter ($\Omega_m$), DE ($\Omega_{DE}$), and radiation ($\Omega_r$) as functions of redshift $z$. The right panel shows the deceleration parameter $q$ and the EoS parameter $\omega$ as functions of redshift $z$. These two panels illustrate the trajectory for the positive coupling parameter, which can transfer energy from DM to DE.}
\label{fig10}
\end{figure}

\begin{figure}[H]
\centering
\includegraphics[scale=0.55]{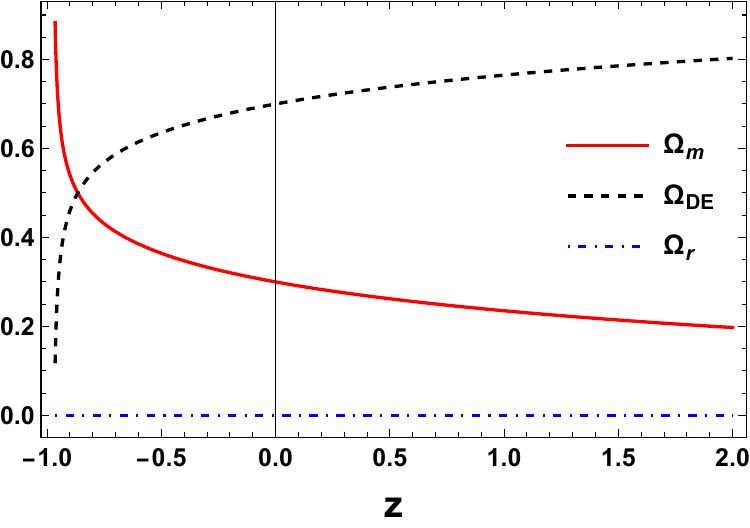} \hspace{0.2in}
\includegraphics[scale=0.55]{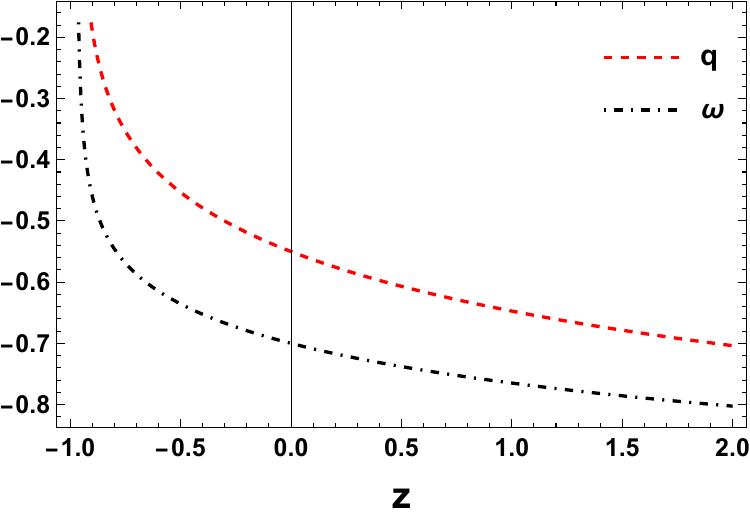}
\caption{\justifying The left panel displays density parameters for matter ($\Omega_m$), DE ($\Omega_{DE}$), and radiation ($\Omega_r$) as functions of redshift $z$. The right panel shows the deceleration parameter $q$ and the EoS parameter $\omega$ as functions of redshift $z$. These two panels illustrate the trajectory for the negative coupling parameter, which can transfer energy from DE to DM.}
\label{fig11}
\end{figure}

Corresponding to the fixed point $R_5$, that is
\begin{equation*}
    R_5: (x,y)=\left(\frac{\sqrt{3} \sqrt{-\alpha +3 n^2+2 \alpha  n}-3 n}{2 \alpha  n-\alpha },0\right)
\end{equation*}
the matter density, DE density, and effective EoS parameters are given by
 \begin{eqnarray*}
  &&  \Omega_m=\frac{3 n-\sqrt{-3 \alpha +9 n^2+6 \alpha  n}}{\alpha  (2 n-1)}+1\,, \\
  && \Omega_{DE} = \frac{\sqrt{-3 \alpha +9 n^2+6 \alpha  n}-3 n}{\alpha  (2 n-1)}\,,\,\,\\ \text{and}\,
  && w= \frac{-3 \alpha -4 \alpha  n^2-2 \sqrt{-3 \alpha +9 n^2+6 \alpha  n}+(8 \alpha +6) n}{\alpha  (1-2 n)^2}\,.
\end{eqnarray*}
The conditions for an accelerating Universe and the stability are depicted in Figures \ref{A2} and \ref{B2}. 
In Figures \ref{fig10} and \ref{fig11}, we illustrate the evolution of \(\Omega_m\), \(\Omega_r\), \(\Omega_{DE}\), \(q\), and \(\omega\) for different values of the model parameter \(n\) and the interaction parameter \(\alpha\). \\
For \(n = 3/2\) and \(\alpha = 4\) in Figure \ref{fig10}, \(\alpha > 0\) indicates the coupling term \(\mathcal{U} > 0\), which means the transfer of energy from DM to DE. This figure shows that the Universe was dominated by matter in the early stages and will be dominated by DE in later stages. Currently, the Universe is dominated by DE, with parameters \(\Omega_m = 0.3\), \(\Omega_r = 0.00005\), \(\Omega_{DE} = 0.7\), \(q_0 = -0.548\), and \(\omega_0 = -0.696\). For these values, the fixed points \(R_1\) and \(R_4\) are stable and represent the de Sitter and quintessence acceleration solutions, respectively.\\
In Figure \ref{fig11}, with \(n = 3/2\) and \(\alpha = -4\), \(\alpha < 0\) indicates the coupling term \(\mathcal{U} < 0\), which means energy transfers from DE to DM. This figure shows that the Universe was dominated by DE in the early stages and will be dominated by DM in the later stages. Currently, the Universe remains dominated by DE, with parameters \(\Omega_m = 0.3\), \(\Omega_r = 0.00005\), \(\Omega_{DE} = 0.7\), \(q_0 = -0.55\), and \(\omega_0 = -0.70\). For these values, the fixed points $R_4$ and $R_5$ have imaginary values, and the remaining point cannot exhibit stable behavior.

\begin{figure}[]
\centering
\includegraphics[scale=0.4]{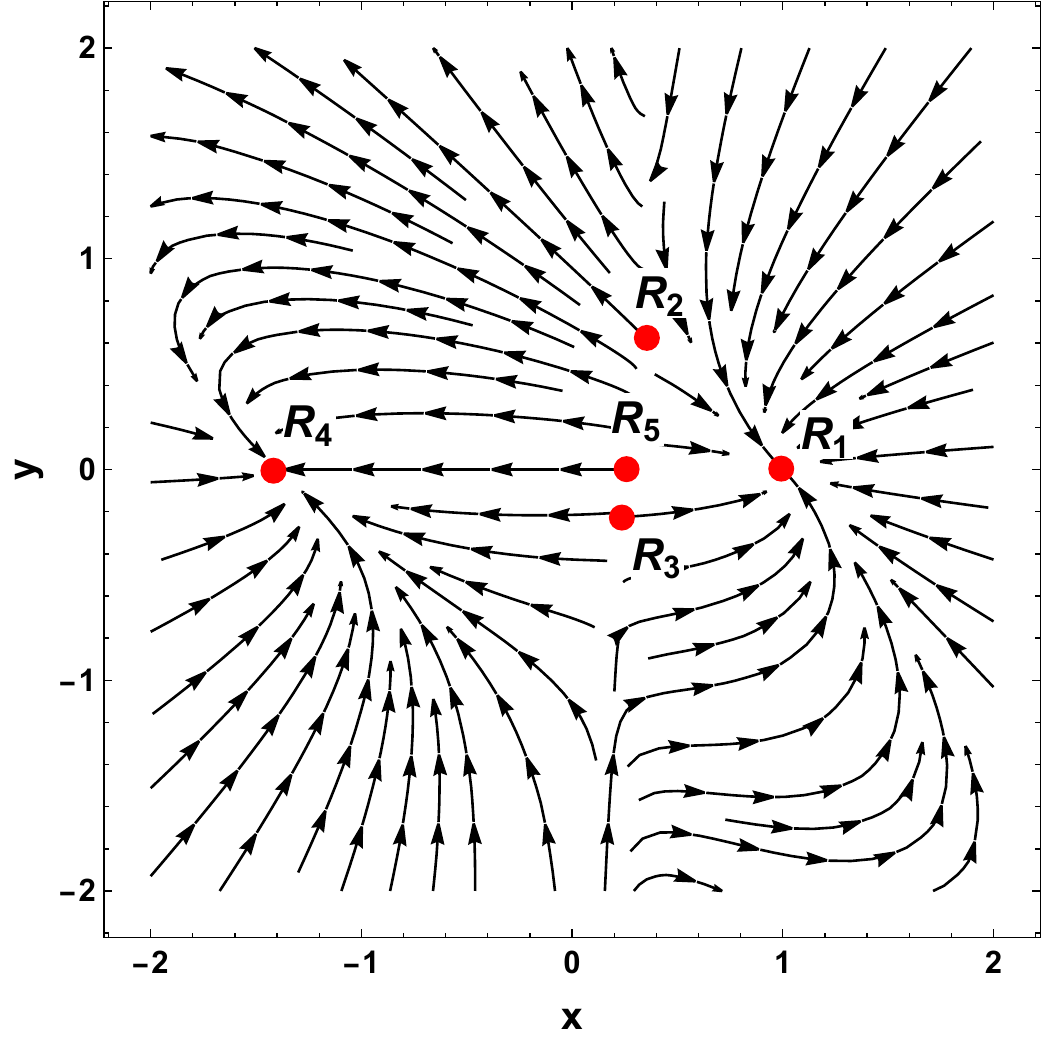} \hspace{0.2in}
\includegraphics[scale=0.4]{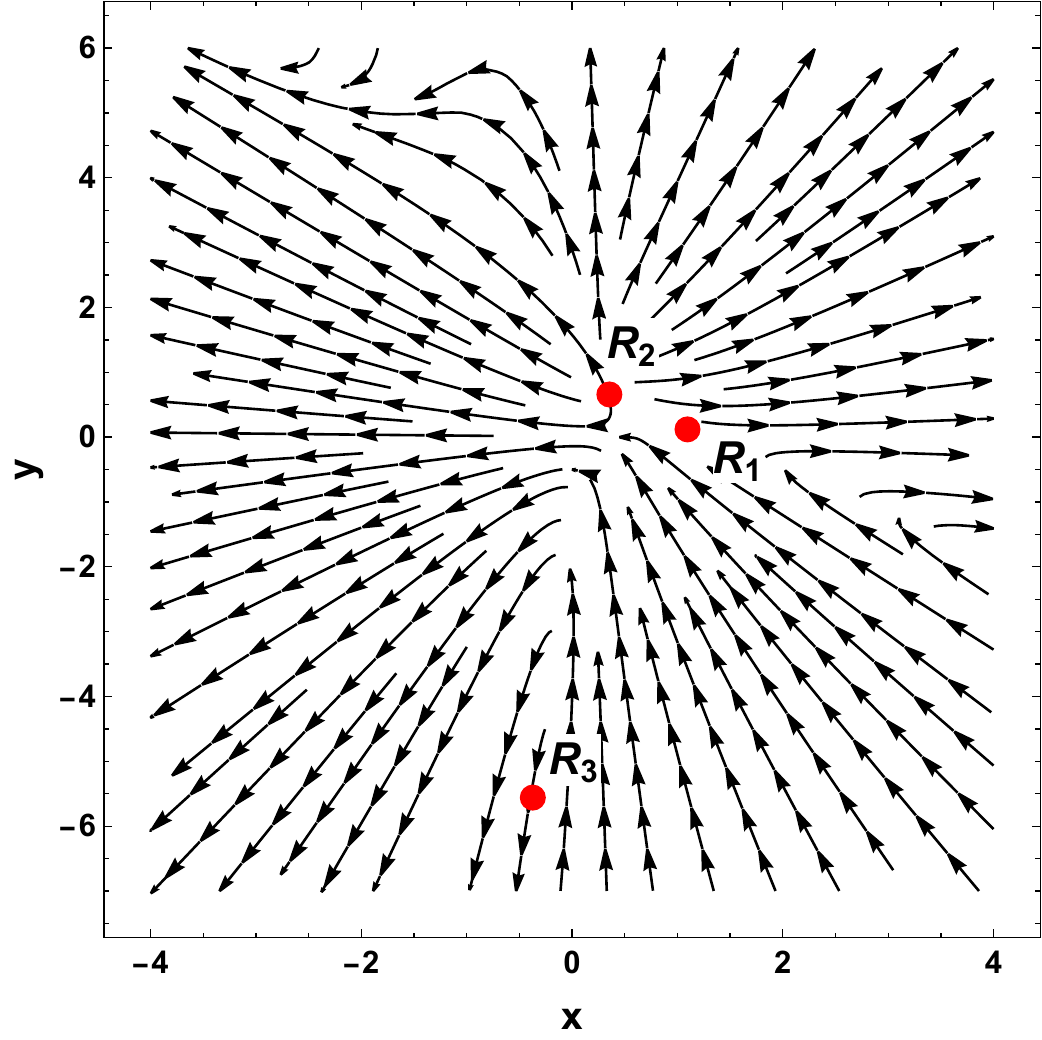}
\caption{Phase plot of $x$ versus $y$. These phase plots depict the stable behavior of $R_1$ and $R_4$ for $\alpha > 0$.}
\label{fig12}
\end{figure}

\section{Conclusion}
\label{5d}
In this chapter, we have explored the behavior of interacting DE and DM within $f(Q)$ gravity, employing a standard framework of dynamical system analysis. We have considered the power-law $f(Q)=6\gamma\, H_0^2\left(\tfrac{Q}{Q_0}\right)^n$ model incorporating with two different forms of interacting DE and DM: $3\alpha H\rho_m$ and $\frac{\alpha}{3H}\rho_m \rho_{DE}$. The parameter $\alpha$ in the interacting terms plays a crucial role in determining viable conditions and estimating the transition from the matter-dominated era (saddle point) to the DE-dominated era (stable node) in viable gravity models at late times. As a result, we have discovered fixed points that can be represented as the late-time accelerating Universe in $f(Q)$ gravity. For the form of ${\cal U}=3\alpha H\rho_m$, we have illustrated the evolution of \(\Omega_m\), \(\Omega_r\), \(\Omega_{DE}\), \(q\), and \(\omega\) for different values of the model parameter \(n\) and the interaction parameter \(\alpha\). For \(n = 3/2\) and \(\alpha = 1/2\) and \(\alpha > 0\), we found the coupling term \(\mathcal{U} > 0\), signifying energy transfer from DM to DE implying that the Universe was dominated by matter in the early stages and would be dominated by DE in later stages. With current data, the fixed point \(P_1\) is stable and represents the de Sitter acceleration solution, while the fixed point \(P_2\) is a saddle node, and \(P_3\) is an unstable node that cannot demonstrate universal acceleration. Moreover, we have considered another situation, of which \(n = 3/2\) and \(\alpha = -2\), \(\alpha < 0\). We discovered for the coupling term \(\mathcal{U} < 0\) that energy is transferred from DE to DM. In this case, the Universe was dominated by DE in the early stages and will be dominated by DM in the later stages. With the current data, the fixed point \(P_3\) is stable and exhibits acceleration for the early and present Universe, but fails to show acceleration for late times, while the fixed point \(P_1\) is a saddle node and \(P_2\) is an unstable node.

For the form of ${\cal U}=\frac{\alpha}{3H}\rho_m \rho_{DE}$, the evolution of \(\Omega_m\), \(\Omega_r\), \(\Omega_{DE}\), \(q\), and \(\omega\) has been examined for different values of the model parameter \(n\) and the interaction parameter \(\alpha\). We have considered \(n = 3/2\) and $\alpha=4$ (\(\alpha > 0\)) and found the coupling term \(\mathcal{U} > 0\), which signifies the transfer of energy from DM to DE. Our results show that the Universe was dominated by matter in the early stages and would be dominated by DE in later stages. Using observational data, the fixed points \(R_1\) and \(R_4\) were stable and would represent the acceleration solutions of de Sitter and quintessence. Additionally, for \(n = 3/2\) and $\alpha=-4$ (\(\alpha < 0\)), we found the coupling term \(\mathcal{U} < 0\), meaning energy transfers from DE to DM implying that the Universe was dominated by DE in the early stages and would be dominated by DM in later stages. Using the current data, the fixed points $R_4$ and $R_5$ have imaginary values, and the remaining point cannot exhibit stable behavior.

Although many viable models $f(Q)$ explain the DE problem in cosmology, our qualitative results from this work can serve as guidelines for more detailed studies. They can also be complementary constraints on viable $f(Q)$ models alongside other cosmological constraints on $f(Q)$ theories. Our framework, based on the cosmological dynamics of interacting DE and DM in $f(Q)$ gravity, constitutes the natural template beyond the standard gravity model, for example, teleparallel gravity and Gauss-Bonnet gravity, or even more generalizations. Advancing our understanding requires developing new theoretical tools to analyze how interactions influence linear and non-linear regimes. Precisely determining an interaction kernel is crucial as it could offer profound insights into fundamental physics, particularly the nature and properties of DM and DE.

\section{Appendix} \label{Appendix}
The stability conditions for the fixed point $R_3$ are as follows:
\begin{itemize}
 \item For Stable Node:
\begin{eqnarray}
&& \frac{1}{8}<n\leq 0.4393\,\, \text{and}\,\, 0<\alpha <\frac{1}{3} (8 n-1).
\end{eqnarray}
\item For Unstable Node:
\begin{eqnarray}
&& n=0\,\, \text{and}\,\, \left(-\frac{1}{3}<\alpha <0\,\,\text{or}\,\, 0<\alpha \leq 1\right),\\
&& n<0\,\,\text{and}\,\,\left(\frac{1}{3} (8 n-1)<\alpha <2 n\right),\\
&& 0<n\leq \frac{1}{8}\,\,\text{and}\,\,\left(\frac{1}{3} (8 n-1)<\alpha <0\right),\\
&& n>\frac{1}{2}\,\,\text{and}\,\, 2 n<\alpha <\frac{1}{3} (8 n-1).
\end{eqnarray}
\item However, the saddle point is otherwise.
\end{itemize}

The stability conditions for the fixed point $R_4$ are as follows:
\begin{itemize}
\item For Stable node:
\begin{eqnarray}
&& 0<n\leq \frac{1}{8}\,\,\text{and}\,\,0<\alpha <-\frac{3 n^2}{2 n-1},\\ 
&& \frac{1}{8}<n<\frac{1}{5}\,\, \text{and}\,\, \frac{1}{3} (8 n-1)<\alpha <-\frac{3 n^2}{2 n-1},\\
&& \frac{1}{2}<n\leq 1\,\, \text{and}\,\, \alpha >0,\\
&& n>1\,\, \text{and}\,\, \left(-\frac{3 n^2}{2 n-1}<\alpha <-3\,\, \text{or}\,\, \alpha >0\right).
 \end{eqnarray}
 
\item For Unstable node:
\begin{eqnarray}
&& n\leq \frac{1}{8}\,\,\text{and}\,\,\alpha <\frac{1}{3} (8 n-1),\\
&& \frac{1}{8}<n<\frac{1}{2}\,\,\text{and}\,\, \alpha <0.
\end{eqnarray}

\item However, the saddle point is otherwise.
\end{itemize}

The stability conditions for the fixed point $R_5$ are as follows:

\begin{itemize}
    \item For Stable node:
    \begin{eqnarray}
    && n\leq 0\,\,\text{and}\,\, \left(\alpha <-3 \,\, \text{or}\,\, 0<\alpha <-\frac{3 n^2}{2 n-1}\right),\\
    && 0<n<\frac{1}{2}\,\,\text{and}\,\, \alpha <-3,\\
    && \frac{1}{2}<n<1\,\,\text{and}\,\, -\frac{3 n^2}{2 n-1}<\alpha <-3.
    \end{eqnarray}

    \item For Unstable node:
    \begin{eqnarray}
       && \frac{1}{5}<n<\frac{1}{2}\,\,\text{and}\,\, \frac{1}{3} (8 n-1)<\alpha <-\frac{3 n^2}{2 n-1},\\
       && n>\frac{1}{2}\,\,\text{and}\,\, \alpha >\frac{1}{3} (8 n-1).
    \end{eqnarray}
 \item However, the saddle point is otherwise. 
\end{itemize}




\chapter{Conclusions and Future Perspectives} 

\label{Chapter6} 
\epigraph{\justifying \textit{``A theory can be proved by experiment; but no path leads from experiment to the birth of the theory."}}{\textit{Albert Einstein}}
\lhead{Chapter 6. \emph{Conclusions}} 


This thesis explored various aspects of modified gravity theory based on nonmetricity, i.e. $f(Q)$ gravity, as a promising framework to explain late-time cosmic acceleration. Our research spans multiple methodologies and reconstruction techniques, combining observational datasets with theoretical developments to investigate the viability and cosmological implications of $f(Q)$ gravity.

Chapter \ref{Chapter1} provided a detailed introduction to the foundational concepts and challenges in modern cosmology, emphasizing the motivation to explore alternative gravity theories. We outlined the mathematical framework of GR and the \( \Lambda \)CDM model while highlighting their limitations, such as the cosmological constant problem, the coincidence problem, and tensions such as the Hubble and \( \sigma_8 \) discrepancies. To address these challenges, we introduced a nonmetricity-based modification of GR as a promising alternative framework. Additionally, we present the use of GP for model-independent reconstruction of cosmological functions from OHD data. 

In chapter \ref{Chapter2}, we demonstrated that $f(Q)$ gravity can successfully mimic the $\Lambda$CDM expansion history of the Universe. To achieve this, we employed several cosmological reconstruction techniques. In the first approach, by assuming different fluid components, we derived a class of $f(Q)$ theories. In the second approach, using e-folding as a reconstruction parameter, we obtained a class of $f(Q)$ theories capable of describing various cosmological scenarios, including transitions to phantom phases, Big Rip singularities, and the standard $\Lambda$CDM epoch. This method offers significant flexibility in reconstructing any desired FLRW cosmology while maintaining consistency with observed cosmic evolution. Our findings reveal that these $f(Q)$ theories replicate the $\Lambda$CDM expansion history but remain indistinguishable from GR at the background level. This highlights the need to investigate perturbative effects, such as growth factor, structure formation, and gravitational waves, to distinguish these theories from GR. Furthermore, solar system tests could provide constraints on the $f(Q)$ models, ensuring their consistency across both local and cosmological scales.

In chapter \ref{Chapter3}, we independently reconstructed the $f(Q)$ function using observational Hubble measurements, including CC and BAO, through GPs analysis. Unlike traditional approaches that assume specific functional forms for $f(Q)$, our model-independent methodology derives $f(Q)$ directly from observational data without arbitrary assumptions. We reconstructed $H(z)$ and its derivative $H'(z)$, which allowed us to express the Friedmann equations in terms of the nonmetricity scalar $Q$. The reconstructed function $f(Q)$ resolves the $H_0$ tension in a model-independent manner, aligning well with recent precise estimates. Our analysis reveals a deviation from $\Lambda$CDM, suggesting a quadratic behavior for $f(Q)$, expressed as $f(Q) = -2\Lambda + \epsilon\,Q^2$, where the deviation parameter $\epsilon$ is constrained to $-4.809 \times 10^{-9} < \epsilon < 5.658 \times 10^{-10}$. We also tested two popular functional forms of $f(Q)$, tightening their parameter constraints compared to traditional observational limits. The cosmological implications, including deceleration parameters, dimensionless DE, and the DE equation of state, confirm the current accelerated expansion of the Universe consistent with recent studies.

In chapter \ref{Chapter4}, we extended our study to the early Universe, reconstructing the scalar potential $V(\phi)$ associated with early DE within $f(Q)$ gravity. Using GP analysis with CC and BAO data, we demonstrated a unified framework for both early-time inflationary dynamics and late-time acceleration. Our findings suggest that in the early Universe, the scalar field drives inflation, whereas at late times, cosmic acceleration is purely governed by the geometrical modifications of $f(Q)$ gravity. This distinction highlights the dynamical evolution of DE within the modified gravity framework.

In chapter \ref{Chapter5}, we explore the interacting DE and DM scenario within the power-law model $f(Q)$. Using dynamical system analysis, we examined two specific forms of interaction terms, $3\alpha H \rho_m$ and $\frac{\alpha}{3H} \rho_m \rho_{DE}$, and analyzed their cosmological consequences. Our results showed the existence of stable fixed points representing the late-time accelerated Universe, while the interaction parameter $\alpha$ plays a pivotal role in determining the energy transfer dynamics between DE and DM. These results provide qualitative insights into the evolution of cosmic components within interacting $f(Q)$ gravity and offer complementary constraints on viable models.

Overall, this thesis highlights the capability of $f(Q)$ gravity to describe various phases of the Universe's evolution, from early inflation to late-time acceleration. By adopting a combination of analytical and model-independent reconstruction techniques, we have derived novel insights into the functional forms of $f(Q)$, the behavior of scalar potentials, and the interplay between cosmic components in modified gravity.

Although the present work provides a comprehensive foundation for $f(Q)$ gravity and its cosmological implications, several open questions remain, paving the way for future research. A critical direction is the study of cosmological perturbations and structure formation, which can help distinguish $f(Q)$ gravity from GR by predicting unique observational signatures such as the growth rates of cosmic structures and gravitational waves. Extending our reconstruction techniques to non-flat FLRW cosmology is another promising avenue, as curvature effects introduce additional complexity into the nonmetricity scalar $Q$ and its governing equations. Furthermore, a rigorous analysis of the reconstructed scalar potential $V(\phi)$ in the early Universe, including its slow-roll dynamics and predictions for inflationary observables, will enhance compatibility with CMB data. Incorporating a wider range of observational datasets, such as Type Ia supernovae, gamma-ray bursts, and weak lensing surveys, will provide stronger constraints on $f(Q)$ models and help address persistent tensions in cosmological parameters.






\addtocontents{toc}{\vspace{2em}} 

\backmatter

%
\cleardoublepage
\pagestyle{fancy}
\label{References}
\lhead{\emph{References}}

\cleardoublepage
\clearpage




\cleardoublepage
\pagestyle{fancy}

\label{Publications}
\lhead{\emph{List of Publications}}

\chapter{List of Publications}
\section*{Thesis Publications}
\begin{enumerate}

\item \textbf{Gaurav N. Gadbail}, Sanjay Mandal, P.K. Sahoo, \textit{Reconstruction of $\Lambda$CDM Universe in $f(Q)$ Gravity}, \textcolor{blue}{Physics Letters B} \textbf{835}, 137509 (2022).

\item \textbf{Gaurav N. Gadbail}, Sanjay Mandal, P.K. Sahoo, \textit{Gaussian Process Approach for Model-Independent Reconstruction of $f(Q)$ Gravity with Direct Hubble Measurements}, \textcolor{blue}{Astrophysical Journal} \textbf{972}, 174 (2024).

\item \textbf{Gaurav N. Gadbail}, Sanjay Mandal, P.K. Sahoo, K. Bamba, \textit{Reconstruction of the Scalar Field Potential in Nonmetricity Gravity through Gaussian Processes}, \textcolor{blue}{Physics Letters B} \textbf{860}, 139232 (2024).

\item \textbf{Gaurav N. Gadbail}, Simran Arora, P. Channuie, P.K. Sahoo, \textit{Cosmological Dynamics of Interacting Dark Energy and Dark Matter in $f(Q)$ Gravity}, \textcolor{blue}{Fortschritte der Physik} \textbf{73}, 2400205 (2025).
\end{enumerate}

\section*{Other Publications}
\begin{enumerate}
\item \textbf{Gaurav N. Gadbail}, Simran Arora, P. K. Sahoo, K. Bamba, \textit{Reconstruction of the Singularity-free $f(\mathcal{R})$ Gravity via Raychaudhuri Equations,} \textcolor{blue}{European Physical Journal C} \textbf{84(7)}, 752 (2024).

\item \textbf{Gaurav N. Gadbail}, P. K. Sahoo, \textit{Modified $f(Q)$ Gravity Models and their Cosmological Consequences,} \textcolor{blue}{Chinese Journal of Physics} \textbf{89}, 1754-1762 (2024).

\item \textbf{Gaurav N. Gadbail}, Himanshu Chaudhary, Amine Bouali, P. K. Sahoo, \textit{Statistical and Observation Comparison of Weyl-Type  $f(Q,T)$ Models with the $\Lambda$CDM Paradigm,} \textcolor{blue}{Nuclear Physics B} \textbf{1009}, 116727 (2024).

\item \textbf{Gaurav N. Gadbail}, Avik De, P. K. Sahoo, \textit{Cosmological Reconstruction and $\Lambda$CDM Universe in  $f(Q,C)$ Gravity,} \textcolor{blue}{European Physical Journal C} \textbf{83(12)}, 1099 (2023).

\item \textbf{Gaurav N. Gadbail}, Ameya Kolhatkar, Sanjay Mandal, P. K. Sahoo, \textit{Correction to Lagrangian for Bouncing Cosmologies in  $f(Q)$ Gravity,} \textcolor{blue}{European Physical Journal C} \textbf{83(7)}, 595 (2023).

\item \textbf{Gaurav N. Gadbail}, Simran Arora, P. K. Sahoo, \textit{Reconstruction of $f(Q,T)$ Lagrangian for Various Cosmological Scenario,} \textcolor{blue}{Physics Letters B} \textbf{838}, 137710 (2023).

\item \textbf{Gaurav N. Gadbail}, Simran Arora, P. K. Sahoo, \textit{Cosmology with viscous Generalized Chaplygin Gas in $f(Q)$ Gravity,} \textcolor{blue}{Annals of Physics} \textbf{451}, 169269 (2023).

\item \textbf{Gaurav N. Gadbail}, Simran Arora, P. K. Sahoo, \textit{Dark Energy Constraint on Equation of State Parameter in the Weyl-type $f(Q,T)$ Gravity,} \textcolor{blue}{Annals of Physics} \textbf{451}, 169244 (2023).

\item \textbf{Gaurav N. Gadbail}, Simran Arora, P. K. Sahoo, \textit{Generalized Chaplygin Gas and Accelerating Universe in $f(Q,T)$ Gravity,} \textcolor{blue}{Physics of the Dark Universe} \textbf{37}, 101074 (2022).

\item \textbf{Gaurav N. Gadbail}, Sanjay Mandal, P.K. Sahoo, \textit{Parametrization of Deceleration Parameter in $f(Q)$ Gravity}, \textcolor{blue}{Physics} \textbf{4(4)}, 1403-1412 (2022).

\item \textbf{Gaurav N. Gadbail}, Simran Arora, Praveen Kumar, P. K. Sahoo, \textit{Interaction of Divergence-free Deceleration Parameter in Weyl-type $f(Q, T)$ Gravity}, \textcolor{blue}{Chinese Journal of Physics} \textbf{79}, 246-255 (2022).

\item \textbf{Gaurav N. Gadbail}, Simran Arora, P. K. Sahoo, \textit{Viscous Cosmology in the Weyl-type $f(Q,T)$ Gravity,} \textcolor{blue}{European Physics Journal C} \textbf{81}, 1088 (2021).

\item \textbf{Gaurav N. Gadbail}, Simran Arora, P. K. Sahoo, \textit{Power Law Cosmology in Weyl-type $f(Q,T)$ Gravity,} \textcolor{blue}{European Physical Journal Plus} \textbf{136}, 1040 (2021).

\end{enumerate}

\cleardoublepage
\pagestyle{fancy}

\label{Conferences/Workshops Attended}
\lhead{\emph{Conferences/Workshops Attended}}

\chapter{Conferences/Workshops Attended}
\section*{Paper presentated}
\begin{enumerate}
\item Presented a research paper entitled ``\textit{Gaussian Process Approach for Model-Independent Reconstruction of $f(Q)$ Gravity with Direct Hubble Measurements}” at the 33rd workshop on \textbf{General Relativity and Gravitation in Japan (JGRG33)} organized by the \textbf{Department of Physics, Kindai University, HigashiOsaka Campus, Osaka, Japan}, during \textcolor{blue}{2nd-6th December 2024}.

\item Presented a poster entitled ``\textit{Gaussian Process Approach for Model-Independent Reconstruction of $f(Q)$ Gravity with Direct Hubble Measurements}” at the program \textbf{3rd IAGRG School on Gravitation and Cosmology} organized in \textbf{ICTS Bengaluru, India}, during \textcolor{blue}{14th-25th October 2024}.

\item Presented a poster entitled ``\textit{Reconstruction of $\Lambda$CDM Universe in $f(Q)$ Gravity}" at the international conference of the \textbf{ Indian Mathematical Society} organized by the \textbf{Department of Mathematics, Birla Institute of Technology and Science, Pilani, Hyderabad Campus}, during \textcolor{blue}{22nd-25th December 2023}.

\item Presented a poster entitled ``\textit{Power law cosmology in Weyl-type $f(Q,T)$ gravity}”, at the international conference “\textbf{Indian Association for General Relativity and Gravitation (IAGRG22)}” organized by the \textbf{Indian Institute of Science Education and Research Kolkata}, during \textcolor{blue}{19th-21th December 2022}.

\item Presented a research paper entitled ``\textit{Viscous cosmology in the Weyl-type $f(Q,T)$ gravity}”, at the international conference “\textbf{International Conference on Mathematical Sciences and its Applications}” organized by the \textbf{Department of Mathematics, SRTM University, Nanded}, during \textcolor{blue}{28th-30th July 2022}.

\item Presented a research paper entitled ``\textit{Power law cosmology in Weyl-type $f(Q,T)$ gravity}” at the international conference “\textbf{Differential Geometry and its Applications}”, jointly organized by the \textbf{Department of Mathematics, Kuvempu University, Shivamogga, Karnataka}, and \textbf{The Tensor Society, Lucknow}, during \textcolor{blue}{4th-5th March 2022}.

\end{enumerate}

\section*{Workshop attended}
\begin{enumerate}
\item Participated in a national workshop on ``\textbf{Contemporary Issues in Astronomy and Astrophysics}” organized by the \textbf{Department of Physics, Shivaji University, Kolhapur, Maharashtra,} in collaboration with \textbf{IUCAA, Pune}, during \textcolor{blue}{13th–15th September 2024}.

\item Participated in an international workshop on ``\textbf{Astronomy Data Analysis with Python}”, jointly organized by the Department of Physics and the \textbf{Department of Mathematics, Maulana Azad National Urdu University (MANUU), Hyderabad}, and sponsored by \textbf{IUCAA, Pune}, during \textcolor{blue}{5th–8th September 2023}.

\end{enumerate}

\cleardoublepage
\pagestyle{fancy}
\lhead{\emph{Biography}}

\chapter{Biography}

\section*{Brief Biography of the Candidate:}
\textbf{Gaurav N. Gadbail} completed his Master of Science in Mathematics (4th Merit) in 2019 and Bachelor of Science in Mathematics, Physics, and Electronics in 2017 from Sant Gadge Baba Amravati University, Amravati, Maharashtra. He qualified for the Council of Scientific and Industrial Research (CSIR) National Eligibility Test (NET) and was awarded the Junior Research Fellowship (JRF) in February 2021 with an All India Rank of 148 in Mathematical Sciences. Additionally, he cleared the Graduate Aptitude Test in Engineering (GATE) in Mathematics in 2021 and the Maharashtra State Eligibility Test (MH-SET) for Assistant Professor in Mathematical Sciences, held on 27th December 2020. In 2024, Gaurav was awarded the BITS International Travel Grant to present his research at the 33rd Workshop on General Relativity and Gravitation (JGRG33) held in Japan. He has an impressive record of research publications in prestigious journals, including \textit{Astrophysical Journal, Physics Letters B, European Physical Journal C, Nuclear Physics B}, and \textit{Annals of Physics}. Gaurav has also actively presented his work at numerous international conferences.

\section*{Brief Biography of the Supervisor:}

\textbf{Prof. Pradyumn Kumar Sahoo} has over 24 years of immense research experience in Theoretical Cosmology, Astrophysical Objects, and Modified Theories of Gravity. He obtained his Ph.D. from Sambalpur University, Odisha, India, in January 2004. In 2009, he joined the Department of Mathematics at BITS Pilani, Hyderabad Campus, as an Assistant Professor and is currently a Professor. He is also an Associate Member of IUCAA, Pune. He is a recipient of the “Prof. S. Venkateswaran Faculty Excellence Award” 
 for the year 2022 from BITS Pilani. He has been awarded a visiting professor fellowship at Transilvania University of Brasov, Romania. According to a survey by researchers from Stanford University, he has been ranked among the top 2\% of scientists worldwide in the field of Nuclear and Particle Physics in the last five years. Throughout his career, he has published more than 250 research articles in various renowned national and international journals. As a visiting scientist, he had the opportunity to visit the European Organization for Nuclear Research (CERN) in Geneva, Switzerland, a renowned center for scientific research. He has participated in numerous national and international conferences, often presenting his work as an invited speaker. Prof. Sahoo has engaged in various research collaborations at both the national and international levels. He has contributed to BITS through five sponsored research projects: University Grants Commission
(UGC 2012-2014), DAAD Research Internships in Science and Engineering (RISE) Worldwide (2018, 2019, 2023, and 2024), Council of Scientific and Industrial Research (CSIR 2019-2022), National Board for Higher Mathematics (NBHM 2022-2025), Anusandhan National Research Foundation (ANRF), and the Department of Science and Technology (DST 2023-2026). He also serves as an expert reviewer for Physical Science Projects for SERB, DST (Government of India), and UGC research schemes. Additionally, he is an editorial board member for various reputable journals, contributing to the research community.

\end{document}